\documentclass[aps,prx,twocolumn,10pt,superscriptaddress,longbibliography,floatfix,footinbib]{revtex4-2}

\usepackage[T1]{fontenc}
\usepackage[utf8]{inputenc}
\usepackage{amsmath,amsthm,mathtools}
\usepackage{newtxtext}
\usepackage{newtxmath}
\usepackage{graphicx}
\usepackage{xcolor}
\usepackage{microtype}
\usepackage{hyperref}
\hypersetup{colorlinks=true,linkcolor=blue!45!black,citecolor=blue!45!black,urlcolor=blue!45!black,
pdfauthor={Yixuan Ma, Weikang Li, Dong-Ling Deng},
pdftitle={Static Bounds on Thermal Fast Scrambling}}
\newtheorem{maintheorem}{Theorem}
\newtheorem{mainproposition}{Proposition}
\newtheorem{lemma}{Lemma}[section]

\newtheorem{proposition}[lemma]{Proposition}
\newtheorem{corollary}[lemma]{Corollary}
\theoremstyle{definition}

\theoremstyle{remark}
\newtheorem{remark}[lemma]{Remark}
\theoremstyle{plain}
\newcommand{\Tr}{\operatorname{Tr}}
\newcommand{\E}{\mathbb E}
\renewcommand{\Re}{\operatorname{Re}}
\renewcommand{\Im}{\operatorname{Im}}
\newcommand{\ii}{\mathrm i}
\newcommand{\dd}{\mathrm d}
\newcommand{\ad}{\operatorname{ad}}
\newcommand{\cJ}{\mathcal J}
\newcommand{\cN}{\mathcal N}
\newcommand{\cE}{\mathcal E}
\newcommand{\Cr}{C^{\mathrm r}}
\newcommand{\Cu}{C^{\mathrm u}}
\newcommand{\Cbar}{\overline{C}}
\newcommand{\Cdbar}{\overline{\overline{C}}}
\newcommand{\one}{\mathbf 1}
\newcommand{\norm}[1]{\lVert #1\rVert}
\newcommand{\mainfigureone}{%
\begin{figure*}[!t]
\centering
\includegraphics[width=0.99\textwidth]{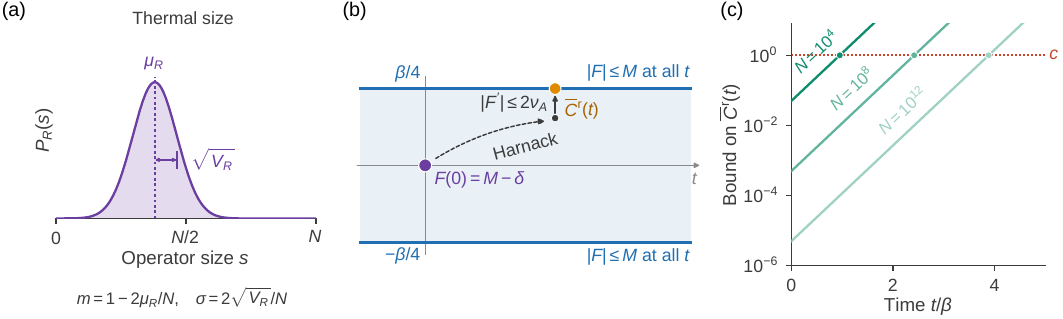}
\caption{%
Thermal size and the chaos strip.
(a)~The thermal-size distribution $P_R(s)$ of $R=\sqrt\rho$ (schematic), with mean $\mu_R$ and standard deviation $\sqrt{V_R}$. For equally weighted Majorana targets and an even Hamiltonian these give $m=1-2\mu_R/N$ and $\sigma=2\sqrt{V_R}/N$.
(b)~The chaos strip $|\Im z|\le\beta/4$ of the quartered OTOC $F$ (schematic). On both edges $|F|$ stays below the edge bound $M$ at all real times. At the origin $F$ falls short of $M$ by the static defect $\delta$. The Harnack inequality carries this defect into the strip, and a bound on the derivative of $F$ carries it to the upper edge. There the raw regularized OTOC equals $2T-2\Re F$, where the edge function $T(t)$ is at most $M$.
(c)~The bound of Theorem~\ref{thm:static} with $\beta \nu_A=1$ and the illustrative choice $\delta=1/N$ for three system sizes, against time in units of $\beta$. The dotted line is the threshold $c=1$, and dots mark the first times at which the bound reaches it. These times are $[\beta/(2\pi)]\log N-(\beta/\pi)\log5$ and move by $[\beta/(2\pi)]\log10^4$ from one curve to the next.
}\label{fig:strip}
\end{figure*}}

\newcommand{\mainfiguretwo}{%
\begin{figure*}[!t]
\centering
\includegraphics[width=0.99\textwidth]{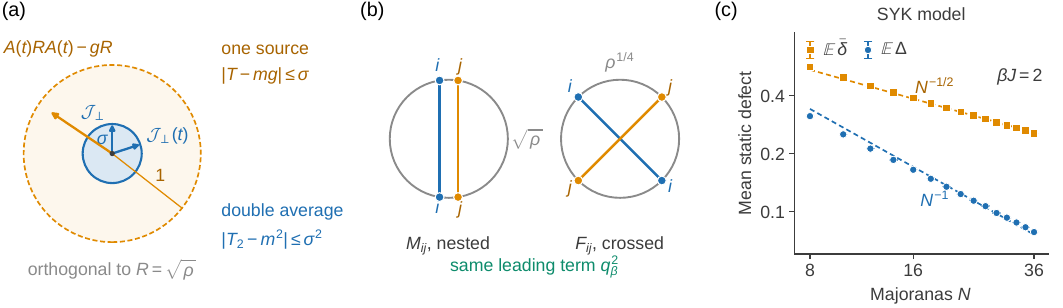}
\caption{%
The double average and static defects in the SYK model.
(a)~Edge functions in the Hilbert--Schmidt space orthogonal to $R=\sqrt\rho$ (schematic). For a single source, the deviation of $T(t)$ from $mg$ is the overlap of $\cJ_\perp=\cJ(R)-mR$, of norm $\sigma$, with the component of $A(t)RA(t)$ orthogonal to $R$, which stays within the unit ball (orange), allowing $|T(t)-mg|$ up to $\sigma$. For the double average, the evolving operator is $\cJ_\perp(t)$ on the sphere of radius $\sigma$ (blue), so the autocorrelation $T_2(t)$ deviates from $m^2$ by at most $\sigma^2$ and is maximized at $t=0$.
(b)~Euclidean four-point functions entering $\Delta=\norm{\cJ(R)}_2^2-F_2(0)$ for $i\ne j$ on the thermal circle of circumference $\beta$ (schematic). Equal Majorana labels are paired in nested order in $M_{ij}$ and in crossed order in $F_{ij}$. In the SYK model, each pair spans half the circle ($\beta/2$), so the annealed averages $\Phi(M_{ij})$ and $\Phi(F_{ij})$ (Sec.~\ref{sec:syk}) share the same leading term $q_\beta^2$, the square of the large-$N$ two-point function at separation $\beta/2$, which cancels in $\Phi(\Delta)$.
(c)~Finite-temperature static defects of the SYK model (Eq.~\eqref{eq:syk}) at $\beta J=2$ with equal source and target weights $1/N$ on all $N$ Majorana modes: disorder averages of the mean single-source defect $\bar\delta=N^{-1}\sum_i\delta_i$ (orange squares) and the double-averaged defect $\Delta$ (blue circles) for $N=8,\dots,36$. Data are obtained by exact diagonalization for $N\le24$ ($64$ independent samples at $N=8,\dots,20$ and $32$ at $N=22,24$) and by preassembled sparse Hamiltonians with batched Chebyshev imaginary-time propagation and stochastic trace estimation for $N=26,\dots,36$ ($16$ samples per size), with standard errors. Dashed $N^{-1/2}$ and $N^{-1}$ guides are anchored at $N=24$. Computational details are given in Appendix~\ref{app:sec-numerics}.
}\label{fig:double}
\end{figure*}}

\newcommand{\mainfigurethree}{%
\begin{figure*}[!t]
\centering
\includegraphics[width=0.99\textwidth]{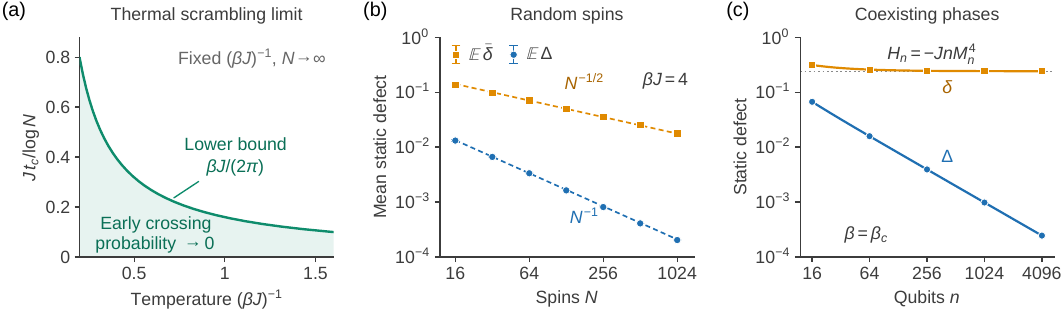}
\caption{%
Thermal scrambling limits and static defects in spin systems.
(a)~Leading scrambling-time lower bound $\beta J/(2\pi)$ on $Jt_c/\log N$ for random $k$-local spin systems (Eq.~\eqref{eq:spins}, Theorem~\ref{thm:spins}) against temperature $(\beta J)^{-1}$, where $t_c$ is the first crossing time of a fixed threshold $c>0$ by the target-averaged regularized OTOC of a fixed single-site source (green curve). For each fixed $(\beta J)^{-1}>0$ and $0<\epsilon<1$, the probability of crossing before $(1-\epsilon)[\beta/(2\pi)]\log N$ tends to zero as $N\to\infty$ (shaded region).
(b)~Finite-temperature static defects of the random two-body spin model in Eq.~\eqref{eq:spins} with $k=d=2$ and $H_{ij}=(\epsilon_{ij}Z_iZ_j+\eta_{ij}X_i+\xi_{ij}X_j)/3$ (independent equiprobable signs) at $\beta J=4$, with equal source and target weights $1/N$ on single-site probes $A_i=B_i=Z_i$: disorder averages of the mean single-source defect $\bar\delta=N^{-1}\sum_i\delta_i$ (orange squares) and the double-averaged defect $\Delta$ (blue circles) for $N=16,\dots,1024$. Data are obtained by worldline quantum Monte Carlo ($32$ independent samples per size), with standard errors. Dashed $N^{-1/2}$ and $N^{-1}$ guides are anchored at $N=64$.
(c)~Static defects at phase coexistence ($\beta_cJ\simeq0.688801$) in the mean-field spin model $H_n=-JnM_n^4$ of $n=16,\dots,4096$ qubits, computed by exact Gibbs sums with equal source and target weights $1/(2n)$ on all $N=2n$ Majorana probes: the single-source defect $\delta$ (orange) tends to $\sqrt v-v>0$ (dotted line), while the double-averaged defect $\Delta$ (blue) obeys $\Delta=1/n+O(n^{-2})$ (Appendix~\ref{app:sec-coexist}). Computational details are given in Appendix~\ref{app:sec-numerics}.
}\label{fig:spins}
\end{figure*}}

\makeatletter
\def\frontmatter@thefootnote{\altaffilletter@sw{\@alph}{\@fnsymbol}{\numexpr\csname c@\@mpfn\endcsname+1\relax}}
\makeatother

\begin{document}
\title{Static Bounds on Thermal Fast Scrambling}

\newcommand{\AffTsinghua}{Center for Quantum Information, IIIS, Tsinghua University, Beijing 100084, China}
\newcommand{\AffTsinghuaCollegeAI}{College of Artificial Intelligence, Tsinghua University, Beijing 100084, China}
\newcommand{\AffQizhi}{Shanghai Qi Zhi Institute, Shanghai 200232, China}
\newcommand{\AffHefei}{Hefei National Laboratory, Hefei 230088, China}

\author{Yixuan Ma}
\affiliation{\AffTsinghua}

\author{Weikang Li}
\email{weikang\_li@tsinghua.edu.cn}
\affiliation{\AffTsinghua}
\affiliation{\AffHefei}

\author{Dong-Ling Deng}
\email{dldeng@tsinghua.edu.cn}
\affiliation{\AffTsinghua}
\affiliation{\AffQizhi}
\affiliation{\AffHefei}

\begin{abstract}
\noindent Scrambling, the process by which initially localized quantum information spreads across an interacting many-body system, is a hallmark of quantum chaos and plays a central role in understanding black hole dynamics. In thermal systems, the chaos bound of Maldacena, Shenker and Stanford limits the exponential growth rate of out-of-time-order correlations to $2\pi/\beta$, assuming that time-ordered correlation functions factorize after a dissipation time. Establishing the corresponding rigorous logarithmic lower bounds on scrambling times at arbitrary fixed temperature has remained an open problem, most notably in disordered systems, where the Gibbs state depends on the same random couplings that drive the dynamics. Here we prove such bounds at every fixed temperature for the Sachdev--Ye--Kitaev (SYK) model and for random $k$-local spin systems, without assuming real-time factorization. Our approach uses the operator-size distribution of the square root of the Gibbs state to quantify how local probes modify its structure. This static information bounds the thermal correlator uniformly in time along the edges of its analytic strip in complex time, and analyticity then limits how rapidly scrambling can develop. For both families of models, we prove that with probability tending to one the scrambling time is at least $(1-\epsilon)[\beta/(2\pi)]\log N$ for any $0<\epsilon<1$, with the coefficient $\beta/(2\pi)$ suggested by the chaos bound. The bound holds for each fixed pair of distinct Majorana modes in the SYK model and for the target average of any fixed single-site source in the spin model. These results place the thermal fast-scrambling timescale on a rigorous footing and show how static properties of the Gibbs state can constrain real-time information spreading.

\end{abstract}
\maketitle

\section{Introduction}

Scrambling is the process in which initially localized information spreads across $N$ interacting degrees of freedom until it becomes hidden in global correlations and invisible to local probes. The fast-scrambling conjecture posits that even the fastest scramblers in nature, such as black holes, require a time of order $\beta\log N$ at inverse temperature $\beta$~\cite{Hayden2007Black,Sekino2008Fast,Lashkari2013Towards}. The spread of a local perturbation is diagnosed by the growth of its squared commutators (or anticommutators) with other local operators, measured by out-of-time-order correlators (OTOCs)~\cite{Larkin1969Quasiclassical,Shenker2014Black,Kitaev2015Simple,Swingle2016Measuring,Nahum2018Operator,vonKeyserlingk2018Operator,Xu2024Scrambling}. In thermal states, Maldacena, Shenker and Stanford (MSS) bounded the Lyapunov exponent extracted from a regularized OTOC by $2\pi/\beta$, assuming that time-ordered thermal correlators factorize after a dissipation time~\cite{Maldacena2016Bound}, and the Sachdev--Ye--Kitaev (SYK) model approaches this rate in the strong-coupling limit~\cite{Kitaev2015Simple,Sachdev1993Gapless,Maldacena2016Remarks,Kitaev2018Soft}. For random all-to-all models, concentration methods establish logarithmic lower bounds on scrambling times defined by local OTOCs in coupling-independent states, including the infinite-temperature state~\cite{Chen2021Concentration}. Yet both the chaos bound and the fast-scrambling conjecture concern thermal states. Do analogous scrambling-time bounds hold in the Gibbs state of the same random Hamiltonian at every fixed temperature, without assuming real-time factorization? Can one rigorously establish such a bound with the coefficient $\beta/(2\pi)$ suggested by the MSS bound?

At the microscopic level, scrambling is driven by operator growth, yet directly bounding this growth in thermal states is obstructed by the coupling between the Gibbs state and the dynamics. Under Heisenberg evolution $e^{\ii tH}$, a simple operator branches into products of increasingly many elementary operators; at infinite temperature, probe-averaged OTOCs measure its growing average size, whereas at finite temperature thermal weights dress the correlator~\cite{Roberts2018Operator,Qi2019Quantum}. Lieb--Robinson-type arguments provide state-independent bounds on operator growth~\cite{Lieb1972Finite,Chen2021OperatorGrowth,Chen2023Speed} and control infinite-temperature OTOCs in two-body all-to-all spin models~\cite{Yin2020Bound}, where the extension to finite temperature was singled out as an important and challenging open problem. A universal finite-temperature speed limit constrains sustained scrambling through return probabilities and entanglement, but does not directly bound local thermal OTOCs~\cite{Vikram2026Proof}. Likewise, the concentration bounds available for random all-to-all systems control OTOCs in coupling-independent states, and do not directly apply to the Gibbs state $\rho = e^{-\beta H}/Z$ of the same Hamiltonian, whose thermal weights are correlated with the couplings~\cite{Chen2021Concentration}.

Complex-time analyticity offers a complementary route to controlling local thermal OTOCs. The MSS argument~\cite{Maldacena2016Bound} uses analyticity in an imaginary-time strip, which we call the chaos strip, to limit real-time growth. To bound the strip edges, it relies on approximate factorization of time-ordered correlators after a dissipation time. An alternative derivation obtains the chaos bound without factorization under an exponential-growth ansatz for a family of OTOC regularizations~\cite{Tsuji2018Bound}. Bounded analyticity in a strip of fixed width, together with inverse-polynomial suppression on a fixed real-time interval, guarantees smallness along the real axis only up to times of order $\log\log N$ (Appendix~\ref{app:sec-strip-limit}). Meanwhile, rigorous results on the SYK Gibbs state cover the free energy~\cite{Gamarnik2026Free}, thermal expectations~\cite{Zlokapa2026Rigorous}, and fast mixing of Gibbs samplers~\cite{Cai2026Fast} at high temperature. Equilibrium control now reaches every fixed temperature as well, yielding free-energy concentration~\cite{Anschuetz2025Strongly}, thermodynamics~\cite{He2026Free}, and a quantum cavity method for local thermal expectations~\cite{Schmidhuber2026Polynomial}. Yet rigorous logarithmic lower bounds with coefficient $\beta/(2\pi)$ on the first-crossing times of local thermal OTOCs at arbitrary fixed temperature remain open for these random models.

In this work, we show that static properties of the Gibbs state at $t = 0$ suffice to bound real-time scrambling at all times, without assuming real-time factorization. Given a local perturbation (the source) and local probes (the targets), we characterize the equilibrium state by its \emph{thermal size}, the operator-size distribution of $\sqrt{\rho}$ as seen by these probes. Along the edges of the chaos strip, the mean and width of this distribution bound the analytically continued four-point correlator uniformly for all real times. For every finite-dimensional Hamiltonian, analyticity in the strip then converts the gap $\delta$ between this edge bound and the four-point correlator at $t = 0$, which we call the \emph{static defect}, into a scrambling-time lower bound $[\beta/(2\pi)]\log(1/\delta)$, up to a shift set by the threshold and the thermal local energy of the source (Theorem~\ref{thm:static}). Moreover, averaging the OTOC over the same probes and weights for both sources and targets (Proposition~\ref{prop:double}) sharpens this edge bound so that the width of the thermal-size distribution enters only through its square. The resulting double-averaged defect $\Delta$ is again purely static, given at $t = 0$ by the difference between two Euclidean four-point functions on the thermal circle.

This static-to-dynamic framework yields rigorous logarithmic scrambling-time lower bounds at every fixed temperature for both the SYK model and random $k$-local spin systems. First, for the SYK model of $N$ Majorana fermions (Theorem~\ref{thm:syk}), Euclidean four-point estimates from the quantum cavity method~\cite{Schmidhuber2026Polynomial} show that the leading contributions to $\Delta$ cancel, and free-energy concentration~\cite{Anschuetz2025Strongly} transfers the resulting estimate to the disorder average of the same-sample Gibbs defect, giving $\E\Delta = O(1/N)$. Combining this static estimate with local-energy bounds~\cite{Cai2026Fast}, we prove that, with probability tending to one, both regularized and unregularized OTOCs between any two fixed distinct Majorana modes stay below any fixed threshold up to time $(1-\epsilon)[\beta/(2\pi)]\log N$ for every $0 < \epsilon < 1$, matching the scrambling-time coefficient suggested by the MSS bound. Second, in all-to-all random $k$-local spin models (Theorem~\ref{thm:spins}), we bound the double-averaged defect directly in the full Gibbs state: a boundary derivative on the chaos strip relates this defect to local double commutators, and independence and centering of the interactions give $\E\Delta=O(1/N)$. Isolating a fixed source from the bath and comparing its dynamics with the bath correlators controlled by Proposition~\ref{prop:double} then yields the same leading coefficient $\beta/(2\pi)$ for the target-averaged OTOC of the source. When the coupling distribution is invariant under permuting the other sites, this lower bound holds for each fixed target site without averaging over targets.

\section{Thermal OTOCs and main results}\label{sec:results}

Let $H$ be a Hamiltonian on a finite-dimensional space, $\rho = e^{-\beta H}/Z$ its Gibbs state at inverse temperature $\beta$, and $X(z) = e^{\ii zH}Xe^{-\ii zH}$ the Heisenberg evolution of an operator $X$, defined for complex times $z$ as well. A source $A$ is perturbed and targets $B_j$ are probed. Both are Hermitian contractions, $\norm{A}, \norm{B_j} \le 1$, and $w_j \ge 0$ are weights with $\sum_j w_j = 1$. We take the commutator $K_j(t) = \ii[A(t), B_j]$ whenever at least one probe is bosonic (a spin or an even fermionic operator), and the anticommutator $K_j(t) = \{A(t), B_j\}$ for two odd Majorana probes. Within any target or double average below, all terms use the same convention. Both are Hermitian, and both vanish at $t=0$ for distinct degrees of freedom. We bound two thermal OTOCs,
\begin{equation}
\Cr_j(t)=\Tr\bigl(\sqrt\rho K_j(t)\sqrt\rho K_j(t)\bigr),\qquad
\Cu_j(t)=\Tr\bigl(\rho K_j(t)^2\bigr),
\label{eq:otocs}
\end{equation}
the regularized OTOC, in which the thermal state is split into two halves, and the unregularized one~\cite{Liao2018Nonlinear,RomeroBermudez2019Regularization}. Both are raw, that is, not divided by thermal norms of the probes, and $0 \le \Cr_j \le \Cu_j \le 4$. We write $\Cbar^{\mathrm r}(t) = \sum_j w_j \Cr_j(t)$ for the target average. For a fixed threshold $c > 0$, we define the scrambling time as the first crossing time $t_c = \inf\{t \ge 0 : \Cbar^{\mathrm r}(t) \ge c\}$.

\textit{Thermal size.} For $N$ Majorana operators $\gamma_a$, with $\{\gamma_a,\gamma_b\}=2\delta_{ab}$, every operator is a combination of products $\gamma_S$ of distinct Majorana operators, and its size distribution weights $|S|$ by the squared coefficient of $\gamma_S$~\cite{Roberts2018Operator,Qi2019Quantum}. We call the size distribution of $\sqrt\rho$ the thermal size. Viewed as a state of two copies of the system, $\sqrt\rho$ is the thermofield double state, so the thermal size is the operator size of the thermofield double~\cite{Qi2019Quantum}. The targets read off the thermal size through the conjugation average
\begin{equation}
\cJ(X)=\sum_jw_jB_jXB_j .
\label{eq:conjugation}
\end{equation}
With $B_j=\gamma_j$ and equal weights, conjugation $\gamma_j\gamma_S\gamma_j$ yields $(-1)^{|S|-1}\gamma_S$ for the $|S|$ indices $j\in S$ and $(-1)^{|S|}\gamma_S$ for the $N-|S|$ indices $j\notin S$; averaging over $j$ therefore multiplies $\gamma_S$ by $(-1)^{|S|}(1-2|S|/N)$, which on even operators is $1-2\cN/N$, where the size superoperator $\cN$ multiplies $\gamma_S$ by $|S|$. For general targets we define, with $R=\sqrt\rho$,
\begin{equation}
m=\Tr\bigl(R\,\cJ(R)\bigr),\qquad\sigma=\norm{\cJ(R)-mR}_2,
\label{eq:size-width}
\end{equation}
where $\norm{X}_2=(\Tr X^\dagger X)^{1/2}$. Since $\cJ$ is self-adjoint for the Hilbert--Schmidt inner product and $\norm{R}_2=1$, $m$ and $\sigma$ are the mean and standard deviation of the spectral distribution of $\cJ$ in $R$; this is the thermal size as seen by the targets. For Majorana fermions and an even Hamiltonian, $m=1-2\mu_R/N$ and $\sigma=2\sqrt{V_R}/N$, where $\mu_R$ and $V_R$ are the mean and the variance of the thermal size, so $\sigma$ is the size width in units of $N/2$ (Fig.~\ref{fig:strip}(a)).

\textit{Chaos strip.} The quartered OTOC splits the thermal state into four quarters,
\begin{equation}
F(z)=\tau\sum_jw_j\Tr\bigl(yA(z)yB_jyA(z)yB_j\bigr),\qquad y=\rho^{1/4},
\label{eq:quartered}
\end{equation}
where $\tau = +1$ for commutators and $\tau = -1$ for anticommutators; this sign accounts for the different signs of the out-of-time-order cross terms in $\Cr_j(t)$. The function $F(z)$ is analytic in the chaos strip $|\Im z| \le \beta/4$, the strip of the MSS argument. With $g = \Tr(RARA)$, which fixes the projection of $A(t)RA(t)$ onto $R$, we call $M = mg + \sigma$ the edge bound and
\begin{equation}
\delta=M-F(0)
\label{eq:defect}
\end{equation}
the static defect. Like $g$, both are static quantities, computed from the Gibbs state and the probes at $t=0$. The source enters the dynamics through its thermal local energy $\nu_A = \norm{\ii[H,A]}_\rho$, with $\norm{X}_\rho = (\Tr\,\rho X^\dagger X)^{1/2}$, which controls the derivative of $F$ along the strip (Appendix~\ref{app:sec-strip}). Because $[H, A]$ involves only the interactions meeting the support of $A$, we have $\nu_A=O(J)$ whenever this local commutator is bounded on an $N$-independent scale $J$. This holds deterministically when the interactions meeting $A$ have $O(J)$ total norm, on a high-probability event in the SYK model, and in mean square ($\E \nu_A^2=O(J^2)$) for random spins.

\begin{maintheorem}[Thermal size bounds the chaos strip]\label{thm:static}
For every Hamiltonian on a finite-dimensional space, every $\beta>0$, Hermitian contractions $A$ and $B_j$ and weights $w_j$, with $\tau$ matched to the commutator or anticommutator, the static defect is nonnegative and
\begin{equation}
\Cbar^{\mathrm r}(t)\le(4+\beta \nu_A)\sqrt\delta\,e^{\pi|t|/\beta}
\label{eq:main-static}
\end{equation}
for all real $t$. In particular, the first crossing time of a threshold $c$ obeys
\begin{equation}
t_c\ge\frac\beta{2\pi}\log\frac1\delta-\frac\beta\pi\log\frac{4+\beta \nu_A}c .
\label{eq:main-crossing}
\end{equation}
\end{maintheorem}

\noindent Theorem~\ref{thm:static} is deterministic: it reduces a lower bound on the scrambling time to a static question, namely how small the defect $\delta$ is.

\textit{Double average.} When the sources range over the targets with the same weights, a smaller static quantity suffices. We write $\Cr_{ij}$ for the regularized OTOC of the source $B_i$ and the target $B_j$, and $\Cdbar^{\mathrm r}(t)=\sum_{i,j}w_iw_j\Cr_{ij}(t)$ for the double average. With the quartered OTOC averaged over both probes,
\begin{equation}
F_2(z)=\tau\sum_{i,j}w_iw_j\Tr\bigl(yB_i(z)yB_jyB_i(z)yB_j\bigr),
\label{eq:quartered-double}
\end{equation}
we call $\Delta=\norm{\cJ(R)}_2^2-F_2(0)$ the double-averaged defect.

\begin{mainproposition}[Double average]\label{prop:double}
For every Hamiltonian on a finite-dimensional space, every $\beta>0$, Hermitian contractions $B_j$ and weights $w_j$, with $\tau$ matched to the commutator or anticommutator, the double-averaged defect is nonnegative and
\begin{equation}
\Cdbar^{\mathrm r}(t)\le(4+\beta\bar\nu)\sqrt\Delta\,e^{\pi|t|/\beta}
\label{eq:main-double}
\end{equation}
for all real $t$, where $\bar\nu=\sum_iw_i\nu_{B_i}$ is the average thermal local energy of the sources.
\end{mainproposition}

\noindent Averaged over the sources, the edge bound of Theorem~\ref{thm:static} is $m^2+\sigma$. Proposition~\ref{prop:double} replaces it by $\norm{\cJ(R)}_2^2=m^2+\sigma^2$, so the size width enters only through its square, turning a width $\sigma$ of order $N^{-1/2}$ into a contribution of order $1/N$; moreover, $\Delta$ never exceeds the average static defect (Appendix~\ref{app:sec-strip}).

\textit{SYK model.} The SYK Hamiltonian couples an even number $N$ of Majorana operators in groups of four~\cite{Sachdev1993Gapless,Kitaev2015Simple,Maldacena2016Remarks},
\begin{equation}
H_N=J\sqrt{\frac6{N^3}}\sum_{|E|=4}g_E\Gamma_E,
\label{eq:syk}
\end{equation}
where $\Gamma_E$ is the product of the four Majorana operators in $E$, in increasing order, and the couplings $g_E$ are independent standard Gaussians.

\begin{maintheorem}[SYK model]\label{thm:syk}
Fix $\beta J>0$, $c>0$ and $0<\epsilon<1$. For fixed distinct $i,j$ and $K_{ij}(t)=\{\gamma_i(t),\gamma_j\}$,
\begin{equation}
\Pr\Bigl[\sup_{0\le t\le(1-\epsilon)\frac\beta{2\pi}\log N}C^x_{ij}(t)\ge c\Bigr]\to0,\qquad x\in\{\mathrm r,\mathrm u\}.
\label{eq:main-syk}
\end{equation}
The same holds for the target average $N^{-1}\sum_jC^x_{ij}$ of a fixed source $i$, and for the commutators of an even source $\ii\gamma_a\gamma_b$ with an odd target $\gamma_j$ or an even target $\ii\gamma_j\gamma_k$, all indices distinct.
\end{maintheorem}

\textit{Random spins.} The second ensemble has $N$ sites of fixed finite local dimension $d$ and independent $k$-body interactions~\cite{Chen2021Concentration},
\begin{equation}
H = b_N\sum_{|e|=k} H_e, \qquad b_N = J\sqrt{\frac{(k-1)!}{k N^{k-1}}}, \label{eq:spins}
\end{equation}
where the $H_e$ are independent, centered, Hermitian, supported on the $k$ sites of $e$, and bounded by $\norm{H_e} \le 1$. They need not be identically distributed. The source $A_i$ and the targets $B_j$ are deterministic single-site Hermitian contractions, independent of the couplings, and $K_{ij}(t) = \ii[A_i(t), B_j]$.

\begin{maintheorem}[Random spins]\label{thm:spins}
Fix finite $d$, $k\ge2$, $\beta J>0$, $c>0$ and $0<\epsilon<1$. For any fixed source site $i$,
\begin{equation}
\Pr\Bigl[\sup_{0\le t\le(1-\epsilon)\frac\beta{2\pi}\log N}\frac1N\sum_{j=1}^NC_{ij}^x(t)\ge c\Bigr] \to 0, \quad x \in \{\mathrm r,\mathrm u\}. \label{eq:main-spins}
\end{equation}
If the law of the couplings is invariant under permutations of the sites that fix $i$, with each target moved along with its site, the same holds for each fixed pair $i\ne j$ without the average.
\end{maintheorem}

In both theorems the Gibbs state belongs to the same sample of couplings as the dynamics, and the supremum covers the whole window, so the theorems bound the probability of any crossing within it. The temperature, the threshold, $\epsilon$ and the labels are fixed as $N\to\infty$. For the regularized SYK OTOC the failure probability is at most $CN^{-\epsilon}$, with $C$ depending on $\beta J$, $c$ and $\epsilon$.

The rest of the paper is organized as follows. Section~\ref{sec:strip} derives Theorem~\ref{thm:static} and Proposition~\ref{prop:double} and explains the role of the double average, and Secs.~\ref{sec:syk} and~\ref{sec:spins} apply this framework to the two ensembles; the complete proofs and the numerical methods are collected in the appendices.

\mainfigureone

\section{Chaos-strip bounds and the double average}\label{sec:strip}

The proof of Theorem~\ref{thm:static} follows the geometry of the MSS argument, with a static bound on the edges of the strip in place of factorization (Fig.~\ref{fig:strip}(b)). Write $h=\beta/4$. On the upper edge of the strip, because $y=\rho^{1/4}$ commutes with the real-time evolution, the imaginary-time shift acts as $yA(t+\ii h)y=y(yA(t)y^{-1})y=RA(t)$, merging the four quarters into two halves. Hence
\begin{equation}
F(t+\ii h)=\tau\sum_jw_j\Tr\bigl(RA(t)B_jRA(t)B_j\bigr),
\label{eq:edge}
\end{equation}
and the lower edge carries the complex conjugate. To decouple the evolving source $A(t)$ from the targets $B_j$, we write each term as $\tau\Tr X_j^2$ with $X_j=R^{1/2}A(t)B_jR^{1/2}$ and apply the Cauchy--Schwarz inequality, $|\Tr X_j^2|\le\Tr X_j^\dagger X_j=\Tr[A(t)RA(t)B_jRB_j]$. Summing over $j$ reassembles the conjugation average $\cJ(R)$, giving $|F(t\pm\ii h)|\le T(t)$ with the edge function
\begin{equation}
T(t)=\Tr\bigl[A(t)RA(t)\,\cJ(R)\bigr].
\label{eq:edge-function}
\end{equation}

The edge function separates the dynamics from the thermal size in the Hilbert--Schmidt space, where $R=\sqrt{\rho}$ serves as a stationary unit vector (Fig.~\ref{fig:double}(a)). All time dependence sits in $A(t)RA(t)$, an operator in the unit ball ($\norm{A(t)RA(t)}_2\le1$) whose projection onto $R$ remains fixed at $\Tr(A(t)RA(t)R)=g$ for all times, because $R$ commutes with the evolution. The conjugation average, in contrast, acts on $R$ alone. Splitting $\cJ(R)=mR+[\cJ(R)-mR]$ into components parallel and orthogonal to $R$ (the latter of norm $\sigma$) and taking the Hilbert--Schmidt inner product with $A(t)RA(t)$ gives
\begin{equation}
|T(t)-mg|\le\sigma,\qquad 0\le T(t)\le M,
\label{eq:edge-bound}
\end{equation}
for all real $t$. The thermal size thus bounds the quartered OTOC on both edges, uniformly in time and without any input about the dynamics. This is the key step that replaces factorization. Since $F$ is analytic and bounded in the strip, the Hadamard three-lines theorem extends the bound to the interior, $|F(z)|\le M$. On the real axis $F$ is real, so $F(0)\le M$ and $\delta\ge0$.

The function $u=M-\Re F$ is harmonic and nonnegative in the strip and equals $\delta$ at the origin. The Harnack inequality for the strip of width $\beta/2$, mapped to a half-plane by $e^{2\pi z/\beta}$, gives
\begin{equation}
u(t)\le\delta\,e^{2\pi|t|/\beta}.
\label{eq:harnack}
\end{equation}
Equation~\eqref{eq:harnack} is a full-strip analogue of the MSS growth estimate for the quartered OTOC, with the static defect as its initial value. Toward the edge, the Harnack factor from $t$ to $t+\ii(h-\eta)$ is $\cot(\pi\eta/\beta)\le\beta/(\pi\eta)$.

Whereas the Harnack inequality~\eqref{eq:harnack} controls the interior of the strip, the regularized OTOC lives on its upper edge. Expanding Eq.~\eqref{eq:otocs} for either sign $\tau$ gives the exact identity
\begin{equation}
\Cbar^{\mathrm r}(t)=2T(t)-2\Re F(t+\ii h),
\label{eq:identity}
\end{equation}
so Eq.~\eqref{eq:edge-bound} yields $\Cbar^{\mathrm r}(t) \le 2u(t+\ii h)$. To reach the edge from the interior, we use that the derivative of Eq.~\eqref{eq:edge} is bounded by $2\nu_A$, since moving the source costs its thermal local energy; the three-lines theorem extends this bound to $F'$ in the whole strip. Comparing the edge with the height $h-\eta$ then gives
\begin{equation}
\Cbar^{\mathrm r}(t)\le\frac{2\beta}{\pi\eta}\,\delta\,e^{2\pi|t|/\beta}+4\nu_A\eta .
\label{eq:eta}
\end{equation}
Choosing $\eta = (\beta/4)\sqrt{\delta}e^{\pi|t|/\beta}$ when this is at most $h = \beta/4$ makes both terms in Eq.~\eqref{eq:eta} proportional to $\sqrt{\delta}e^{\pi|t|/\beta}$, giving $\Cbar^{\mathrm r}(t) \le (8/\pi + \beta \nu_A)\sqrt{\delta}e^{\pi|t|/\beta} \le (4 + \beta \nu_A)\sqrt{\delta}e^{\pi|t|/\beta}$, while the bound $\Cbar^{\mathrm r}(t) \le 4$ covers the remaining times, where $\sqrt{\delta}e^{\pi|t|/\beta} > 1$. This proves Eq.~\eqref{eq:main-static}, and comparing its right-hand side with the threshold $c$ yields Eq.~\eqref{eq:main-crossing}, completing the proof of Theorem~\ref{thm:static} (Appendix~\ref{app:sec-strip}). Approaching the edge turns the prefactor $\delta$ into $\sqrt{\delta}$ and halves the exponent of the bound, so the lower bound on the first crossing time keeps the coefficient $\beta/(2\pi)$ of $\log(1/\delta)$.

\mainfiguretwo

The MSS argument has the same structure~\cite{Maldacena2016Bound}. There, a time-ordered four-point function is assumed to stay close to its factorized value after a dissipation time, which bounds the quartered OTOC on the strip. A small gap $\varepsilon$ between the OTOC and this bound at a reference time after dissipation then delays any fixed departure of the quartered OTOC by $[\beta/(2\pi)]\log(1/\varepsilon)$ at leading logarithmic order, which is of order $\log N$ when $\varepsilon$ is polynomially small in $N$. In Theorem~\ref{thm:static}, the edge bound is instead proved for every Hamiltonian, the static defect supplies the gap already at zero time, and Eq.~\eqref{eq:main-crossing} is the corresponding lower bound, with the threshold correction made explicit. Figure~\ref{fig:strip}(c) illustrates this for the representative choice $\delta=1/N$, the scale of the SYK bound $\E\Delta=O(1/N)$ (Sec.~\ref{sec:syk}): the times at which the bound reaches the threshold grow by $[\beta/(2\pi)]\log10^4$ with every factor $10^4$ in $N$.

\textit{The double average.} For a single source, stationarity fixes only the overlap of $A(t)RA(t)$ with $R$, so the edge function can deviate from $mg$ by up to $\sigma$ at all times (Fig.~\ref{fig:double}(a)). When $\sigma=\Theta(N^{-1/2})$ and $mg-F(0)=O(1/N)$, the single-source defect $\delta = mg + \sigma - F(0)$ is of order $N^{-1/2}$, and for bounded $\nu_A$ Theorem~\ref{thm:static} yields only $[\beta/(4\pi)]\log N$. When the sources range over the targets with the same weights, however, the operators $B_i(t)RB_i(t)$ add up to $e^{\ii tH}\cJ(R)e^{-\ii tH}$, and the averaged edge function becomes the autocorrelation
\begin{equation}
T_2(t)=\Tr\bigl[e^{\ii tH}\cJ(R)e^{-\ii tH}\cJ(R)\bigr]\le\norm{\cJ(R)}_2^2=m^2+\sigma^2,
\label{eq:autocorrelation}
\end{equation}
which deviates from $m^2$ by at most $\sigma^2$ and is largest at $t=0$ (Fig.~\ref{fig:double}(a)). Replacing $M$ by $\norm{\cJ(R)}_2^2$ in the proof of Theorem~\ref{thm:static} gives Proposition~\ref{prop:double} (Appendix~\ref{app:sec-strip}): the double-averaged OTOC obeys $\Cdbar^{\mathrm r}(t) \le (4+\beta\bar{\nu})\sqrt{\Delta}e^{\pi|t|/\beta}$ with the double-averaged defect $\Delta = \norm{\cJ(R)}_2^2 - F_2(0)$. Because $\sigma$ enters only through $\sigma^2$, a bound $\E\Delta = O(1/N)$, together with the local-energy control established below, yields a lower bound with the full coefficient $\beta/(2\pi)$ of $\log N$ for the double-averaged OTOC. The mechanism that makes $\Delta$ small depends on the model. In the SYK model, $\norm{\cJ(R)}_2^2$ and $F_2(0)$ place the same four insertions around the thermal circle in nested and crossed order, and their leading contributions cancel (Fig.~\ref{fig:double}(b), Sec.~\ref{sec:syk}). In random spin systems, the imaginary-time derivative on the boundary of the chaos strip selects the interactions that meet both probes, and centering makes their averaged contribution quadratic in the couplings (Sec.~\ref{sec:spins}).

\textit{Operator-size meaning and phase coexistence.} For Majorana fermions the two defects have a direct operator-size meaning. With $A = \gamma_i$, equal weights on all $N$ targets $\gamma_j$ and an even Hamiltonian, the quartered OTOC on the real axis is $F(t) = g[1 - 2n_\beta(t)/N]$, where $n_\beta(t)$ is the mean size of $\rho^{1/4}\gamma_i(t)\rho^{1/4}$. The static defect of the source $\gamma_i$ and the double-averaged defect become (Appendix~\ref{app:sec-size})
\begin{equation}
\delta_i=\frac2N\Bigl[\sqrt{V_R}+s_i\Bigr],\qquad\Delta=\frac2N\Bigl[\bar s+\frac{2V_R}N\Bigr],
\label{eq:defect-size}
\end{equation}
where $s_i = \langle X_i, \mathcal{N}X_i\rangle - \mu_R g_i$ with $X_i = \rho^{1/4}\gamma_i\rho^{1/4}$, $g_i = \norm{X_i}_2^2$, $\bar{s} = N^{-1}\sum_i s_i$ and $\langle X,Y\rangle = \Tr\,X^\dagger Y$. The single-source defect $\delta_i$ is controlled by the size width $\sqrt{V_R}$ and the insertion shift $s_i$, the $g_i$-weighted change in the mean size of $\sqrt{\rho}$ when the source sits between its two halves. Reference~\cite{Qi2019Quantum} measures the size of an operator at finite temperature by the same kind of shift, with the operator placed on one side of $\sqrt{\rho}$, while Ref.~\cite{Lucas2019Operator} uses a thermal inner product. For extensive variance $V_R = \Theta(N)$ and $s_i = O(1)$, the width $\sqrt{V_R} = \Theta(\sqrt{N})$ dominates $\delta_i$, whereas in $\Delta$ the variance term $4V_R/N^2$ and an $O(1)$ average shift $\bar{s}$ both contribute at most order $1/N$.

A small double-averaged defect can coexist with a macroscopically broad thermal size. Consider the mean-field magnet $H_n=-JnM_n^4$, with $M_n=n^{-1}\sum_{r=1}^n Z_r$ and equal source and target weights on all $N=2n$ Majorana probes. Every local energy is at most $8J$. At the coexistence temperature, where the paramagnet coexists with two ordered phases, the different single-flip overlaps of these phases give $V_{R_n}/n^2\to v$ and $\delta\to\sqrt v-v>0$, with $0<v<1/4$, so Theorem~\ref{thm:static} alone yields no logarithmic bound. Yet the variance term in Eq.~\eqref{eq:defect-size} cancels against the insertion shift, leaving $\Delta=1/n+O(n^{-2})$ (Fig.~\ref{fig:spins}(c)), and Proposition~\ref{prop:double} still gives a logarithmic lower bound for the double-averaged OTOC in the full Gibbs state. A direct calculation gives a stronger lower bound of order $n/J$ for Majorana probes on distinct qubits (Appendix~\ref{app:sec-coexist}).

\section{SYK scrambling needs logarithmic time}\label{sec:syk}

For the SYK model, we deduce Theorem~\ref{thm:syk} from Proposition~\ref{prop:double} in two steps: we establish $\E\Delta=O(1/N)$ in the Gibbs state, and we convert this equilibrium bound into control of fixed probes across the logarithmic time window (Appendix~\ref{app:sec-syk}). All $N$ Majorana operators serve as both sources and targets, with equal weights $w_i=1/N$ and $\tau=-1$.

\textit{Two Euclidean four-point functions.} By cyclicity of the trace, expanding $\norm{\cJ(R)}_2^2$ and $F_2(0)$ writes the double-averaged defect directly as the difference of two Euclidean four-point functions,
\begin{equation}
\begin{aligned}
\Delta&=\frac1{N^2}\sum_{i,j}(M_{ij}-F_{ij}),\\
M_{ij}&=\Tr(R\gamma_i\gamma_jR\gamma_j\gamma_i),\\
F_{ij}&=-\Tr(y\gamma_i y\gamma_j y\gamma_i y\gamma_j),
\end{aligned}
\label{eq:syk-defect}
\end{equation}
with $R=\rho^{1/2}$ and $y=\rho^{1/4}$. Both correlators place the same four Majorana insertions on the thermal circle of circumference $\beta$, in different orders (Fig.~\ref{fig:double}(b)). In $M_{ij}$, the two factors of $R\propto e^{-\beta H_N/2}$ separate $(\gamma_i\gamma_j)$ from $(\gamma_j\gamma_i)$ by imaginary time $\beta/2$ in nested order; in $F_{ij}$, the four factors of $y\propto e^{-\beta H_N/4}$ alternate $\gamma_i,\gamma_j,\gamma_i,\gamma_j$ at spacing $\beta/4$, so equal labels still sit at antipodal points separated by $\beta/2$ in crossed order.

Because $\rho$ is normalized by the random partition function $Z=\Tr e^{-\beta H_N}$, we first pass to the annealed functional $\Phi(f)=\E[Zf]/\E Z$, which averages the unnormalized trace and $Z$ separately. At every fixed temperature, the large-$N$ Wick theorem of Schmidhuber and Zlokapa~\cite{Schmidhuber2026Polynomial} reduces $\Phi$ on Euclidean correlators to a signed sum over equal-label pairings of products of the large-$N$ two-point function, up to $O(1/N)$ corrections. For $i\ne j$, a single pairing matches the labels in $M_{ij}$ and in $F_{ij}$. Each pair spans half the thermal circle, and the fermionic sign of the crossed pairing cancels the minus sign in $F_{ij}$, so both correlators share the leading term $q_\beta^2$, the square of the large-$N$ two-point function at separation $\beta/2$ (Appendix~\ref{app:syk-sec-euclid}). Together with the $O(1/N)$ weight of the diagonal terms $i=j$, this cancellation gives $\Phi(\Delta)=O(1/N)$.

To return from $\Phi$ to the expectation $\E$ in the same-sample Gibbs state, we combine the samplewise bound $0\le\Delta\le2$ with the lower-tail bound $\Pr(Z<\E Z/2)\le2e^{-c_{\beta J}N}$, derived in Appendix~\ref{app:syk-sec-quenched} using the concentration estimates of Ref.~\cite{Anschuetz2025Strongly}. Since $\Delta\le2(Z/\E Z)\Delta$ whenever $Z\ge\E Z/2$, taking expectations bounds $\E\Delta$ by $2\Phi(\Delta)$ up to an exponentially small term, and the Wick estimates for all $i\ne j$ then complete the bound,
\begin{equation}
\begin{aligned}
\Phi(M_{ij}),\ \Phi(F_{ij})&=q_\beta^2+O(1/N)\quad(i\ne j),\\
\E\Delta\le2\Phi(\Delta)+4e^{-c_{\beta J}N}&\le\frac{C_{\beta J}}N.
\end{aligned}
\label{eq:syk-Delta}
\end{equation}
The order of these steps matters. Forming the nonnegative combination $\Delta$ before removing the weight $Z/\E Z$ preserves the $O(1/N)$ estimate, because positivity of $\Delta$ lets the comparison use only the lower tail of $Z$; comparing $M_{ij}$ and $F_{ij}$ separately using only partition-function concentration would give an error bound of order $N^{-1/2}$. Finite-temperature numerical calculations up to $N=36$ in Fig.~\ref{fig:double}(c) are consistent with this $N^{-1}$ decay of $\E\Delta$ and with the slower $N^{-1/2}$ decay of the mean single-source defect $\E\bar\delta$, where $\bar\delta=N^{-1}\sum_i\delta_i$ (see Appendix~\ref{app:sec-numerics} for methods).

\mainfigurethree

\textit{From the double average to a fixed pair across the window.} Applying Proposition~\ref{prop:double} with Eq.~\eqref{eq:syk-Delta} also requires control of the local energies $\nu_{\gamma_i}=\norm{\ii[H_N,\gamma_i]}_\rho$. The Jacobian bound of Cai, Zhan and Zlokapa~\cite{Cai2026Fast} provides a $\beta$-independent, permutation-invariant event $\cE_N$ with $\Pr(\cE_N^c)\le2e^{-N}$ on which $\norm{[H_N,\gamma_i]}\le \nu_*=O(J)$ for every mode (Appendix~\ref{app:syk-sec-strip}). On $\cE_N$, Proposition~\ref{prop:double} and Eq.~\eqref{eq:syk-Delta} give $\E[\one_{\cE_N}\Cdbar^{\mathrm r}(t)]=O(N^{-\epsilon/2})$ uniformly for $0\le t\le T_N=(1-\epsilon)[\beta/(2\pi)]\log N$, and permutation symmetry transfers this bound to each distinct pair $i\ne j$ as well as to the target average of a fixed source.

On $\cE_N$ the time derivative of $\Cr_{ij}$ is bounded by $8\nu_*=O(J)$, so any crossing of a threshold $c$ keeps $\Cr_{ij}\ge c/2$ over an interval of $N$-independent length. A union bound over an $O(\log N)$-point time grid already gives a vanishing crossing probability $O((\log N)N^{-\epsilon/2})$ (Appendix~\ref{app:sec-crossing}). Weighting $\Cr_{ij}(s)$ instead by the strip Poisson kernel $p_\beta(s)=(2/\beta)\operatorname{sech}(2\pi s/\beta)$ removes the logarithmic factor: the Poisson average $\delta_{P,ij}=\frac12\int p_\beta(s)\Cr_{ij}(s)\,\dd s$ is a nonnegative static quantity whose double average is at most $\Delta$ (Appendix~\ref{app:sec-poisson}), giving $\E[\one_{\cE_N}\delta_{P,ij}]=O(1/N)$, while a crossing before $T_N$ forces $\delta_{P,ij}$ above a constant times $e^{-2\pi T_N/\beta}=N^{-(1-\epsilon)}$. Markov's inequality then gives (Appendix~\ref{app:syk-sec-proof})
\begin{equation}
\Pr\!\left[\sup_{0\le t\le T_N}\Cr_{ij}(t)\ge c\right]
\le C_{\beta J,c,\epsilon}N^{-\epsilon}+2e^{-N},
\label{eq:syk-window}
\end{equation}
and the same bound holds for the target average of a fixed source.

\textit{Unregularized and even probes.} To pass from $\Cr_{ij}$ to $\Cu_{ij}$, we compare symmetrized thermal weights in the energy eigenbasis, with $p_n=e^{-\beta E_n}/Z$, where transitions of energy difference $\omega$ carry weight $\sqrt{p_mp_n}\cosh(\beta\omega/2)$ in $\Cu_{ij}$ and $\sqrt{p_mp_n}$ in $\Cr_{ij}$. Splitting transitions at a fixed cutoff $\Omega>0$ bounds the low-frequency part by $\cosh(\beta\Omega/2)\Cr_{ij}$ and the high-frequency part by $\norm{[H_N,K_{ij}]}_\rho^2/\Omega^2$, which is $O(J^2/\Omega^2)$ on $\cE_N$ (Appendix~\ref{app:sec-transfer}). Choosing $\Omega\propto J/\sqrt c$ keeps the high-frequency contribution below $c/2$, so $\Cu_{ij}(t)\ge c$ forces $\Cr_{ij}(t)$ across a positive threshold independent of $N$. For an even source $\ii\gamma_a\gamma_b$ paired with an odd target $\gamma_j$ or an even target $\ii\gamma_j\gamma_k$ (all labels distinct), the Leibniz rule and a Schatten-norm product estimate reduce any crossing on $\cE_N$ to a crossing of a fixed threshold in $\Cu_{vl}$ by one of the underlying odd pairs (Appendices~\ref{app:sec-transfer} and~\ref{app:syk-sec-even}). Finally, the dimensionless coefficient $\beta J/(2\pi)$ in the thermal lower bound on $Jt_c/\log N$ vanishes as $\beta J\to0$. For $\beta J<4/133$, however, the Jacobian bound of Ref.~\cite{Cai2026Fast} reaches the strip edge at imaginary time $\beta/4$ and controls real-time growth directly, giving a window of order $J^{-1}\log N$ whose coefficient remains positive as $\beta J\to0$ (Appendix~\ref{app:syk-sec-hight}).

\section{Random spins need logarithmic time}\label{sec:spins}

Although Theorem~\ref{thm:spins} concerns perturbations of a single spin $i$, its proof also passes through the double average rather than applying Theorem~\ref{thm:static} directly to $A_i$. In the single-source bound, a defect $\delta=\Theta(N^{-1/2})$ from thermal-size fluctuations halves the prefactor to $\beta/(4\pi)$ when $\nu_{A_i}=O(J)$. At phase coexistence, macroscopic size fluctuations can even keep $\delta$ of order one and leave no logarithmic bound at all, as in the mean-field magnet of Sec.~\ref{sec:strip} (Fig.~\ref{fig:spins}(c)). In Proposition~\ref{prop:double}, by contrast, the size width enters only through $\sigma^2$, and cancellations can keep $\Delta=O(1/N)$ even at phase coexistence. Obtaining a leading logarithmic lower bound with coefficient $\beta/(2\pi)$ at every fixed temperature (Fig.~\ref{fig:spins}(a)) then rests on two steps: bounding $\E\Delta=O(1/N)$ directly in the Gibbs state from the boundary of the chaos strip, and separating site $i$ from the remaining spins so that its spreading reduces to the double average of the surrounding bath.

\textit{The double-averaged defect.} In the SYK model, $\E\Delta=O(1/N)$ followed from a leading-order cancellation in the large-$N$ Wick expansion. General $k$-local spin systems need not admit such an expansion, and their Gibbs states may develop low-temperature order. Commuting probes nevertheless give a direct route to $\Delta$ on the chaos strip. Let $B_a$ be mutually commuting Hermitian contractions with weights $w_a$ and $\tau=+1$. At the boundary point $z=\ii\beta/4$ above the origin, the four factors of $\rho^{1/4}$ in $F_2$ merge into two factors of $R=\rho^{1/2}$, giving $F_2(\ii\beta/4)=\sum_{a,b}w_aw_b\Tr(RB_aB_bRB_aB_b)$ (Eq.~\eqref{eq:edge}). Because the probes commute, this boundary value coincides with the static upper bound $\norm{\cJ(R)}_2^2$.

The harmonic function $u(z)=\norm{\cJ(R)}_2^2-\Re F_2(z)$ is nonnegative on the strip $|\Im z|\le\beta/4$, equals $\Delta$ at the origin, and vanishes at $z=\pm\ii\beta/4$. Since its boundary values on the two strip edges agree, the Poisson representation bounds its central value by $\beta/(2\pi)$ times its inward normal derivative at $z=\ii\beta/4$. Differentiating $F_2$ in imaginary time expresses this derivative through double commutators of $H$ with pairs of probes (Appendix~\ref{app:spin-sec-boundary}),
\begin{equation}
0\le\Delta\le{}\frac{\beta}{2\pi}\sum_{a,b}w_aw_b
\times\Tr\!\left(R[B_a,[H,B_b]]RB_aB_b\right).
\label{eq:spin-swap}
\end{equation}
Instead of explicit four-point estimates, Eq.~\eqref{eq:spin-swap} bounds the double-averaged defect through the local terms in $H$ that fail to commute with two probes simultaneously, regardless of the phase of $\rho$.

Write $H=\sum_e D_e$ with $D_e=b_NH_e$, and place one probe on each site with weight $w_a=1/N$. By the Jacobi identity, $[B_a,[D_e,B_b]]$ vanishes unless the $k$-site support $e$ contains both $a$ and $b$. Omitting $D_e$ makes the Gibbs state of $H-D_e$ independent of $D_e$, so the contribution linear in $D_e$ vanishes in expectation because $\E D_e=0$. Restoring $D_e$ perturbs $R=\rho^{1/2}$ at order $\beta D_e$ and leaves a second-order contribution of at most $8\beta\,\E\norm{D_e}^2$ for each probe pair in $e$. Summing $\E\norm{D_e}^2$ over all $\binom Nk$ terms gives a total interaction variance of at most order $NJ^2$, that is, $O(J^2)$ per spin, while only a fraction $(k/N)^2$ of all probe pairs lie inside a given $k$-site support $e$. Combining centering with this two-site dilution yields, in the full Gibbs state at every fixed temperature (Appendix~\ref{app:spin-sec-response}),
\begin{equation}
\E\Delta\le\frac{4\beta^2}{\pi}\binom Nk b_N^2\left(\frac{k}{N}\right)^2\le\frac{4\beta^2J^2}{\pi N}.
\label{eq:spin-defect}
\end{equation}
Quantum Monte Carlo simulations of a random transverse-field Ising model (Fig.~\ref{fig:spins}(b)) are consistent with this $N^{-1}$ decay of $\E\Delta$ and with the $N^{-1/2}$ decay of $\E\bar\delta$ (Appendix~\ref{app:sec-numerics}).

\textit{Spreading from a fixed source.} Combined with bounded local-energy moments (Appendix~\ref{app:spin-sec-moments}), Proposition~\ref{prop:double} and Eq.~\eqref{eq:spin-defect} bound the disorder expectation $\E[N^{-2}\sum_{a,b}\Cr_{ab}(t)]$ by $O(N^{-1/2}e^{\pi|t|/\beta})$. Under site-permutation symmetry with identical probe choices, every source spin shares this average in expectation. In an inhomogeneous ensemble, however, we must rule out faster scrambling from the chosen site $i$. The relevant comparison comes from the centered random interactions connecting spin $i$ to the bath: their disorder-averaged linear response in the decoupled state is bounded by a constant times a uniform bath double average.

Specifically, we split $H=K+V$ into the bath interactions $K$ avoiding site $i$ and the couplings $V$ meeting site $i$. Under $K$ alone, $A_i$ is stationary ($[K,A_i]=0$) and commutes with every bath probe $B_j$ ($j\ne i$). In the interaction picture, $A_i(t)=W(t)A_iW(t)^\dagger$ with $W(t)=e^{\ii tH}e^{-\ii tK}$, so any commutator with $B_j$ enters through $[W(t),B_j]$. Differentiating $W(s)B_jW(s)^\dagger$ gives $\ii W(s)[V(s),B_j]W(s)^\dagger$, where $V(s)=e^{\ii sK}Ve^{-\ii sK}$ evolves under the bath $K$ alone, and integrating over $s\in[0,t]$ yields (Appendix~\ref{app:spin-sec-duhamel})
\begin{equation}
\begin{aligned}
[A_i(t),B_j]
&=[W(t),B_j]A_iW(t)^\dagger\\
&\quad-A_i(t)[W(t),B_j]W(t)^\dagger,\\
[W(t),B_j]
&=\ii\int_0^t W(s)\,[V(s),B_j]\,W(s)^\dagger W(t)\,\dd s.
\end{aligned}
\label{eq:spin-duhamel}
\end{equation}
Equation~\eqref{eq:spin-duhamel} makes the causal structure explicit: any influence of $A_i$ on $B_j$ must enter the bath through $V$. The integrand isolates the commutator of the bath-evolved coupling $V(s)$ with the target, while the surrounding factors $W$ still carry the source coupling.

In the decoupled bath Gibbs state $\rho_K=e^{-\beta K}/\Tr e^{-\beta K}$, which is independent of $V$, each term in $V$ couples site $i$ to $k-1$ bath sites with variance at most $b_N^2=O(N^{-(k-1)})$, matching the weight of a uniform average over all $\binom{N-1}{k-1}$ bath subsets of size $k-1$. Averaging over the centered source couplings cancels cross terms between distinct subsets. For a Gaussian source $V_g$ matching the covariance of $V$, the target-averaged regularized commutator $C_{\rho_K}^{\mathrm r}(X,Y)=\norm{\rho_K^{1/4}\ii[X,Y]\rho_K^{1/4}}_2^2$ is therefore bounded by a double average within the bath itself (Appendix~\ref{app:spin-sec-reference}),
\begin{equation}
\begin{aligned}
\E\frac1N\sum_{j\ne i}
C_{\rho_K}^{\mathrm r}\!\left(V_g(s),B_j\right)
&\le C\,\E\mathcal B_{k-1}(s)\\
&\le C'N^{-1/2}e^{\pi|s|/\beta}.
\end{aligned}
\label{eq:spin-reference}
\end{equation}
Here $\mathcal B_{k-1}(s)$ is the double-averaged regularized OTOC of an operator basis on $(k-1)$-site bath subsets under $K$, whose disorder expectation satisfies the same $O(N^{-1/2}e^{\pi|s|/\beta})$ bound once Eq.~\eqref{eq:spin-swap} and Proposition~\ref{prop:double} are extended to $(k-1)$-site probes (Appendix~\ref{app:spin-sec-frame}). The root-mean-square thermal norm of the Gaussian reference commutator is the square root of the left-hand side of Eq.~\eqref{eq:spin-reference}. It is therefore $O(N^{-1/4}e^{\pi|s|/(2\beta)})$, and uniformly $O(N^{-\epsilon/4})$ for $0\le s\le T_N=(1-\epsilon)[\beta/(2\pi)]\log N$.

\textit{Returning to the full system.} Passing from this decoupled estimate to Theorem~\ref{thm:spins} relies on the fact that $V$ couples only a single spin to an extensive bath, with $O(J^2)$ total coupling variance. We first replace $V$ by a spectrally clipped Gaussian coupling $D_\Lambda$, with $H_\Lambda=K+D_\Lambda$, so that Eq.~\eqref{eq:spin-duhamel} places $[D_\Lambda(s),B_j]$ inside the interaction-picture unitaries $W$ (Appendix~\ref{app:spin-sec-duhamel}). We then bound how the surrounding factors $W$ affect its regularized norm in the bath state. Moving an operator of bath energy transfer $\omega$ across $\rho_K^{1/4}\propto e^{-\beta K/4}$ can multiply its norm by $e^{\beta|\omega|/4}$, but $W$ changes the bath energy only through the source coupling, so bath-frequency estimates control its high-frequency tails. With the cutoffs in Appendix~\ref{app:spin-sec-scales}, filtering these transfers on a scale $o(J\log N)$ keeps the low-frequency amplification below any fixed positive power of $N$; the $N^{-\epsilon/4}$ decay from Eq.~\eqref{eq:spin-reference} then dominates, while the high-frequency error vanishes over $[0,T_N]$ (Appendices~\ref{app:sec-meas} and~\ref{app:sec-cond}). Finally, we return from $\rho_K$ and $D_\Lambda$ to the original coupling $V$ and its Gibbs state $\rho$. Bounded thermal moments of the source coupling keep the expected relative entropies between $\rho_K$ and the relevant Gibbs states bounded. Combined with spectral transfer between regularized and unregularized correlators, this makes vanishing of the expected window maximum equivalent in $\rho_K$ and in the Gibbs state of the same dynamics, and source-coupling comparisons undo the clipping and Gaussian replacement (Appendix~\ref{app:sec-dyn}). Markov's inequality then gives Eq.~\eqref{eq:main-spins} for $x\in\{\mathrm r,\mathrm u\}$ in the original Hamiltonian and its Gibbs state, and replicating a given target across the bath yields the fixed-pair bound when the coupling law is invariant under permutations fixing $i$ (Appendix~\ref{app:spin-sec-completion}).

\section{Discussion}\label{sec:discussion}

Theorems~\ref{thm:syk} and~\ref{thm:spins} show that, with probability tending to one, thermal scrambling times obey the same leading lower bound $[\beta/(2\pi)]\log N$ at every fixed temperature in both the SYK model and random $k$-local spin systems, despite their very different microscopic dynamics. Earlier derivations of the thermal chaos bound relied on real-time assumptions, such as factorization of time-ordered correlators after a dissipation time~\cite{Maldacena2016Bound} or an exponential-growth ansatz~\cite{Tsuji2018Bound}. Here, premature scrambling is instead ruled out directly by equilibrium structure at $t = 0$. Because the same Hamiltonian generates both Heisenberg evolution and the Gibbs weights $e^{-\beta H}$, shifting time to the edges of the chaos strip converts a dynamical problem over long real times into a static estimate on the Euclidean thermal circle. Analyticity across the strip sets the Planckian prefactor $\beta/(2\pi)$~\cite{Maldacena2016Bound,Lucas2019Operator}, while for controlled thermal local energies, the $O(1/N)$ expected difference between nested and crossed Euclidean correlators gives the $\log N$ growth with system size.

This perspective also clarifies how finite-temperature scrambling differs from the infinite-temperature case, especially near a phase transition. At infinite temperature, $\sqrt{\rho} \propto I$ has zero operator size, so nontrivial operator strings appear only through Heisenberg evolution~\cite{Roberts2018Operator,Lucas2020Nonperturbative,Yin2020Bound,Chen2021Concentration}. At finite temperature, by contrast, $\sqrt{\rho}$ can already carry a nontrivial operator-size distribution at $t = 0$ encoding equilibrium correlations~\cite{Qi2019Quantum}. Near a thermal transition, these static correlations can become macroscopic: at first-order coexistence (Fig.~\ref{fig:spins}(c)), mixing phases of different mean thermal sizes makes the size variance of $\sqrt{\rho}$ grow as $N^2$ and keeps the single-source defect $\delta$ of order one. In this example, averaging symmetrically over sources and targets cancels these static ensemble fluctuations against the probe-insertion shift, leaving $\Delta = O(1/N)$. Macroscopic equilibrium fluctuations therefore leave the thermal scrambling-time lower bound intact even when the single-source defect stays of order one, and Theorem~\ref{thm:spins} establishes the same $[\beta/(2\pi)]\log N$ bound across any finite-temperature phase transition in random spin ensembles.

Comparing $[\beta/(2\pi)]\log N$ with actual scrambling dynamics across temperature further highlights what governs a system's approach to the bound. Toward high temperature ($\beta J \to 0$), the chaos strip narrows and $\beta/(2\pi)$ vanishes, whereas the scrambling time remains at least of order $J^{-1}\log N$, as proved directly for the high-temperature SYK model in Appendix~\ref{app:syk-sec-hight}. At generic finite $\beta J$, large-$N$ analyses of the SYK model find a Lyapunov exponent strictly below $2\pi/\beta$~\cite{Maldacena2016Remarks,Kitaev2018Soft,Kobrin2021Many}, while in the commuting mean-field model of Fig.~\ref{fig:spins}(c), effective two-qubit couplings of order $J/N$ keep inter-qubit scrambling times at least of order $N/J$ at every temperature, even though $\Delta$ scales as $1/N$. At fixed temperature, improving the constant in an estimate $\E\Delta\le C/N$ shifts the lower bound only by an additive time, so the leading prefactor stays $\beta/(2\pi)$; accounting for submaximal rates at finite coupling, or algebraically long delays in ordered and integrable models, therefore calls for finer equilibrium data such as the full imaginary-time profile and spectral structure of Euclidean correlators~\cite{Murthy2019Bounds,Parker2019Universal,Avdoshkin2020Euclidean}. Toward strong coupling ($\beta J \gg 1$), by contrast, low-energy SYK dynamics is governed by the collective Schwarzian mode dual to nearly-$\mathrm{AdS}_2$ gravity~\cite{Maldacena2016Remarks,Stanford2017Fermionic,Kitaev2018Soft}, the Lyapunov exponent approaches $2\pi/\beta$, and our prefactor $\beta/(2\pi)$ matches the predicted asymptotic timescale. Extending rigorous equilibrium control from fixed $\beta J$ into the low-temperature regime where $\beta J$ grows with $N$, as well as proving matching upper bounds that certify fast scrambling, remain important open problems.

Finally, because Theorem~\ref{thm:static} and Proposition~\ref{prop:double} hold deterministically for any finite-dimensional Hamiltonian, the static-defect framework applies beyond disordered all-to-all models. In spatially structured systems such as lattices~\cite{Lieb1972Finite}, sparse graphs~\cite{Bentsen2019Fast}, and power-law ensembles~\cite{Kuwahara2020Strictly,Tran2020Hierarchy,Tran2021LiebRobinson,Kuwahara2021Absence,Chen2023Speed}, operator spreading is constrained simultaneously by interaction geometry and by the Gibbs state~\cite{Roberts2016LiebRobinson,Han2019Quantum}; combining chaos-strip analyticity with Lieb--Robinson bounds offers a route to temperature-dependent thermal light cones, and to bounds on butterfly velocities whenever propagation is ballistic. A second direction concerns how scrambling is diagnosed. Local out-of-time-order correlators detect when a perturbation first reaches individual probes, whereas global information recovery in Hayden--Preskill protocols~\cite{Hayden2007Black,Sekino2008Fast,Landsman2019Verified} depends on subsystem-averaged correlators and operator entanglement~\cite{Hosur2016Chaos,Yoshida2017Efficient,Styliaris2021Information}. Whereas Proposition~\ref{prop:double} averages sources and targets over the same probe family, Hayden--Preskill recovery pairs operator bases on distinct input and output subsystems; developing static chaos-strip defects for such pairs of subsystem bases, with appropriate thermal weighting, could connect equilibrium bounds on local commutators to macroscopic information decoupling.

\begin{acknowledgments}
We thank Sirui Lu, Qi Ye, and Si Jiang for helpful discussions.
We acknowledge support from the National Natural Science Foundation of China (T2225008), the Quantum Science and Technology-National Science and Technology Major Project (2021ZD0302203), the Fundamental and Interdisciplinary Disciplines Breakthrough Plan of the Ministry of Education of China (JYB2025XDXM112), the Shanghai Qi Zhi Institute Innovation Program, and the Tsinghua University Dushi Program.
We acknowledge the use of GPT-6 Astra and Gemini 4 Carbon to assist with theoretical derivations, numerical calculations, and language polishing. We checked all derivations and take full responsibility for the content.
\end{acknowledgments}

\appendix

\section{Thermal size and the chaos strip}\label{app:sec-static}

This appendix proves Theorem~\ref{thm:static} and Proposition~\ref{prop:double} for an arbitrary Hamiltonian on a finite-dimensional space. It also collects the deterministic estimates that both models use: the Majorana size identities, a spectral transfer between the two raw OTOCs, a product estimate in Schatten norms, a first-crossing lemma and a Poisson average of the regularized OTOC.

All Hamiltonians act on finite-dimensional Hilbert spaces, $\Tr$ is the unnormalized trace and $I$ is the identity. We write $\norm{X}$ for the operator norm, $\norm{X}_p$ for the Schatten norms, $\norm{X}_\rho=(\Tr\rho X^\dagger X)^{1/2}$ for the thermal norm and $\langle X,Y\rangle=\Tr X^\dagger Y$ for the Hilbert--Schmidt inner product. A Hermitian contraction is a Hermitian operator $X$ with $\norm{X}\le1$. The local energy of a probe $A$ is $\norm{[H,A]}$, and its thermal local energy is $\nu_A=\norm{\ii[H,A]}_\rho$. For a random Hamiltonian, $\E$ and $\Pr$ refer to its couplings. The physical OTOCs use the Gibbs state of the sample generating the dynamics; auxiliary correlators specify the state and the Hamiltonian separately.

\subsection{The chaos strip}\label{app:sec-strip}

We use the setting of Sec.~\ref{sec:results}, with $R=\rho^{1/2}$, $y=\rho^{1/4}$ and $h=\beta/4$. Hilbert--Schmidt Cauchy--Schwarz gives
\begin{equation}
\Cr_j=\norm{yK_jy}_2^2=\langle K_jR,RK_j\rangle\le\Tr(\rho K_j^2)=\Cu_j\le4
\label{app:eq-otoc-order}
\end{equation}
for the raw OTOCs of Eq.~\eqref{eq:otocs}. The static quantities are $m$ and $\sigma$ of Eq.~\eqref{eq:size-width}, $g=\Tr(RARA)$ and the edge bound $M=mg+\sigma$. The quartered OTOC $F$ is defined in Eq.~\eqref{eq:quartered}, the static defect is $\delta=M-F(0)$, and the edge function is $T(t)=\Tr(A(t)RA(t)\cJ(R))$.

\begin{lemma}\label{app:lem-strip}
For every Hamiltonian, $\beta>0$, Hermitian contractions $A$ and $B_j$, weights $w_j$ and sign $\tau$, the following hold.
\begin{enumerate}
\item[(i)] $0\le m\le1$, $0\le g\le1$, $m^2+\sigma^2\le1$ and $M\le\sqrt2$.
\item[(ii)] For real $t$, $|F(t\pm\ii h)|\le T(t)$, $|T(t)-mg|\le\sigma$ and $0\le T(t)\le M$.
\item[(iii)] $|F(z)|\le M$ for $|\Im z|\le h$. The function $F$ is real on the real axis, and $\delta\ge0$.
\end{enumerate}
\end{lemma}

\begin{proof}
\emph{Normalization.} $\norm{R}_2^2=\Tr\rho=1$. The map $\cJ$ is completely positive, self-adjoint for $\langle\cdot,\cdot\rangle$, and contracts the Hilbert--Schmidt norm, since $\norm{B_jXB_j}_2\le\norm{X}_2$. Hence $\cJ(R)\ge0$, $m\ge0$ as the trace of a product of two positive operators, and $m\le\norm{R}_2\norm{\cJ(R)}_2\le1$. Since $mR$ is the orthogonal projection of $\cJ(R)$ on the unit vector $R$, $m^2+\sigma^2=\norm{\cJ(R)}_2^2\le1$. Likewise $g=\norm{R^{1/2}AR^{1/2}}_2^2\ge0$ and $g\le\norm{ARA}_2\norm{R}_2\le1$. Cauchy--Schwarz in $\mathbb R^2$ gives $M\le(m^2+\sigma^2)^{1/2}(g^2+1)^{1/2}\le\sqrt2$.

\emph{Edges of the strip.} Since $y=e^{-hH}/(\Tr e^{-\beta H})^{1/4}$ commutes with $e^{\ii tH}$ and $e^{\ii(t\pm\ii h)H}=e^{\ii tH}e^{\mp hH}$,
\begin{equation}
yA(t+\ii h)y=RA(t),\qquad yA(t-\ii h)y=A(t)R
\label{app:eq-edge}
\end{equation}
for real $t$. Thus $F(t+\ii h)$ is given by Eq.~\eqref{eq:edge}, and $F(t-\ii h)$ is its complex conjugate. With $X_j=R^{1/2}A(t)B_jR^{1/2}$ each term of Eq.~\eqref{eq:edge} equals $\tau\Tr X_j^2$, and $|\Tr X_j^2|\le\Tr X_j^\dagger X_j=\Tr(A(t)RA(t)B_jRB_j)$. Summing over $j$ gives $|F(t\pm\ii h)|\le T(t)$. The edge function is nonnegative, as the trace of the product of $A(t)RA(t)\ge0$ and $\cJ(R)\ge0$. Stationarity gives $\Tr(A(t)RA(t)R)=g$ for all $t$, because $R$ commutes with $e^{\ii tH}$. Writing $\cJ(R)=mR+(\cJ(R)-mR)$ and using $\norm{A(t)RA(t)}_2\le1$ gives $|T(t)-mg|\le\sigma$, which proves (ii).

\emph{The strip.} In finite dimension $F$ is entire, and $\norm{A(z)}\le e^{2|\Im z|\norm{H}}$ makes it bounded on the closed strip $S=\{|\Im z|\le h\}$. The Phragm\'en--Lindel\"of principle for a strip (the Hadamard three-lines theorem) bounds a bounded analytic function on $S$ by the supremum of its boundary values, so (ii) gives $|F|\le M$ on $S$. For real $t$ the operators $A(t)$, $B_j$ and $y$ are Hermitian, so cyclicity gives $\overline{\Tr(yA(t)yB_jyA(t)yB_j)}=\Tr(B_jyA(t)yB_jyA(t)y)=\Tr(yA(t)yB_jyA(t)yB_j)$. Hence $F$ is real on the real axis, $F(0)\in[-M,M]$ and $\delta\ge0$.
\end{proof}

\begin{proof}[Proof of Theorem~\ref{thm:static}]
\emph{Harnack estimates.} The function $u=M-\Re F$ is harmonic and nonnegative on the open strip and continuous on its closure. If $u$ vanishes at an interior point, it vanishes identically by the minimum principle. Otherwise $u>0$ inside $S$. The map $\zeta=e^{2\pi z/\beta}$ sends the open strip conformally onto the right half-plane. For two points $\zeta_0,\zeta_1$ of the half-plane, Harnack's inequality reads $u(\zeta_1)\le\frac{1+q}{1-q}u(\zeta_0)$ with the pseudo-hyperbolic distance $q=|\zeta_1-\zeta_0|/|\zeta_1+\bar\zeta_0|$. On the real axis, $\zeta_0=1$ and $\zeta_1=e^{2\pi t/\beta}$ give $\frac{1+q}{1-q}=e^{2\pi|t|/\beta}$, which is Eq.~\eqref{eq:harnack}. For $0<\eta\le h$, the points $t$ and $t+\ii(h-\eta)$ map to $\zeta_0=e^{2\pi t/\beta}$ and $\zeta_1=\zeta_0e^{\ii\theta}$ with $\theta=\pi/2-2\pi\eta/\beta$. Then $q=\tan(\theta/2)=\tan(\pi/4-\pi\eta/\beta)$ and $\frac{1+q}{1-q}=\cot(\pi\eta/\beta)\le\beta/(\pi\eta)$, so
\begin{equation}
u\bigl(t+\ii(h-\eta)\bigr)\le\frac{\beta}{\pi\eta}\,\delta\,e^{2\pi|t|/\beta}.
\label{app:eq-harnack-edge}
\end{equation}

\emph{Derivative.} On the upper edge, $\frac{\dd}{\dd t}F(t+\ii h)=2\tau\sum_jw_j\Tr\bigl(R\dot A(t)B_jRA(t)B_j\bigr)$ with $\dot A(t)=\ii[H,A(t)]$, where cyclicity has combined the two terms of the product rule. Hilbert--Schmidt Cauchy--Schwarz gives
\[
\bigl|\Tr\bigl(R\dot A(t)\cdot B_jRA(t)B_j\bigr)\bigr|\le\norm{R\dot A(t)}_2\,\norm{B_jRA(t)B_j}_2\le \nu_A,
\]
because $\norm{R\dot A}_2^2=\Tr(\rho\dot A^2)$ and stationarity makes it independent of $t$. The lower edge is the complex conjugate. Since $F'$ is analytic and bounded on $S$, the three-lines theorem gives $|F'|\le2\nu_A$ on $S$, and
\begin{equation}
\bigl|F(t+\ii h)-F\bigl(t+\ii(h-\eta)\bigr)\bigr|\le2\nu_A\eta.
\label{app:eq-derivative}
\end{equation}

\emph{The raw regularized OTOC on the edge.} Expanding Eq.~\eqref{eq:otocs} for either sign $\tau$ gives
\[
\begin{split}
\Cr_j(t)={}&2\Tr\bigl(A(t)RA(t)B_jRB_j\bigr)\\
&-2\tau\Re\Tr\bigl(RA(t)B_jRA(t)B_j\bigr),
\end{split}
\]
and the weighted sum is the identity~\eqref{eq:identity}.

\emph{Conclusion.} If $\delta=0$, then $u\equiv0$, and Lemma~\ref{app:lem-strip}(ii) and Eq.~\eqref{eq:identity} give $\Cbar^{\mathrm r}(t)=2T(t)-2M\le0$, so $\Cbar^{\mathrm r}\equiv0$. Otherwise put $X=\delta e^{2\pi|t|/\beta}$. Lemma~\ref{app:lem-strip}(ii), Eqs.~\eqref{eq:identity}, \eqref{app:eq-harnack-edge} and \eqref{app:eq-derivative} give
\[
\begin{split}
\Cbar^{\mathrm r}(t)&\le2u(t+\ii h)\le2u\bigl(t+\ii(h-\eta)\bigr)+4\nu_A\eta\\
&\le\frac{2\beta X}{\pi\eta}+4\nu_A\eta .
\end{split}
\]
For $X\le1$ the choice $\eta=(\beta/4)\sqrt X\le h$ gives $\Cbar^{\mathrm r}(t)\le(8/\pi+\beta \nu_A)\sqrt X$, and $8/\pi<4$. For $X>1$ the bound $\Cbar^{\mathrm r}\le4$ suffices. This proves Eq.~\eqref{eq:main-static}. If $t$ is smaller than the right side of Eq.~\eqref{eq:main-crossing}, then $(4+\beta \nu_A)\sqrt\delta\,e^{\pi t/\beta}<c$, which proves Eq.~\eqref{eq:main-crossing}.
\end{proof}

\begin{proof}[Proof of Proposition~\ref{prop:double}]
For each source $B_i$ let $F_i$, $T_i$ and $g_i=\Tr(RB_iRB_i)$ be the quartered OTOC, the edge function and the overlap of the setting above, with the targets $B_j$ and the weights $w_j$. Then $F_2=\sum_iw_iF_i$ and $\Cdbar^{\mathrm r}=\sum_iw_i\Cbar^{\mathrm r}_i$, and Lemma~\ref{app:lem-strip}(ii) gives $|F_2(t\pm\ii h)|\le T_2(t)=\sum_iw_iT_i(t)$. Since $R$ commutes with $e^{\ii tH}$, $B_i(t)RB_i(t)=e^{\ii tH}B_iRB_ie^{-\ii tH}$, and Hilbert--Schmidt Cauchy--Schwarz gives
\begin{equation}
T_2(t)=\Tr\bigl[e^{\ii tH}\cJ(R)e^{-\ii tH}\cJ(R)\bigr]\le\norm{\cJ(R)}_2^2,
\label{app:eq-autocorrelation}
\end{equation}
with equality at $t=0$. The three-lines theorem gives $|F_2|\le\norm{\cJ(R)}_2^2$ on the strip, $F_2$ is real on the real axis, and $\Delta\ge0$. The proof of Theorem~\ref{thm:static} then applies with $F_2$, $T_2$, $\norm{\cJ(R)}_2^2$, $\Delta$ and $\Cdbar^{\mathrm r}$ in place of $F$, $T$, $M$, $\delta$ and $\Cbar^{\mathrm r}$. The identity~\eqref{eq:identity} holds for each source and hence for the averages, and the derivative of $F_2$ on the edges is bounded by $2\bar\nu$, the average of the bounds $2\nu_{B_i}$. This proves Eq.~\eqref{eq:main-double}.

Since $\sum_iw_ig_i=\Tr(R\cJ(R))=m$, the static defects $\delta_i=mg_i+\sigma-F_i(0)$ of the sources average to $m^2+\sigma-F_2(0)$. With $\norm{\cJ(R)}_2^2=m^2+\sigma^2$ from the proof of Lemma~\ref{app:lem-strip}, this gives
\begin{equation}
\Delta=\sum_iw_i\delta_i-\sigma(1-\sigma)\le\sum_iw_i\delta_i .
\label{app:eq-double-single}
\end{equation}
\end{proof}

\subsection{Majorana size identities}\label{app:sec-size}

Let $\gamma_1,\dots,\gamma_N$, with $N$ even, be Majorana operators, $\{\gamma_a,\gamma_b\}=2\delta_{ab}$, acting irreducibly on $\mathbb C^{2^{N/2}}$. For $S=\{s_1<\dots<s_{|S|}\}$ put $\gamma_S=\gamma_{s_1}\cdots\gamma_{s_{|S|}}$. These monomials are orthogonal, $\langle\gamma_S,\gamma_T\rangle=2^{N/2}\delta_{ST}$, and span all operators. The size superoperator acts as $\cN\gamma_S=|S|\gamma_S$. An operator $X$ with $\norm{X}_2=1$ has the size distribution $P_X(s)=2^{-N/2}\sum_{|S|=s}|\langle\gamma_S,X\rangle|^2$, with mean $\langle X,\cN X\rangle$ and variance $\langle X,\cN^2X\rangle-\langle X,\cN X\rangle^2$. The thermal size is the size distribution of $R=\sqrt\rho$.

\begin{lemma}\label{app:lem-size}
Let $\cJ(X)=N^{-1}\sum_{a=1}^N\gamma_aX\gamma_a$, the conjugation average of the full family with equal weights.
\begin{enumerate}
\item[(i)] $\cJ(\gamma_S)=(-1)^{|S|}(1-2|S|/N)\gamma_S$.
\item[(ii)] If $H$ commutes with the fermion parity, then $m=1-2\mu_R/N$ and $\sigma=2\sqrt{V_R}/N$, where $\mu_R=\langle R,\cN R\rangle$ and $V_R=\langle R,\cN^2R\rangle-\mu_R^2$ are the mean and the variance of the thermal size.
\item[(iii)] If in addition $A=\gamma_i$ and $\tau=-1$, then $F(t)=g-\frac2N\langle X(t),\cN X(t)\rangle$ for real $t$, where $X(t)=y\gamma_i(t)y$ and $\norm{X(t)}_2^2=g$. Thus $F(t)=g\,(1-2n_\beta(t)/N)$, where $n_\beta(t)$ is the mean size of $X(t)/\sqrt g$.
\item[(iv)] Under the assumptions of (iii), the static defect of the source $\gamma_i$ is $\delta_i$ of Eq.~\eqref{eq:defect-size}, with $X_i=X(0)$ and $g_i=g$.
\item[(v)] If $H$ commutes with the fermion parity and every $\gamma_i$ serves as a source with weight $1/N$ and $\tau=-1$, the double-averaged defect is $\Delta$ of Eq.~\eqref{eq:defect-size}.
\end{enumerate}
\end{lemma}

\begin{proof}
(i) If $a\notin S$, $\gamma_a$ anticommutes with every factor of $\gamma_S$, so $\gamma_a\gamma_S\gamma_a=(-1)^{|S|}\gamma_S$. If $a\in S$, it anticommutes with the other $|S|-1$ factors, so $\gamma_a\gamma_S\gamma_a=(-1)^{|S|-1}\gamma_S$. Averaging over $a$ gives $N^{-1}[(N-|S|)-|S|](-1)^{|S|}\gamma_S$. (ii) An even $H$ gives an even $R$, on which $\cJ=1-2\cN/N$. Hence $m=\langle R,\cJ R\rangle=1-2\mu_R/N$ and $\cJ(R)-mR=-\frac2N(\cN-\mu_R)R$, whose squared norm is $4V_R/N^2$. (iii) $X(t)$ is Hermitian and odd, so $F(t)=-N^{-1}\sum_a\Tr(X\gamma_aX\gamma_a)=-\langle X,\cJ X\rangle=\langle X,(1-2\cN/N)X\rangle$. By cyclicity and stationarity, $\norm{X(t)}_2^2=\Tr(R\gamma_i(t)R\gamma_i(t))=g$. (iv) Insert (ii) and (iii) at $t=0$ in $\delta=mg+\sigma-F(0)$. (v) Average (iv) over $i$ and insert $\sigma=2\sqrt{V_R}/N$ in Eq.~\eqref{app:eq-double-single}.
\end{proof}

\subsection{Spectral transfer and a product estimate}\label{app:sec-transfer}

\begin{lemma}\label{app:lem-transfer}
Let $K$ be Hermitian and $\Omega>0$. Then
\begin{equation}
\Tr(\rho K^2)\le\cosh(\beta\Omega/2)\,\Tr(RKRK)+\frac{\norm{[H,K]}_\rho^2}{\Omega^2}.
\label{app:eq-transfer}
\end{equation}
Consequently, if $\norm{[H,K]}\le G$ with $G>0$ and $\Tr(\rho K^2)\ge c>0$, then
\begin{equation}
\Tr(RKRK)\ge c_{\beta,G}=\frac c2\operatorname{sech}\frac{\beta G}{\sqrt{2c}}.
\label{app:eq-kappa}
\end{equation}
The same statements hold for weighted averages over a family $K_j$ with a common bound $\norm{[H,K_j]}\le G$.
\end{lemma}

\begin{proof}
In an eigenbasis $H|n\rangle=E_n|n\rangle$ with $p_n=e^{-\beta E_n}/Z$, Hermiticity gives $|K_{mn}|=|K_{nm}|$ and, after symmetrization in $m,n$,
\[
\Tr(\rho K^2)=\sum_{m,n}\sqrt{p_mp_n}\cosh\frac{\beta(E_m-E_n)}2|K_{mn}|^2.
\]
The pairs with $|E_m-E_n|\le\Omega$ form a symmetric set, and their contribution is at most $\cosh(\beta\Omega/2)\sum_{m,n}\sqrt{p_mp_n}|K_{mn}|^2=\cosh(\beta\Omega/2)\Tr(RKRK)$. The remaining pairs contribute at most $\Omega^{-2}\sum_{m,n}p_n(E_m-E_n)^2|K_{mn}|^2=\Omega^{-2}\norm{[H,K]}_\rho^2$. For the consequence choose $\Omega=\sqrt2G/\sqrt c$, so that the second term of Eq.~\eqref{app:eq-transfer} is at most $c/2$. For a weighted average apply Eq.~\eqref{app:eq-transfer} to each $K_j$ and sum.
\end{proof}

For $K=\ii[A(t),B]$ or $K=\{A(t),B\}$ with Hermitian contractions, the Leibniz rule and stationarity give $\norm{[H,K]}\le2(\norm{[H,A]}+\norm{[H,B]})$, so $G$ is set by the local energies of the two probes, not by the bandwidth of $H$. The threshold $c_{\beta,G}$ depends on $\beta G$ and $c$ only.

\begin{lemma}\label{app:lem-product}
Let $D$ be Hermitian and $V$ a contraction. Then
\begin{equation}
\norm{yDVy}_2^2\le\norm{D}\,\norm{D}_\rho,\qquad\norm{yVDy}_2^2\le\norm{D}\,\norm{D}_\rho.
\label{app:eq-product}
\end{equation}
\end{lemma}

\begin{proof}
Schatten H\"older gives $\norm{yDVy}_2\le\norm{yD}_4\norm{Vy}_4$, and $\norm{Vy}_4\le\norm{y}_4=(\Tr\rho)^{1/4}=1$. The fourth power obeys $\norm{yD}_4^4=\Tr(RD^2RD^2)\le\Tr(\rho D^4)\le\norm{D}^2\Tr(\rho D^2)$, where the middle step is $\Tr(XYXY)\le\Tr(X^2Y^2)$ for Hermitian $X=R$ and $Y=D^2$, which follows from $\norm{[X,Y]}_2^2\ge0$. The second bound is the adjoint of the first.
\end{proof}

\subsection{First crossings}\label{app:sec-crossing}

The following lemma turns bounds at individual times into a bound on the probability of a crossing within a whole window. It applies to a random continuous function $f$ on $[0, T]$, defined on the same probability space as an event $\cE$.

\begin{lemma}[Constant mesh]\label{app:lem-mesh}
Suppose that on $\cE$ the function $f$ is nonnegative and Lipschitz with a deterministic constant $L>0$. Then for $c>0$
\begin{equation}
\begin{split}
\Pr\Bigl[\sup_{0\le t\le T}f(t)\ge c\Bigr]\le{}&\Pr(\cE^c)\\
&+\frac2c\Bigl(2+\frac{2LT}{c}\Bigr)\sup_{0\le t\le T}\E\bigl[\one_{\cE}f(t)\bigr].
\end{split}
\label{app:eq-mesh}
\end{equation}
\end{lemma}

\begin{proof}
Choose a grid in $[0,T]$ containing both endpoints, with spacing at most $c/(2L)$ and at most $2+2LT/c$ points. On $\cE$, if $f(t)\ge c$ at some $t$, then $f\ge c/2$ at the nearest grid point. Markov's inequality at each grid point and a union bound give Eq.~\eqref{app:eq-mesh}.
\end{proof}

For the raw regularized OTOC the Lipschitz constant is $8\nu_A$. With $\dot A=\ii[H,A(t)]$, its derivative is $\frac{\dd}{\dd t}\Cr_j=2\Tr(R\dot K_jRK_j)$, where $\dot K_j=\ii[\dot A,B_j]$ or $\{\dot A,B_j\}$ is a sum of two terms of the forms $\dot AB_j$ and $B_j\dot A$. Cyclicity and Cauchy--Schwarz give $|\Tr(R\dot AB_jRK_j)|\le\norm{R\dot A}_2\norm{B_jRK_j}_2\le2\norm{\dot A}_\rho$ and $|\Tr(RB_j\dot ARK_j)|=|\Tr(\dot AR\,K_jRB_j)|\le\norm{\dot AR}_2\norm{K_jR}_2\le2\norm{\dot A}_\rho$, since $\norm{R\dot A}_2=\norm{\dot AR}_2=\norm{\dot A}_\rho$ and $\norm{RK_j}_2=\norm{K_jR}_2=\Cu_j(t)^{1/2}\le2$. Stationarity gives $\norm{\dot A}_\rho=\nu_A$, so $|\frac{\dd}{\dd t}\Cr_j|\le8\nu_A$ for every $j$, and the same holds for the target average. When an operator-norm bound $\norm{[H,A]}\le \nu_*$ holds on an event, each $\Cr_j$ is Lipschitz with the deterministic constant $8\nu_*$ there, and Lemma~\ref{app:lem-mesh} applies.

\subsection{The Poisson average}\label{app:sec-poisson}

A bounded analytic function on the chaos strip is the Poisson integral of its values on the edges. The map $\zeta=e^{2\pi z/\beta}$ sends the strip onto the right half-plane and the origin to $\zeta=1$, and the Poisson kernel of the half-plane at $\zeta=1$ gives each edge the density $p_\beta(s)/2$, where
\begin{equation}
p_\beta(s)=\frac2\beta\operatorname{sech}\frac{2\pi s}\beta,\qquad\int_{\mathbb R}p_\beta(s)\,\dd s=1 .
\label{app:eq-poisson-kernel}
\end{equation}
In the setting of Appendix~\ref{app:sec-strip} we call
\begin{equation}
\delta_P=\frac12\int_{\mathbb R}p_\beta(s)\,\Cbar^{\mathrm r}(s)\,\dd s
\label{app:eq-poisson-defect}
\end{equation}
the Poisson defect.

\begin{lemma}\label{app:lem-poisson}
For every Hamiltonian, $\beta>0$, Hermitian contractions $A$ and $B_j$, weights $w_j$ and sign $\tau$, the following hold.
\begin{enumerate}
\item[(i)] $0\le\delta_P\le\delta$, and $\delta_P=\Tr(ARA\,Q)-F(0)$, where $Q$ solves $RQ+QR=2y\cJ(R)y$.
\item[(ii)] If the sources range over the targets with the same weights and $\delta_{P,i}$ is the Poisson defect of the source $B_i$, then $\sum_iw_i\delta_{P,i}\le\Delta$.
\item[(iii)] If $\Cbar^{\mathrm r}$ is Lipschitz with a constant $L>0$ and $\Cbar^{\mathrm r}(t_0)\ge c>0$ for some $|t_0|\le T$, then $\delta_P\ge b\,e^{-2\pi T/\beta}$ with $b=\frac c4\int_{-r}^rp_\beta(s)\,\dd s$ and $r=c/(2L)$.
\end{enumerate}
\end{lemma}

\begin{proof}
(i) The function $F$ is bounded and analytic on the open strip and continuous on its closure, and $F(s-\ii h)=\overline{F(s+\ii h)}$. Its Poisson integral at the origin is therefore
\[
F(0)=\int_{\mathbb R}p_\beta(s)\,\Re F(s+\ii h)\,\dd s ,
\]
and the identity~\eqref{eq:identity} gives $\delta_P=\int p_\beta T-F(0)$. Since $\Cbar^{\mathrm r}\ge0$ and $T\le M$, $0\le\delta_P\le M-F(0)=\delta$. In an eigenbasis of $H$, let $r_n$ and $y_n$ be the eigenvalues of $R$ and $y$. The Fourier transform of $p_\beta$ is $\operatorname{sech}(\beta\omega/4)$, and $\operatorname{sech}[\beta(E_m-E_n)/4]=2y_my_n/(r_m+r_n)$. Hence $\int p_\beta T=\sum_{m,n}(ARA)_{mn}Q_{nm}$ with $Q_{nm}=2y_n\cJ(R)_{nm}y_m/(r_n+r_m)$, which is the solution of $RQ+QR=2y\cJ(R)y$.

(ii) By (i) for each source, $\sum_iw_i\delta_{P,i}=\int p_\beta T_2-F_2(0)$, and Eq.~\eqref{app:eq-autocorrelation} with $\int p_\beta=1$ bounds it by $\norm{\cJ(R)}_2^2-F_2(0)=\Delta$.

(iii) On $|s-t_0|\le r$ the Lipschitz bound gives $\Cbar^{\mathrm r}(s)\ge c/2$. Since $\cosh(x+x')\le e^{|x'|}\cosh x$, $p_\beta(s+t_0)\ge e^{-2\pi|t_0|/\beta}p_\beta(s)$, and
\[
\delta_P\ge\frac c4\int_{-r}^rp_\beta(s+t_0)\,\dd s\ge b\,e^{-2\pi T/\beta}.\qedhere
\]
\end{proof}

The Poisson defect is thus a static quantity, like $\delta$ and $\Delta$, and a crossing at any time up to $T$ forces it above $b\,e^{-2\pi T/\beta}$ without a grid of times.

\section{Phase coexistence and the double average}\label{app:sec-coexist}

Take $n\ge2$ qubits, the magnetization $M_n=n^{-1}\sum_{r=1}^nZ_r$, $J>0$, and
\begin{equation}
H_n=-JnM_n^4=-\frac J{n^3}\Bigl(\sum_{r=1}^nZ_r\Bigr)^4 .
\label{app:eq-coexist-H}
\end{equation}
Embed the qubits in $N=2n$ Majorana operators through $Z_r=\ii\gamma_{2r-1}\gamma_{2r}$. Then $H_n$ is even. We use the Gibbs state $\rho_n=e^{-\beta H_n}/Z_n$ on $(\mathbb C^2)^{\otimes n}$, with $R_n=\rho_n^{1/2}$ and $y=\rho_n^{1/4}$. Every Majorana has source and target weight $1/(2n)$, and the double average includes all ordered pairs, with $\tau=-1$.

A Majorana operator $\gamma_a$ with $a\in\{2r-1,2r\}$ anticommutes with $Z_r$ and commutes with every other $Z_s$. Conjugation by $\gamma_a$ therefore maps a diagonal operator with values $f(z)$ to the diagonal operator with values $f(z^{(r)})$, where $z^{(r)}$ is the configuration $z$ with $z_r$ reversed. Since $[H_n,\gamma_a]=(H_n-\gamma_aH_n\gamma_a)\gamma_a$, $|x^4-x'^4|\le4|x-x'|$ on $[-1,1]$, and a single flip changes $M_n$ by $2/n$,
\begin{equation}
\norm{[H_n,\gamma_a]}=\max_z|H_n(z)-H_n(z^{(r)})|\le8J
\label{app:eq-coexist-energy}
\end{equation}
for all $n$ and $a$.

Let $s(x)=-\frac{1+x}2\log\frac{1+x}2-\frac{1-x}2\log\frac{1-x}2$ be the binary entropy, $I(x)=\log2-s(x)$, and
\begin{equation}
a_c=\min_{0<x\le1}\frac{I(x)}{x^4}.
\label{app:eq-coexist-ac}
\end{equation}
Put $Q(x)=x\operatorname{artanh}x-4I(x)$. Since $I'(x)=\operatorname{artanh}x$, one has $\frac{\dd}{\dd x}(I/x^4)=Q/x^5$, $Q'(x)=x/(1-x^2)-3\operatorname{artanh}x$ and $Q''(x)=(4x^2-2)/(1-x^2)^2$. Thus $Q'$ decreases on $(0,1/\sqrt2)$ from $Q'(0)=0$ and then increases to $+\infty$, so it has exactly one positive zero. Since $Q(x)=-x^2+O(x^6)$ and $Q(x)\to+\infty$ as $x\to1$, $Q$ has exactly one positive zero $x_*\in(0,1)$, with $Q'(x_*)>0$. The minimum in Eq.~\eqref{app:eq-coexist-ac} is attained only at $x_*$, and $Q(x_*)=0$ gives $a_c=\operatorname{artanh}(x_*)/(4x_*^3)$.

\begin{proposition}\label{app:prop-coexist}
At $\beta_c=a_c/J$ the Gibbs states $\rho_n=e^{-\beta_cH_n}/Z_n$ satisfy, for the conjugation average of all $N=2n$ Majorana operators and any source $\gamma_a$ with $\tau=-1$,
\begin{equation}
\frac{V_{R_n}}{n^2}\to v,\qquad\sigma_n\to\sqrt v,\qquad\delta\to\sqrt v-v,
\label{app:eq-coexist-result}
\end{equation}
with $v=w_0w_1(1-q_*)^2\in(0,\frac14)$, where $q_*=\sqrt{1-x_*^2}$, $w_0=1/(1+2A_*)$, $w_1=2A_*/(1+2A_*)$ and $A_*=[(1-x_*^2)Q'(x_*)/x_*]^{-1/2}$. The single-source defect stays bounded away from zero, while the double-averaged defect obeys $\Delta=1/n+O(n^{-2})$. The latter estimate holds at every fixed $\beta J$. For $\gamma_a$ and $\gamma_b$ on different qubits $r$ and $s$, $\Cu_{ab}(t)\le16t^2\norm{B_{rs}}^2$, where the coefficient $B_{rs}$ of $Z_rZ_s$ in $H_n$ has $\norm{B_{rs}}=O(J/n)$. A fixed threshold therefore needs at least a time of order $n/J$.
\end{proposition}

\begin{proof}
\emph{Phases.} With $a=\beta J$, the number of configurations with magnetization $x$ and the Boltzmann factor combine to $\exp[n\psi_a(x)+O(\log n)]$ with $\psi_a(x)=s(x)+ax^4$. At $a=a_c$ the global maxima of $\psi_{a_c}$ are exactly $0$ and $\pm x_*$, all with value $\log 2$, and they are nondegenerate, since $\psi_{a_c}''(0)=-1$ and $\psi_{a_c}''(x_*)=-1/(1-x_*^2)+3\operatorname{artanh}(x_*)/x_*=-Q'(x_*)/x_*<0$. Stirling's formula with its prefactor gives each nondegenerate maximum the weight $A(x)=[(1-x^2)|\psi''_{a_c}(x)|]^{-1/2}$, so $A(0)=1$, $A(\pm x_*)=A_*$ and $Z_n/2^n\to1+2A_*$. The law of $M_n$ under $\rho_n$ converges weakly to $w_0\delta_0+\frac{w_1}2(\delta_{x_*}+\delta_{-x_*})$. The mesh $2/n$ resolves the Gaussian width $O(n^{-1/2})$ of each maximum.

\emph{Size as flip overlaps.} On diagonal operators, written as functions of $z$, the reflections $X_rf(z)=f(z^{(r)})$ satisfy $X_rZ_S=(-1)^{\one_{r\in S}}Z_S$ for $Z_S=\prod_{u\in S}Z_u$. Since $Z_S$ has size $2|S|$, $\cN=\sum_r(1-X_r)$ on diagonal operators. The diagonal operator $R_n$ is the function $r_n(z)=\sqrt{p_n(z)}$ with $p_n(z)=e^{-\beta H_n(z)}/Z_n$. Put $a_n=\langle R_n,X_1R_n\rangle$ and $b_n=\langle R_n,X_1X_2R_n\rangle$. Permutation symmetry gives the exact identities
\begin{equation}
\mu_{R_n}=n(1-a_n),\qquad V_{R_n}=n^2(b_n-a_n^2)+n(1-b_n).
\label{app:eq-coexist-moments}
\end{equation}
At $\beta_c$, uniformly in $z$, $r_n(z^{(r)})/r_n(z)=\exp[-\eta(M_n)z_r+O(1/n)]$ with $\eta(x)=4a_cx^3$, and the ratio for two flips is the product of the two single ratios up to $O(1/n)$. Conditioned on $M_n=x$, finitely many spins are asymptotically independent with mean $x$, so a single flip has overlap $\cosh\eta(x)-x\sinh\eta(x)$. At a maximum, $x=\tanh\eta(x)$, and this overlap equals $1/\cosh\eta(x)=\sqrt{1-x^2}$, namely $1$ at $x=0$ and $q_*$ at $\pm x_*$. Two flips on different spins give the square. Hence $a_n\to a_\infty=w_0+w_1q_*$ and $b_n\to b_\infty=w_0+w_1q_*^2$, and Eq.~\eqref{app:eq-coexist-moments} gives $V_{R_n}/n^2\to b_\infty-a_\infty^2=w_0w_1(1-q_*)^2=v$.

\emph{Defect.} Conjugation by either Majorana operator of qubit $r$ acts on diagonal operators as $X_r$. Thus $\cJ(R_n)=n^{-1}\sum_rX_rR_n$, $m_n=g=a_n$ and $\sigma_n^2=V_{R_n}/n^2$. We evaluate the pair terms $M_{ab}$ and $F_{ab}$ of Eq.~\eqref{eq:syk-defect} at any $\beta>0$.

Fix a source $\gamma_a$ on qubit $r$, and let $\gamma_{\bar a}$ be the other Majorana on that qubit. Writing $y^{(r)}(z)=y(z^{(r)})$, one has $y\gamma_a=\gamma_a y^{(r)}$ and $(\gamma_a\gamma_{\bar a})^2=-I$. Therefore
\[
F_{aa}=-\Tr\bigl[y^2(y^{(r)})^2\bigr]=-a_n,\qquad
F_{a\bar a}=a_n.
\]
Both $I$ and $\gamma_a\gamma_{\bar a}$ commute with $R_n$, so $M_{aa}=M_{a\bar a}=\Tr R_n^2=1$. For a target on a different qubit $s$, the Majorana anticommutation sign cancels $\tau=-1$, giving $M_{ab}=b_n$ and
\[
F_{ab}=f_n:=\sum_z\bigl[p_n(z)p_n(z^{(r)})p_n(z^{(s)})p_n(z^{(r,s)})\bigr]^{1/4}.
\]
For each source, the two targets on its own qubit have total weight $1/n$: their crossed terms cancel, while their nested terms contribute $1/n$. The remaining targets have total weight $1-1/n$. Summing gives
\begin{equation}
\begin{aligned}
F(0)=F_2(0)&=\left(1-\frac1n\right)f_n,\\
\norm{\cJ(R_n)}_2^2&=\frac1n+\left(1-\frac1n\right)b_n,\\
\Delta&=\frac1n+\left(1-\frac1n\right)(b_n-f_n).
\end{aligned}
\label{app:eq-coexist-exact}
\end{equation}

To estimate the difference, for distinct $r,s$ put
\[
u_z=[p_n(z)p_n(z^{(r,s)})]^{1/4},\qquad
v_z=[p_n(z^{(r)})p_n(z^{(s)})]^{1/4}.
\]
The bijection $z\mapsto z^{(r)}$ interchanges $u_z^2$ and $v_z^2$, hence $\sum_z u_z^2=\sum_z v_z^2=b_n$ and $\sum_z u_zv_z=f_n$. Write $Y=\sum_{u\notin\{r,s\}}Z_u$ and
\[
B_{rs}=-\frac{J}{n^3}(12Y^2+8).
\]
The mixed energy difference is
\[
H_n(z)+H_n(z^{(r,s)})-H_n(z^{(r)})-H_n(z^{(s)})
=4z_rz_sB_{rs}(z),
\]
so $u_z/v_z=e^{-\beta z_rz_s B_{rs}(z)}$ and
\begin{equation}
\begin{aligned}
b_n-f_n
&=\frac12\sum_z(u_z-v_z)^2\\
&=\sum_z u_zv_z\bigl[\cosh(\beta B_{rs}(z))-1\bigr].
\end{aligned}
\label{app:eq-coexist-positive}
\end{equation}
Cauchy--Schwarz and $\sum_z p_n(z)=1$ give $f_n\le b_n\le1$. Since $\norm{B_{rs}}\le J[12(n-2)^2+8]/n^3\le12J/n$ for $n\ge2$,
\begin{equation}
\begin{aligned}
0\le b_n-f_n&\le\cosh(\beta\norm{B_{rs}})-1\\
&\le\cosh(12\beta J/n)-1=O_{\beta J}(n^{-2}).
\end{aligned}
\label{app:eq-coexist-rate}
\end{equation}
The bound is uniform in $n$ at fixed $\beta J$, and Eq.~\eqref{app:eq-coexist-exact} proves $\Delta=1/n+O(n^{-2})$.

At $\beta_c$, $\sigma_n^2\to v$ and $F_2(0)\to b_\infty=a_\infty^2+v$. The identity $\delta=\Delta+\sigma_n-\sigma_n^2$ then gives $\delta\to\sqrt v-v$. Here $w_0,w_1>0$, $w_0w_1\le1/4$ and $0<q_*<1$, so $0<v<1/4$ and this limit is positive. In Eq.~\eqref{eq:defect-size}, the same cancellation reads $\bar s+V_{R_n}/n=n\Delta=1+O(n^{-1})$.

\emph{Slow scrambling.} With $Y$ and $B_{rs}$ as above, expanding $(Z_r+Z_s+Y)^4$ gives $H_n=H'+Z_r\Phi$, with $\Phi=\Phi_0+B_{rs}Z_s$, where $H'$ and $\Phi_0$ are diagonal and do not contain $Z_r$, and $\Phi_0$ does not contain $Z_s$. Since $\gamma_a$ anticommutes with $Z_r$ and commutes with the other diagonal operators, $\gamma_a(t)=\gamma_ae^{-2\ii tZ_r\Phi}$. Since $\gamma_b$ anticommutes with $Z_s$ only, $\{\gamma_a(t),\gamma_b\}=-2\ii\gamma_ae^{-2\ii tZ_r\Phi_0}\sin(2tB_{rs}Z_rZ_s)\gamma_b$, whose square is $4\sin^2(2tB_{rs})$. Hence $\Cu_{ab}(t)\le16t^2\norm{B_{rs}}^2$ with $\norm{B_{rs}}\le J[12(n-2)^2+8]/n^3=O(J/n)$.
\end{proof}

\section{Limits of strip continuation from short times}\label{app:sec-strip-limit}

Smallness on a fixed interval of real times can be continued along real time by analyticity, but only up to times of order $\log\log N$. Suppose $\varphi$ is analytic with $|\varphi|\le1$ in the strip $|\Im z|<4r$ and $|\varphi(x)|\le\varepsilon$ for $|x|\le4r$. On the half-disks of radius $4r$ above and below this interval, the two-constants theorem bounds $|\varphi|$ by $\varepsilon^{\omega}$, where the harmonic measure $\omega$ of the diameter is at least $\frac12$ on the disk $|z|\le r$. Hadamard's three-circles theorem with radii $r$, $2r$ and $4r$ then passes a bound $\varepsilon_k$ on a disk of radius $r$ to the concentric disk of radius $2r$, as $\varepsilon_k^{1/2}$. This disk contains the disk of radius $r$ whose center lies a distance $r$ further along the real axis. Hence $\sup_{|x|\le T}|\varphi(x)|\le\varepsilon^{2^{-(\lceil T/r\rceil+1)}}$, which stays small only up to $T$ of order $r\log\log(1/\varepsilon)$. With $\varepsilon$ a power of $1/N$, this continuation argument therefore reaches only times of order $\log\log N$. The entire functions
\begin{equation}
G_n(z)=\frac{1-e^{-k^2z^2}}{2B}\Bigl[e^{-b_ne^{-kz}}+e^{-b_ne^{kz}}\Bigr],\qquad b_n=e^{ka}\log n,
\label{eq:Gn}
\end{equation}
show that no argument using only these properties can do better.

\begin{proposition}\label{app:prop-striplimit}
Fix $r_0>0$, $a>0$ and $0<k<\pi/(2r_0)$, and let $G_n$, $n\ge2$, be the functions of Eq.~\eqref{eq:Gn}, with $B=1+e^{k^2r_0^2}$. Each $G_n$ is entire and even, $|G_n(z)|\le1$ for $|\Im z|\le r_0$, $0\le G_n(x)\le1$ for real $x$, $G_n(0)=0$, and $\sup_{|x|\le a}G_n(x)\le1/(Bn)$. Yet at $T_n=a+k^{-1}\log\log n$,
\begin{equation}
G_n(T_n)\to\frac1{2eB}>0 .
\label{app:eq-Gn-limit}
\end{equation}
\end{proposition}

\begin{proof}
For $z=x+\ii y$ with $|y|\le r_0$, $\cos(ky)>0$ and $|\exp(-b_ne^{\pm kz})|=\exp[-b_ne^{\pm kx}\cos(ky)]\le1$. The bracket in Eq.~\eqref{eq:Gn} divided by two is therefore at most one in modulus. The prefactor obeys $|e^{-k^2z^2}|=e^{-k^2(x^2-y^2)}\le e^{k^2r_0^2}$, hence $|1-e^{-k^2z^2}|\le B$ and $|G_n|\le1$ in the strip. On the real axis both factors are nonnegative, and the first, smaller than one, vanishes at $x=0$. For $|x|\le a$, $b_ne^{\pm kx}\ge\log n$, so each exponential is at most $1/n$ and $G_n(x)\le1/(Bn)$. At $T_n$, $b_ne^{-kT_n}=1$ and $b_ne^{kT_n}=e^{2ka}(\log n)^2\to\infty$. The bracket therefore tends to $e^{-1}$, while $1-e^{-k^2T_n^2}\to1$.
\end{proof}

For any fixed $0<c<1/(2eB)$ and large $n$, the first time at which $G_n$ reaches $c$ is at most $T_n$, which is of order $\log\log n$. The functions $G_n$ show that analyticity in a strip of fixed width, a uniform bound, positivity, time-reversal symmetry, an exact zero at $x=0$ and smallness $1/n$ on a fixed interval are compatible with crossing a fixed threshold at times of order $\log\log n$. Theorem~\ref{thm:static} and Proposition~\ref{prop:double} use a small static defect relative to a uniform edge bound, together with derivative control linking the strip interior to the physical OTOC on its boundary.

\section{The SYK model}\label{app:sec-syk}

This appendix proves the inputs to Theorem~\ref{thm:syk} that are specific to the SYK model with quartic Gaussian couplings. Three of them come from recent work. The local modular strip and a bound on local energies follow from the random Jacobian estimate of Cai, Zhan and Zlokapa~\cite{Cai2026Fast}. Concentration of the partition function, which rests on the commutation index of Anschuetz, Chen, Kiani and King~\cite{Anschuetz2025Strongly}, converts annealed Gibbs averages into averages of same-sample Gibbs expectations. The large-$N$ Wick theorem for Euclidean correlators of Schmidhuber and Zlokapa~\cite{Schmidhuber2026Polynomial} controls the double-averaged defect. The appendix also reduces even probes to odd ones and gives an explicit logarithmic window at high temperature. Appendix~\ref{app:syk-sec-proof} combines these inputs with Proposition~\ref{prop:double} and Lemma~\ref{app:lem-poisson} into the proof of Theorem~\ref{thm:syk}.

\subsection{Model and Gibbs functional}\label{app:syk-sec-model}

Let $N\ge4$ be even and let $\gamma_1,\dots,\gamma_N$ be Majorana operators, $\{\gamma_a,\gamma_b\}=2\delta_{ab}$, acting irreducibly on $\mathbb C^{2^{N/2}}$ as in Appendix~\ref{app:sec-size}. For $E=\{e_1<e_2<e_3<e_4\}$ the product $\Gamma_E=\gamma_{e_1}\gamma_{e_2}\gamma_{e_3}\gamma_{e_4}$ is a Hermitian unitary, and the SYK Hamiltonian of Eq.~\eqref{eq:syk} is $H_N=\kappa_N\sum_{|E|=4}g_E\Gamma_E$ with $\kappa_N=J\sqrt{6/N^3}$ and independent standard Gaussian coefficients $g_E$. Equivalently, $H_N=\sum_{a<b<c<d}J_{abcd}\gamma_a\gamma_b\gamma_c\gamma_d$ with independent centered Gaussian couplings of variance $6J^2/N^3$, and we extend $J_{abcd}$ to an antisymmetric tensor that vanishes when two indices coincide. Another sign convention for the monomials gives the same law of $H_N$. We write $H=H_N$ when $N$ is fixed. The inverse temperature $\beta>0$ does not depend on $N$, and $Z=\Tr e^{-\beta H}$, $\rho=e^{-\beta H}/Z$, $R=\rho^{1/2}$ and $y=\rho^{1/4}$ are formed with the same sample of $H$ that generates the dynamics.

Averages over the couplings are first bounded for the Gibbs functional
\begin{equation}
\Phi(X)=\frac{\E\Tr\bigl(e^{-\beta H}X\bigr)}{\E Z},
\label{app:syk-eq-Phi}
\end{equation}
where the expectation also covers the dependence of $X$ on the couplings. For a scalar random variable $f$ we write $\Phi(f)=\E[Zf]/\E Z$, which is Eq.~\eqref{app:syk-eq-Phi} for $X=fI$. The functional $\Phi$ is positive with $\Phi(I)=1$. It differs from $\E\Tr(\rho X)$ because the random weight $Z$ enters the average. It is the annealed functional of Ref.~\cite{He2026Free} and, up to the normalization of Appendix~\ref{app:syk-sec-euclid}, the functional $\Phi_N$ of Eq.~(5.1) of Ref.~\cite{Schmidhuber2026Polynomial}. Lemma~\ref{app:syk-lem-quenched} converts bounds on $\Phi$ into bounds on same-sample Gibbs expectations.

For complex $z$ we write $X(z)=e^{\ii zH}Xe^{-\ii zH}$, as in Sec.~\ref{sec:results}. Physical time $t$ is real. For labels $i,j$ the source $\gamma_i$ and the target $\gamma_j$ are odd, $\tau=-1$, and
\[
\begin{gathered}
K_{ij}(t)=\{\gamma_i(t),\gamma_j\},\qquad
\Cu_{ij}(t)=\Tr\bigl(\rho K_{ij}(t)^2\bigr),\\
\Cr_{ij}(t)=\Tr\bigl(RK_{ij}(t)RK_{ij}(t)\bigr).
\end{gathered}
\]
By Eq.~\eqref{app:eq-otoc-order}, $0\le\Cr_{ij}\le\Cu_{ij}\le4$.

\subsection{The local modular strip}\label{app:syk-sec-strip}

The local modular strip and the bounds on local energies follow from the random Jacobian estimate of Ref.~\cite{Cai2026Fast}. For real $s$ let $a_k(s)=e^{-sH}\gamma_ke^{sH}$. In the complex-time convention above, $a_k(s)=\gamma_k(\ii s)$ and $\gamma_k(x+\ii s)=e^{\ii xH}a_k(s)e^{-\ii xH}$ for real $x$, so that $\norm{\gamma_k(x+\ii s)}=\norm{a_k(s)}$. Following Ref.~\cite{Cai2026Fast}, let $D\gamma(s)$ be the $N\times N$ block operator on $\mathbb C^N\otimes\mathbb C^{2^{N/2}}$ with entries $(D\gamma(s))_{kl}=\frac12\{\gamma_l,a_k(s)\}$. For $N$-tuples of operators $\boldsymbol b=(b_m)$ and $\boldsymbol b'=(b'_m)$, the block superoperator $\mathsf M(\boldsymbol b,\boldsymbol b')$ acts on $N$-tuples $Y=(Y_l)$ of operators by
\begin{equation}
\begin{split}
\bigl(\mathsf M(\boldsymbol b,\boldsymbol b')Y\bigr)_k=\frac13\sum_{l,m,n}J_{klmn}\bigl(&b_mb_nY_l-b_mY_lb'_n\\
&+Y_lb'_mb'_n\bigr).
\end{split}
\label{app:syk-eq-M}
\end{equation}
Tuples carry the norm $\norm{Y}^2=\sum_l\norm{Y_l}_2^2$, and $\norm{\mathsf M}$ is the corresponding operator norm. With $\boldsymbol a(s)=(a_1(s),\dots,a_N(s))$, let $\cE_N$ be the event on which
\begin{equation}
\begin{aligned}
\norm{D\gamma(s)-I}&\le\frac{133J|s|}{1-133J|s|},\\
\norm{\mathsf M\bigl(\boldsymbol a(s),\boldsymbol a(-s)\bigr)}&\le\frac{133J}{(1-133J|s|)^2}
\end{aligned}
\label{app:syk-eq-czz}
\end{equation}
hold for all real $s$ with $133J|s|<1$.

The event is compatible with relabeling the Majorana operators. For a permutation $\pi$ of the labels, the map $\gamma_k\mapsto\gamma_{\pi(k)}$ preserves the anticommutation relations, so it extends to an automorphism of the full matrix algebra, which is implemented by a unitary $U_\pi$ with $U_\pi\gamma_kU_\pi^\dagger=\gamma_{\pi(k)}$. Then $U_\pi H(g)U_\pi^\dagger=H(g^\pi)$, where $g^\pi_{\pi(E)}=\pm g_E$ with the sign of the permutation that orders $\pi(E)$. The coefficients $g^\pi$ are again independent standard Gaussians. We call an event $\cE$ of the couplings permutation invariant if
\begin{equation}
g\in\cE\iff g^\pi\in\cE\qquad\text{for every }\pi .
\label{app:syk-eq-perm}
\end{equation}

\begin{lemma}[Local modular strip]\label{app:syk-lem-strip}
The event $\cE_N$ has the following properties.
\begin{enumerate}
\item[(i)] $\Pr(\cE_N^c)\le2e^{-N}$. The event is defined by the couplings alone, so it does not depend on $\beta$, and it is permutation invariant.
\item[(ii)] On $\cE_N$, $\norm{\gamma_k(x+\ii s)}\le2$ for all $k$, all real $x$ and all $|s|\le r_0=1/(532J)$.
\item[(iii)] On $\cE_N$, $\norm{[H,\gamma_k]}\le1064J$ and $\norm{[H,\ii\gamma_u\gamma_v]}\le2128J$ for all $k$ and all $u\ne v$.
\end{enumerate}
\end{lemma}

\begin{proof}
(i) Corollary~15 of Ref.~\cite{Cai2026Fast} provides an event of probability at least $1-2e^{-N}$ on which both bounds of Eq.~\eqref{app:syk-eq-czz} hold for all real $s$ with $133J|s|<1$. This event is contained in $\cE_N$. Under $g\mapsto g^\pi$ the operators $a_k(s)$ become $U_\pi a_{\pi^{-1}(k)}(s)U_\pi^\dagger$, and the antisymmetric couplings $J_{klmn}$ become $J_{\pi^{-1}(k)\pi^{-1}(l)\pi^{-1}(m)\pi^{-1}(n)}$. Both block operators of Eq.~\eqref{app:syk-eq-czz} are therefore conjugated by isometries that permute the blocks and conjugate every entry by $U_\pi$. Their norms do not change, which proves Eq.~\eqref{app:syk-eq-perm} for $\cE_N$.

(ii) Fix real $s$ with $266J|s|<1$ and define the mixed Gram matrix $G(s)_{kl}=\frac12\{a_k(s),a_l(-s)\}$. Since $a_l(-s)=a_l(s)^\dagger$, the block operator $G(s)$ is self-adjoint. The entry $(k,l)$ of
\[
(I_N\otimes e^{sH})\,D\gamma(2s)\,(I_N\otimes e^{-sH})
\]
is $\frac12\{e^{sH}\gamma_le^{-sH},e^{sH}a_k(2s)e^{-sH}\}=\frac12\{a_l(-s),a_k(s)\}$, so this operator equals $G(s)$. Equation~(4.4) of Ref.~\cite{Cai2026Fast} is this identity at $s=\beta/4$. A similarity transformation preserves the spectrum, and the norm of the self-adjoint $G(s)$ equals its spectral radius. On $\cE_N$, Eq.~\eqref{app:syk-eq-czz} at $2s$ gives
\[
\begin{split}
\norm{G(s)}&\le\norm{D\gamma(2s)}\le1+\frac{266J|s|}{1-266J|s|}\\
&=\frac1{1-266J|s|}.
\end{split}
\]
With $a_k=a_k(s)$, the diagonal blocks give the operator inequality
\[
a_ka_k^\dagger\le a_ka_k^\dagger+a_k^\dagger a_k=2G(s)_{kk}\le2\norm{G(s)}I .
\]
Hence $\norm{a_k(s)}^2\le2/(1-266J|s|)$, which is at most $4$ for $|s|\le r_0$ because then $266J|s|\le1/2$. Since $\norm{\gamma_k(x+\ii s)}=\norm{a_k(s)}$, this proves (ii).

(iii) The function $z\mapsto\gamma_k(z)$ is entire, with derivative $\ii[H,\gamma_k]$ at $z=0$. On the circle $|z|=r_0$ we have $|\Im z|\le r_0$, so $\norm{\gamma_k(z)}\le2$ by (ii), and the Cauchy estimate gives $\norm{[H,\gamma_k]}\le2/r_0=1064J$. The Leibniz rule gives $\norm{[H,\ii\gamma_u\gamma_v]}\le\norm{[H,\gamma_u]}+\norm{[H,\gamma_v]}\le2128J$.
\end{proof}

\begin{remark}\label{app:syk-rem-lipschitz}
On $\cE_N$ the functions $t\mapsto C^x_{ij}(t)$, $x\in\{\mathrm r,\mathrm u\}$, are Lipschitz with the deterministic constant $8512J$ for all labels $i,j$. For $x=\mathrm r$ this is the constant $8\nu_*$ of Appendix~\ref{app:sec-crossing} with $\nu_*=1064J$. For $x=\mathrm u$, the derivative $\{\ii[H,\gamma_i(t)],\gamma_j\}$ of $K_{ij}(t)$ has norm at most $2128J$ by Lemma~\ref{app:syk-lem-strip}(iii), and $|\frac{\dd}{\dd t}\Cu_{ij}|\le2\norm{\frac{\dd}{\dd t}K_{ij}}\norm{K_{ij}}\le8512J$. Lemma~\ref{app:lem-mesh} therefore applies with the Lipschitz constant $8512J$ and $\cE=\cE_N$.
\end{remark}

\subsection{Concentration of the partition function}\label{app:syk-sec-quenched}

The OTOC and the Gibbs state depend on the same couplings, so a bound on $\Phi$ does not directly bound the average over the couplings of a same-sample Gibbs expectation. The following lemma makes this transfer at every fixed $\beta$, through a lower tail bound on $Z$. It rests on the commutation index of the quartic Majorana monomials,
\[
\Delta_N=\sup_\omega\binom N4^{-1}\sum_{|E|=4}\bigl(\Tr\omega\Gamma_E\bigr)^2,
\]
where the supremum runs over all density matrices $\omega$. Theorem~4 of Ref.~\cite{Anschuetz2025Strongly} shows that $\Delta_N\le C_\Delta N^{-2}$ with an absolute constant $C_\Delta$.

\begin{lemma}[Partition-function bridge]\label{app:syk-lem-quenched}
Let $c_{\beta J}=(\log2)^2/[2C_\Delta(\beta J)^2]$. For every even $N\ge4$ the following hold.
\begin{enumerate}
\item[(i)] The function $f=\log Z$ of the couplings $g=(g_E)$ is Lipschitz with constant $L_N=\sup_g\norm{\nabla f(g)}_2$, and $L_N^2\le C_\Delta(\beta J)^2/(4N)$.
\item[(ii)] $\Pr(Z<\E Z/2)\le2e^{-c_{\beta J}N}$.
\item[(iii)] Every random variable $F=F(g)$ with $0\le F\le F_{\max}$ obeys
\begin{equation}
\E F\le2\Phi(F)+2F_{\max}e^{-c_{\beta J}N},\qquad\Phi(F)=\frac{\E(ZF)}{\E Z}.
\label{app:syk-eq-quenched}
\end{equation}
In particular, $\E F\le2\E(ZF)/\E Z+8e^{-c_{\beta J}N}$ when $0\le F\le4$.
\end{enumerate}
\end{lemma}

\begin{proof}
(i) Duhamel's formula and cyclicity of the trace give $\partial_{g_E}Z=-\beta\kappa_N\Tr(\Gamma_Ee^{-\beta H})$, so $\partial_{g_E}f=-\beta\kappa_N\Tr(\rho\Gamma_E)$. For every sample,
\[
\begin{split}
\norm{\nabla f}_2^2&=\beta^2\kappa_N^2\sum_{|E|=4}\bigl(\Tr\rho\Gamma_E\bigr)^2\le\beta^2\kappa_N^2\binom N4\Delta_N\\
&=\frac{(\beta J)^2(N-1)(N-2)(N-3)}{4N^2}\,\Delta_N\\
&\le\frac{C_\Delta(\beta J)^2}{4N}.
\end{split}
\]
The gradient bound holds at every $g$, so $f$ is Lipschitz with constant $L_N$.

(ii) The Gaussian concentration inequality for Lipschitz functions~\cite{Boucheron2013Concentration} gives $\Pr(|f-\E f|\ge x)\le2e^{-x^2/(2L_N^2)}$ for $x>0$. Applying the moment-generating-function argument of Sec.~III of the Supplemental Material to  Ref.~\cite{Anschuetz2025Strongly} with Lipschitz constant $L_N$ gives $\log\E Z-\E f\le4L_N^2$. Jensen's inequality gives the lower bound $0\le\log\E Z-\E f$. If $N\ge2C_\Delta(\beta J)^2/\log2$, then $4L_N^2\le\frac12\log2$, and $Z<\E Z/2$ forces $f-\E f<-\log2+4L_N^2\le-\frac12\log2$. Hence
\[
\Pr\bigl(Z<\tfrac12\E Z\bigr)\le2\exp\Bigl(-\frac{(\log2)^2}{8L_N^2}\Bigr)\le2e^{-c_{\beta J}N}.
\]
For smaller $N$, $c_{\beta J}N<\log2$ and the right side exceeds one.

(iii) On the event $Z\ge\E Z/2$ we have $F\le2ZF/\E Z$, and on its complement $F\le F_{\max}$. Taking expectations and using (ii) proves Eq.~\eqref{app:syk-eq-quenched}.
\end{proof}

\subsection{Euclidean four-point functions}\label{app:syk-sec-euclid}

The double-averaged defect is controlled by the large-$N$ Wick theorem of Schmidhuber and Zlokapa for Euclidean correlators~\cite{Schmidhuber2026Polynomial}. We first match normalizations.

\emph{Normalizations.} Reference~\cite{Schmidhuber2026Polynomial} writes the Gibbs weight as $e^{\beta H}$ and the Hamiltonian as $H_{\rm SZ}=\sqrt{3/(8N^3)}\sum_EJ_E\Gamma^{\rm SZ}_E$ with standard Gaussian $J_E$, where $\Gamma^{\rm SZ}_E=-\Gamma_E$ for $|E|=4$ (its Eq.~(2.1) and the remark after its Eq.~(2.10)). Its Majorana operators $\psi_a$ are our $\gamma_a$. Since $\sqrt{6/N^3}=4\sqrt{3/(8N^3)}$, the choice $J_E=g_E$ in Eq.~\eqref{eq:syk} gives $-\beta H_N=bH_{\rm SZ}$ with $b=4\beta J$. The Gibbs state of each sample is therefore the state of Ref.~\cite{Schmidhuber2026Polynomial} at inverse temperature $b$, and $\Phi$ is its functional $\Phi_N$ of Eq.~(5.1).

\emph{Wick theorem.} Reference~\cite{Schmidhuber2026Polynomial} works with Grassmann variables $\chi_a(u)$, $u\in[0,1]$, and the normalized functional $\langle\cdot\rangle_N$ of its Eq.~(2.19) at inverse temperature $b$. For distinct times $1>t_1>\dots>t_r>0$ its Eq.~(2.20) reads
\begin{multline*}
\bigl\langle\chi_{a_1}(t_1)\cdots\chi_{a_r}(t_r)\bigr\rangle_N\\
=\frac1{\E Z}\,\E\Tr\bigl(e^{-(1-t_1)\beta H}\gamma_{a_1}e^{-(t_1-t_2)\beta H}\gamma_{a_2}\\
\times\cdots e^{-(t_{r-1}-t_r)\beta H}\gamma_{a_r}e^{-t_r\beta H}\bigr),
\end{multline*}
where consecutive Majorana operators $\gamma_{a_q}$ and $\gamma_{a_{q+1}}$ are separated by $e^{-(t_q-t_{q+1})\beta H}$. For $2k$ insertions at such times and every fixed $b$, its Eq.~(6.23) is the large-$N$ Wick theorem
\begin{multline}
\bigl\langle\chi_{a_1}(t_1)\cdots\chi_{a_{2k}}(t_{2k})\bigr\rangle_N\\
=\sum_\pi(-1)^{\operatorname{cr}(\pi)}\prod_{\{r,s\}\in\pi}2\delta_{a_ra_s}G_*(t_r-t_s)+O_{b,k}(N^{-1}),
\label{app:syk-eq-wick}
\end{multline}
where $\pi$ runs over the pairings of $\{1,\dots,2k\}$, $\operatorname{cr}(\pi)$ is the number of crossings of its pairs, $r<s$ in each pair, and $G_*$ is the solution of the saddle-point equations, Eq.~(2.21) of Ref.~\cite{Schmidhuber2026Polynomial}, normalized by $G_*(0^+)=\frac12$. The error comes from the uniform bound of its Eq.~(6.25) on centered moments, which rests on its Corollary~7, and from its Eq.~(6.27), a bound on a time-independent matrix. It is therefore uniform in the times and the labels. Equation~(6.27) holds for $N$ above some $N_b$, and we enlarge the constant to cover smaller $N$, since the quantities compared below are bounded by one. By Lemma~40 of Ref.~\cite{Schmidhuber2026Polynomial}, $G_*$ extends holomorphically to a neighborhood of $[0,1]$.

\begin{lemma}[Euclidean four-point functions]\label{app:syk-lem-wick}
Let $q_\beta=2G_*(\frac12)$ at inverse temperature $b=4\beta J$, and let $M_{ij}$ and $F_{ij}$ be the four-point functions of Eq.~\eqref{eq:syk-defect}. There is a constant $C_b$ such that for every even $N\ge4$ and all distinct labels $i,j$
\begin{equation}
\bigl|\Phi(M_{ij})-q_\beta^2\bigr|\le\frac{C_b}N,\qquad\bigl|\Phi(F_{ij})-q_\beta^2\bigr|\le\frac{C_b}N .
\label{app:syk-eq-MF}
\end{equation}
\end{lemma}

\begin{proof}
\emph{Crossed pairs.} With the times $(\frac78,\frac58,\frac38,\frac18)$, Eq.~(2.20) of Ref.~\cite{Schmidhuber2026Polynomial} and cyclicity of the trace give $\Phi(F_{ij})=-\langle\chi_i(\frac78)\chi_j(\frac58)\chi_i(\frac38)\chi_j(\frac18)\rangle_N$. For $i\ne j$ only the pairing $\{1,3\},\{2,4\}$ matches the labels. Its pairs cross once, and each spans the time $\frac12$, so Eq.~\eqref{app:syk-eq-wick} gives $-q_\beta^2$ up to $O_b(N^{-1})$. This proves the second bound.

\emph{Nested pairs.} The function $M_{ij}$ places two operators at each of two antipodal points of the thermal circle. For $0<\epsilon<\frac14$ let
\begin{multline*}
f(\epsilon)=\frac1{\E Z}\,\E\Tr\bigl(e^{-\beta H/4}\gamma_ie^{-\epsilon\beta H}\gamma_j\\
\times e^{-(1/2-2\epsilon)\beta H}\gamma_je^{-\epsilon\beta H}\gamma_ie^{-\beta H/4}\bigr),
\end{multline*}
which Eq.~(2.20) of Ref.~\cite{Schmidhuber2026Polynomial} identifies with $\langle\chi_i(\frac34)\chi_j(\frac34-\epsilon)\chi_j(\frac14+\epsilon)\chi_i(\frac14)\rangle_N$. For $i\ne j$ only the nested pairing $\{1,4\},\{2,3\}$ matches the labels, without crossings, and Eq.~\eqref{app:syk-eq-wick} gives $|f(\epsilon)-q_\beta\,2G_*(\frac12-2\epsilon)|\le C_b/N$ uniformly in $\epsilon$. As $\epsilon\to0$ the trace in $f(\epsilon)$ converges for every sample, and H\"older's inequality bounds its modulus by $Z$, with $\E Z<\infty$. Dominated convergence and cyclicity give $f(\epsilon)\to\Phi(M_{ij})$, and the continuity of $G_*$ at $\frac12$ proves the first bound.
\end{proof}

\begin{proposition}[Double-averaged defect]\label{app:syk-prop-defect}
Let every Majorana operator serve as a source and as a target, with equal weights and $\tau=-1$. Then for every even $N\ge4$
\begin{equation}
\E\,\Delta\le\frac{4+4C_b}N+4e^{-c_{\beta J}N},
\label{app:syk-eq-Delta}
\end{equation}
with $C_b$ of Lemma~\ref{app:syk-lem-wick} and $c_{\beta J}$ of Lemma~\ref{app:syk-lem-quenched}.
\end{proposition}

\begin{proof}
By cyclicity, $\norm{\cJ(R)}_2^2=N^{-2}\sum_{i,j}\Tr(\gamma_iR\gamma_i\gamma_jR\gamma_j)=N^{-2}\sum_{i,j}M_{ij}$, and $F_2(0)=N^{-2}\sum_{i,j}F_{ij}$, which is Eq.~\eqref{eq:syk-defect}. The diagonal terms have $M_{ii}=\Tr R^2=1$ and $|F_{ii}|\le1$ by H\"older's inequality, so they contribute at most $2/N$. Lemma~\ref{app:syk-lem-wick} bounds $|\Phi(M_{ij})-\Phi(F_{ij})|$ by $2C_b/N$ for each of the $N(N-1)$ other terms, hence $\Phi(\Delta)\le(2+2C_b)/N$. Proposition~\ref{prop:double} gives $0\le\Delta\le2$ for every sample, since $|F_2(0)|\le\norm{\cJ(R)}_2^2\le1$. Lemma~\ref{app:syk-lem-quenched}(iii) with $F=\Delta$ and $F_{\max}=2$ proves Eq.~\eqref{app:syk-eq-Delta}.
\end{proof}

The bound uses the positivity of $\Delta$ for every sample. The partition-function bridge applied to $M_{ij}$ and $F_{ij}$ separately would have to control $\E[(Z/\E Z-1)X]$ for signed $X$, and the concentration of $\log Z$ bounds the fluctuations of $Z/\E Z$ only at order $N^{-1/2}$.

\subsection{Even probes}\label{app:syk-sec-even}

Even probes reduce to pairs of odd Majorana operators. The reduction holds at each time on the event $\cE_N$, at every temperature, and costs only fixed thresholds. Fix distinct labels $a,b,j,k$ and put $A=\ii\gamma_a\gamma_b$ and $B=\ii\gamma_j\gamma_k$. For $l\in\{j,k\}$ let $K_{A,l}(t)=\ii[A(t),\gamma_l]$, and let $K_{A,B}(t)=\ii[A(t),B]$. These are Hermitian operators of norm at most $2$, and we write $C^x_{A,l}$ and $C^x_{A,B}$, $x\in\{\mathrm r,\mathrm u\}$, for their raw OTOCs, Eq.~\eqref{eq:otocs}. The OTOCs $\Cu_{vl}$ with $v\in\{a,b\}$ are those of Appendix~\ref{app:syk-sec-model}. For $G>0$ and $x>0$ put
\[
\varphi(x)=\frac{x^2}{64},\qquad\lambda_G(x)=\frac x2\operatorname{sech}\frac{\beta G}{\sqrt{2x}},
\]
so that the threshold of Lemma~\ref{app:lem-transfer} is $c_{\beta,G}=\lambda_G(c)$, and define the fixed thresholds
\begin{equation}
\begin{aligned}
\tilde d^{\,\mathrm r}_{\beta,c}&=\varphi(c),&\tilde d^{\,\mathrm u}_{\beta,c}&=\varphi\bigl(\lambda_{6384J}(c)\bigr),\\
d^{\,\mathrm r}_{\beta,c}&=\tilde d^{\,\mathrm u}_{\beta,\varphi(c)},&d^{\,\mathrm u}_{\beta,c}&=d^{\,\mathrm r}_{\beta,\lambda_{8512J}(c)}.
\end{aligned}
\label{app:syk-eq-thresholds}
\end{equation}
Explicitly, $d^{\,\mathrm r}_{\beta,c}=\varphi(\lambda_{6384J}(\varphi(c)))$ and $d^{\,\mathrm u}_{\beta,c}=\varphi(\lambda_{6384J}(\varphi(\lambda_{8512J}(c))))$. All four are positive and depend only on $\beta J$ and $c$.

\begin{lemma}[Even probes]\label{app:syk-lem-even}
Let $\beta>0$ and $c>0$. On $\cE_N$ the following hold for every real $t$.
\begin{enumerate}
\item[(i)] $\norm{[H,K]}\le G$ with $G=4256J$ for $K=K_{vl}(t)=\{\gamma_v(t),\gamma_l\}$, $G=6384J$ for $K=K_{A,l}(t)$ and $G=8512J$ for $K=K_{A,B}(t)$.
\item[(ii)] If $C^x_{A,l}(t)\ge c$, then $\Cu_{vl}(t)\ge\tilde d^{\,x}_{\beta,c}$ for some $v\in\{a,b\}$.
\item[(iii)] If $C^x_{A,B}(t)\ge c$, then $\Cu_{vl}(t)\ge d^{\,x}_{\beta,c}$ for some $v\in\{a,b\}$ and $l\in\{j,k\}$.
\end{enumerate}
Consequently, for $T\ge0$ and $x\in\{\mathrm r,\mathrm u\}$,
\begin{equation}
\begin{split}
\Pr\Bigl[\sup_{0\le t\le T}C^x_{A,B}(t)\ge c\Bigr]\le{}&2e^{-N}\\
&+\sum_{v,l}\Pr\Bigl[\sup_{0\le t\le T}\Cu_{vl}(t)\ge d^{\,x}_{\beta,c}\Bigr],
\end{split}
\label{app:syk-eq-even-window}
\end{equation}
with the sum over the four odd pairs $v\in\{a,b\}$, $l\in\{j,k\}$, and the same bound holds for $C^x_{A,l}$ with the two pairs $v\in\{a,b\}$ and the thresholds $\tilde d^{\,x}_{\beta,c}$.
\end{lemma}

\begin{proof}
(i) For $K=\ii[X(t),Y]$ or $K=\{X(t),Y\}$ with Hermitian contractions $X$ and $Y$, the Leibniz rule and stationarity give $\norm{[H,K]}\le2(\norm{[H,X]}+\norm{[H,Y]})$, as noted after Lemma~\ref{app:lem-transfer}. Lemma~\ref{app:syk-lem-strip}(iii) bounds the local energies by $1064J$ for odd and $2128J$ for even probes, which gives the three constants.

We use Lemma~\ref{app:lem-product} in the following form. If $K$ is Hermitian with $\norm{K}\le2$ and $V$ is unitary, then
\[
\norm{yKVy}_2^2\le2\norm{K}_\rho,\qquad\norm{yVKy}_2^2\le2\norm{K}_\rho,
\]
where $\norm{K}_\rho=\Tr(\rho K^2)^{1/2}$ is the square root of the unregularized OTOC of $K$.

(ii) The identity $[XY,Z]=X\{Y,Z\}-\{X,Z\}Y$ with $X=\gamma_a(t)$, $Y=\gamma_b(t)$ and $Z=\gamma_l$ gives
\[
K_{A,l}(t)=-\gamma_a(t)K_{bl}(t)+K_{al}(t)\gamma_b(t).
\]
The product estimate, with $V=\gamma_a(t)$ and $V=\gamma_b(t)$, and $(\alpha+\alpha')^2\le2\alpha^2+2\alpha'^2$ give
\[
\Cr_{A,l}(t)=\norm{yK_{A,l}(t)y}_2^2\le4\Bigl(\sqrt{\Cu_{al}(t)}+\sqrt{\Cu_{bl}(t)}\Bigr).
\]
If $\Cr_{A,l}(t)\ge c$, the larger square root is at least $c/8$, so $\Cu_{vl}(t)\ge\varphi(c)$ for some $v$. If $\Cu_{A,l}(t)\ge c$, Lemma~\ref{app:lem-transfer} with $G=6384J$ from (i) gives $\Cr_{A,l}(t)\ge\lambda_{6384J}(c)$, and the first case applies with this threshold.

(iii) The Leibniz rule gives $[A(t),\ii\gamma_j\gamma_k]=\ii([A(t),\gamma_j]\gamma_k+\gamma_j[A(t),\gamma_k])$, and $[A(t),\gamma_l]=-\ii K_{A,l}(t)$, so
\[
K_{A,B}(t)=\ii\bigl(K_{A,j}(t)\gamma_k+\gamma_jK_{A,k}(t)\bigr).
\]
The product estimate gives
\[
\Cr_{A,B}(t)\le4\Bigl(\sqrt{\Cu_{A,j}(t)}+\sqrt{\Cu_{A,k}(t)}\Bigr).
\]
If $\Cr_{A,B}(t)\ge c$, some $l\in\{j,k\}$ has $\Cu_{A,l}(t)\ge\varphi(c)$, and (ii) with the threshold $\varphi(c)$ gives some $v$ with $\Cu_{vl}(t)\ge\tilde d^{\,\mathrm u}_{\beta,\varphi(c)}=d^{\,\mathrm r}_{\beta,c}$. If $\Cu_{A,B}(t)\ge c$, Lemma~\ref{app:lem-transfer} with $G=8512J$ gives $\Cr_{A,B}(t)\ge\lambda_{8512J}(c)$, and the regularized case applies with this threshold, which gives $d^{\,\mathrm u}_{\beta,c}$ by Eq.~\eqref{app:syk-eq-thresholds}.

For the window bounds, the OTOCs are continuous in $t$, so a supremum over $[0,T]$ that reaches $c$ is attained at some time. On $\cE_N$, (ii) or (iii) at that time shows that one of the odd pairs reaches its threshold within the window. A union bound over the pairs and $\Pr(\cE_N^c)\le2e^{-N}$ give Eq.~\eqref{app:syk-eq-even-window} and its analogue for $C^x_{A,l}$.
\end{proof}

Equation~\eqref{app:syk-eq-even-window} carries every first-crossing bound for fixed odd pairs over to even sources with even or odd targets, with the thresholds of Eq.~\eqref{app:syk-eq-thresholds} in place of $c$.

\subsection{\texorpdfstring{Proof of Theorem~\ref{thm:syk}}{Proof of Theorem 2}}\label{app:syk-sec-proof}

Fix $0<\epsilon<1$ and put $T_N=(1-\epsilon)[\beta/(2\pi)]\log N$.

\begin{proposition}\label{app:syk-prop-window}
Fix $\beta J>0$, $c>0$ and $0<\epsilon<1$. Let $f$ be either $\Cr_{ij}$ for distinct labels $i$ and $j$, with $K_{ij}(t)=\{\gamma_i(t),\gamma_j\}$, or the target average $\Cbar^{\mathrm r}_i=N^{-1}\sum_l\Cr_{il}$ of a source $i$. There is a constant $C_{\beta J,c,\epsilon}$ such that for all even $N\ge4$
\begin{equation}
\Pr\Bigl[\sup_{0\le t\le T_N}f(t)\ge c\Bigr]\le C_{\beta J,c,\epsilon}N^{-\epsilon}+2e^{-N}.
\label{app:syk-eq-rate}
\end{equation}
The same probability for $\Cu_{ij}$ or $N^{-1}\sum_l\Cu_{il}$ in place of $f$ tends to zero as $N\to\infty$. Both conclusions hold, with other constants, for $K(t)=\ii[\ii\gamma_a(t)\gamma_b(t),\gamma_j]$ and for $K(t)=\ii[\ii\gamma_a(t)\gamma_b(t),\ii\gamma_j\gamma_k]$ with distinct labels.
\end{proposition}

\noindent Proposition~\ref{app:syk-prop-window} proves Theorem~\ref{thm:syk}.

\begin{proof}
Take all Majorana operators as sources and targets, with equal weights and $\tau=-1$. Here and below the constants depend only on $\beta J$, $c$ and $\epsilon$.

\emph{The double average up to the window.} On $\cE_N$ every thermal local energy is at most $1064J$ by Lemma~\ref{app:syk-lem-strip}(iii), and Proposition~\ref{prop:double} gives, for every sample in $\cE_N$ and every real $t$,
\[
\Cdbar^{\mathrm r}(t)\le(4+1064\beta J)\sqrt\Delta\,e^{\pi|t|/\beta}.
\]
Since $e^{\pi T_N/\beta}=N^{(1-\epsilon)/2}$, Jensen's inequality and Proposition~\ref{app:syk-prop-defect} give
\begin{equation}
\sup_{0\le t\le T_N}\E\bigl[\one_{\cE_N}\Cdbar^{\mathrm r}(t)\bigr]\le C_1N^{-\epsilon/2}.
\label{app:syk-eq-average}
\end{equation}

\emph{Permutations.} Let $\pi$ be a permutation of the labels. The sample $g^\pi$ of Appendix~\ref{app:syk-sec-strip} has the law of $g$, the event $\cE_N$ is permutation invariant, and $\Cr_{kl}(t;g^\pi)=\Cr_{\pi^{-1}(k)\pi^{-1}(l)}(t;g)$. Hence $\E[\one_{\cE_N}\Cr_{kl}(t)]$ is the same for all ordered pairs $k\ne l$, and $\E[\one_{\cE_N}\Cbar^{\mathrm r}_k(t)]$ is the same for all sources $k$. Since every $\Cr_{kl}$ is nonnegative,
\begin{equation}
\E\bigl[\one_{\cE_N}f(t)\bigr]\le\frac N{N-1}\,\E\bigl[\one_{\cE_N}\Cdbar^{\mathrm r}(t)\bigr],
\label{app:syk-eq-perm-mean}
\end{equation}
and the same comparison holds for the Poisson defects of these functions, which are nonnegative and linear in them.

\emph{The window by a grid.} On $\cE_N$, $f$ is Lipschitz with the constant $8512J$ (Remark~\ref{app:syk-rem-lipschitz}). Lemma~\ref{app:lem-mesh} with $L=8512J$ and $T=T_N$, together with Eqs.~\eqref{app:syk-eq-average} and~\eqref{app:syk-eq-perm-mean}, gives
\[
\Pr\Bigl[\sup_{0\le t\le T_N}f(t)\ge c\Bigr]\le2e^{-N}+C_2(\log N)\,N^{-\epsilon/2},
\]
which already tends to zero.

\emph{The window by the Poisson defect.} Let $\delta_P(f)=\frac12\int p_\beta(s)f(s)\,\dd s$ be the Poisson defect of $f$, Eq.~\eqref{app:eq-poisson-defect}. Lemma~\ref{app:lem-poisson}(ii) gives $N^{-2}\sum_{k,l}\delta_P(\Cr_{kl})\le\Delta$ for every sample, and Eq.~\eqref{app:syk-eq-perm-mean} with Proposition~\ref{app:syk-prop-defect} gives $\E[\one_{\cE_N}\delta_P(f)]\le\frac N{N-1}\E\,\Delta\le C_3/N$. On $\cE_N$ a crossing of $c$ by $f$ at a time $t\le T_N$ forces $\delta_P(f)\ge b\,e^{-2\pi T_N/\beta}=bN^{-(1-\epsilon)}$ by Lemma~\ref{app:lem-poisson}(iii), with $L=8512J$ and $b>0$ depending only on $\beta J$ and $c$. Markov's inequality gives
\[
\Pr\Bigl[\sup_{0\le t\le T_N}f(t)\ge c\Bigr]\le2e^{-N}+\frac{C_3}bN^{-\epsilon},
\]
which is Eq.~\eqref{app:syk-eq-rate}.

\emph{Unregularized OTOC.} On $\cE_N$, the Leibniz rule and Lemma~\ref{app:syk-lem-strip}(iii) give $\norm{[H,K_{il}(t)]}\le4256J$ for all $t$ and all labels $l$, including $l=i$, as in Lemma~\ref{app:syk-lem-even}(i). Suppose that $\Cu_{ij}(t)\ge c$ or $N^{-1}\sum_l\Cu_{il}(t)\ge c$. Lemma~\ref{app:lem-transfer}, for one operator or for the average with the common bound $G=4256J$, then gives the same inequality for the regularized OTOC with the threshold $c_{\beta,4256J}$, which depends only on $\beta J$ and $c$. Therefore
\[
\begin{split}
\Pr\Bigl[\sup_{0\le t\le T_N}\Cu_{ij}(t)\ge c\Bigr]\le{}&\Pr(\cE_N^c)\\
&+\Pr\Bigl[\sup_{0\le t\le T_N}\Cr_{ij}(t)\ge c_{\beta,4256J}\Bigr],
\end{split}
\]
which tends to zero by Eq.~\eqref{app:syk-eq-rate}, and likewise for the target average.

\emph{Even probes.} Equation~\eqref{app:syk-eq-even-window} with $T=T_N$ bounds the probability of a crossing by an even probe by $2e^{-N}$ plus the probabilities that the unregularized OTOCs of at most four odd pairs cross fixed positive thresholds within the window. The previous step bounds each of these, which gives both conclusions for even probes, with the same window $T_N$ and the same rate.
\end{proof}

The proof uses the dynamics only through the bound on local energies, which sets $\bar\nu$ and the Lipschitz constant. The decay of $\Delta$ comes from Euclidean four-point functions at fixed imaginary times.

\subsection{An explicit rate at high temperature}\label{app:syk-sec-hight}

At high temperature the random Jacobian estimate of Ref.~\cite{Cai2026Fast} also controls the dynamics directly and gives an explicit logarithmic window. The proof uses an exact gradient identity of Ref.~\cite{Cai2026Fast} and the bound on the block superoperator $\mathsf M$ at the imaginary time $\beta/4$. Let $P_f$ be the fermion parity, a Hermitian unitary that anticommutes with every $\gamma_k$ and commutes with $H$. Let $A$ be a fixed Hermitian contraction in the algebra generated by $\gamma_s$, $s\in S$, with $|S|=\ell$, and of parity $p\in\{0,1\}$, $P_fAP_f=(-1)^pA$. For $j=1,\dots,N$ put $K_j(t)=\ii[A(t),\gamma_j]$ if $p=0$ and $K_j(t)=\{A(t),\gamma_j\}$ if $p=1$, and let $C^x_{A,j}$ be the corresponding raw OTOCs, Eq.~\eqref{eq:otocs}. For $A=\gamma_i$ these are the OTOCs $C^x_{ij}$ of Appendix~\ref{app:syk-sec-model}.

\begin{proposition}[High temperature]\label{app:syk-prop-hight}
Let $0\le\beta J<4/133$ and $B_\beta=133J/(1-133\beta J/4)^2$.
\begin{enumerate}
\item[(i)] On $\cE_N$, for all real $t$ and all $T\ge0$,
\begin{equation}
\begin{aligned}
\sum_{j=1}^N\Cr_{A,j}(t)&\le4\ell e^{2B_\beta|t|},\\
\sum_{j\notin S}\,\sup_{0\le t'\le T}\Cr_{A,j}(t')&\le4\ell\bigl(e^{2B_\beta T}-1\bigr).
\end{aligned}
\label{app:syk-eq-hight-sum}
\end{equation}
\item[(ii)] For $j\notin S$, $c>0$, $T\ge0$ and the first crossing time $t_{j,c}=\inf\{t\ge0:\Cr_{A,j}(t)\ge c\}$,
\begin{equation}
\Pr(t_{j,c}\le T)\le2e^{-N}+\frac{4\ell\bigl(e^{2B_\beta T}-1\bigr)}{c\,(N-\ell)}.
\label{app:syk-eq-hight-cross}
\end{equation}
\item[(iii)] Let $A=\gamma_u$, with $\ell=1$ and $G=4256J$, or $A=\ii\gamma_u\gamma_v$, with $\ell=2$ and $G=6384J$. For $j\notin S$ and $t^{\mathrm u}_{j,c}=\inf\{t\ge0:\Cu_{A,j}(t)\ge c\}$,
\begin{equation}
\Pr(t^{\mathrm u}_{j,c}\le T)\le2e^{-N}+\frac{4\ell\bigl(e^{2B_\beta T}-1\bigr)}{c_{\beta,G}\,(N-\ell)},
\label{app:syk-eq-hight-u}
\end{equation}
with the threshold $c_{\beta,G}$ of Lemma~\ref{app:lem-transfer}.
\end{enumerate}
\end{proposition}

For fixed support size $\ell$ and $0<\epsilon<1$, with $T=(1-\epsilon)\log(N-\ell)/(2B_\beta)$, the right sides of Eqs.~\eqref{app:syk-eq-hight-cross} and~\eqref{app:syk-eq-hight-u} are at most $2e^{-N}+4\ell(N-\ell)^{-\epsilon}/c$ and $2e^{-N}+4\ell(N-\ell)^{-\epsilon}/c_{\beta,G}$, and they tend to zero. On $\cE_N$, part (i) also keeps the target average $N^{-1}\sum_j\Cr_{A,j}(t)$ below $c$ for $|t|<(2B_\beta)^{-1}\log[cN/(4\ell)]$.

\begin{proof}
\emph{Tilted Majorana operators and the gradient identity.} Put $\tilde\gamma_k=y\gamma_ky^{-1}=a_k(\beta/4)$, so that $\tilde\gamma_k^\dagger=y^{-1}\gamma_ky=a_k(-\beta/4)$, and $c_k(X)=\frac12(\tilde\gamma_kX-X\tilde\gamma_k^\dagger)$ for every operator $X$. Lemma~4 of Ref.~\cite{Cai2026Fast}, Eqs.~(3.40) and~(3.49), gives for every operator $X$
\begin{equation}
c_k([H,X])-[H,c_k(X)]=\sum_l\mathsf M(\tilde{\boldsymbol\gamma},\tilde{\boldsymbol\gamma}^\dagger)_{kl}\bigl(c_l(X)\bigr),
\label{app:syk-eq-gradient}
\end{equation}
where $\mathsf M(\tilde{\boldsymbol\gamma},\tilde{\boldsymbol\gamma}^\dagger)_{kl}$ is the block $(k,l)$ of Eq.~\eqref{app:syk-eq-M} with $b_m=\tilde\gamma_m$ and $b'_m=\tilde\gamma_m^\dagger$. The identity follows from a direct expansion. Since $y$ commutes with $H$, $H=\frac1{24}\sum J_{klmn}\tilde\gamma_k\tilde\gamma_l\tilde\gamma_m\tilde\gamma_n$, and the tilted operators obey the anticommutation relations of the $\gamma_k$. Hence $[\tilde\gamma_k,H]=\frac13\sum_{l,m,n}J_{klmn}\tilde\gamma_l\tilde\gamma_m\tilde\gamma_n$, and likewise with $\tilde\gamma^\dagger$ in place of $\tilde\gamma$. The left side of Eq.~\eqref{app:syk-eq-gradient} equals $\frac12([\tilde\gamma_k,H]X-X[\tilde\gamma_k^\dagger,H])$. Expanding the right side with Eq.~\eqref{app:syk-eq-M}, the terms in which $X$ stands between tilted operators cancel in pairs by the antisymmetry of $J_{klmn}$, and the remaining terms coincide with the left side. On $\cE_N$, the second bound of Eq.~\eqref{app:syk-eq-czz} at $s=\beta/4$, allowed since $133J\beta/4<1$, gives
\begin{equation}
\norm{\mathsf M(\tilde{\boldsymbol\gamma},\tilde{\boldsymbol\gamma}^\dagger)}\le B_\beta .
\label{app:syk-eq-Mbound}
\end{equation}
This is Corollary~15 of Ref.~\cite{Cai2026Fast}, its Eq.~(5.28), at $s=\beta/4$.

\emph{The OTOC as a gradient.} Put $O(t)=A(t)P_f^p$ and $X(t)=yO(t)y$. Since $P_f$ anticommutes with $\gamma_j$ and commutes with $H$, $[\gamma_j,O(t)]=\hat K_j(t)P_f^p$ with the graded commutator $\hat K_j(t)=\gamma_jA(t)-(-1)^pA(t)\gamma_j$, which is $K_j(t)$ times a phase. The definition of $c_j$ gives $c_j(X(t))=\frac12y[\gamma_j,O(t)]y=\frac12y\hat K_j(t)y\,P_f^p$, because $P_f$ commutes with $y$. The unitary $P_f^p$ does not change the Hilbert--Schmidt norm, so
\begin{equation}
\Cr_{A,j}(t)=\norm{yK_j(t)y}_2^2=4\norm{c_j(X(t))}_2^2 .
\label{app:syk-eq-Cgrad}
\end{equation}

\emph{Growth.} Since $y$ commutes with $H$, $X(t)=e^{\ii tH}X(0)e^{-\ii tH}$. Define the tuple $w(t)=(w_k(t))$ with $w_k(t)=e^{-\ii tH}c_k(X(t))e^{\ii tH}$. By Eq.~\eqref{app:syk-eq-gradient},
\[
w_k'(t)=\ii e^{-\ii tH}\sum_l\mathsf M(\tilde{\boldsymbol\gamma},\tilde{\boldsymbol\gamma}^\dagger)_{kl}\bigl(e^{\ii tH}w_l(t)e^{-\ii tH}\bigr)e^{\ii tH}.
\]
Conjugation of every component by a fixed unitary is an isometry of the tuple space, so Eq.~\eqref{app:syk-eq-Mbound} gives $\norm{w'}\le B_\beta\norm{w}$ on $\cE_N$, and $\norm{w(t)}\le e^{B_\beta|t|}\norm{w(0)}$. This is the growth argument in the proof of Lemma~5 of Ref.~\cite{Cai2026Fast}, Eqs.~(3.54)--(3.59), applied to the physical real-time evolution. By Eq.~\eqref{app:syk-eq-Cgrad}, $\sum_j\Cr_{A,j}(t)=4\norm{w(t)}^2\le e^{2B_\beta|t|}S_0$ with $S_0=\sum_j\Cr_{A,j}(0)$. If $j\notin S$, then $\gamma_j$ anticommutes with every $\gamma_s$, $s\in S$, and $K_j(0)=0$ because $A$ has parity $p$. If $j\in S$, then $\norm{yK_j(0)y}_2\le\norm{K_j(0)}\le2$ by H\"older's inequality. Hence $S_0\le4\ell$, which proves the first bound of Eq.~\eqref{app:syk-eq-hight-sum}.

\emph{Total variation.} Put $q_k(t)=\Cr_{A,k}(t)=4\norm{w_k(t)}_2^2$. Then $|q_k'|\le8\norm{w_k}_2\norm{w_k'}_2$, and the Cauchy--Schwarz inequality over $k$ gives, for $t\ge0$,
\[
\sum_k|q_k'(t)|\le8\norm{w}\norm{w'}\le8B_\beta\norm{w(0)}^2e^{2B_\beta t}=2B_\beta S_0e^{2B_\beta t}.
\]
For $k\notin S$, $q_k(0)=0$ and $\sup_{[0,T]}q_k\le\int_0^T|q_k'|\,\dd t$. Summing over $k\notin S$,
\[
\sum_{k\notin S}\sup_{0\le t\le T}q_k(t)\le S_0\bigl(e^{2B_\beta T}-1\bigr)\le4\ell\bigl(e^{2B_\beta T}-1\bigr),
\]
which is the second bound of Eq.~\eqref{app:syk-eq-hight-sum}.

\emph{Fixed target.} Let $\pi$ be a permutation of the labels that fixes $S$ pointwise, with $U_\pi$ as in Appendix~\ref{app:syk-sec-strip}. Then $U_\pi AU_\pi^\dagger=A$ and $H(g^\pi)=U_\pi H(g)U_\pi^\dagger$, so the OTOCs of the sample $g^\pi$ are $\Cr_{A,j}(t;g^\pi)=\Cr_{A,\pi^{-1}(j)}(t;g)$. Since $g^\pi$ has the law of $g$ and $\cE_N$ is permutation invariant, the expectation $\E[\one_{\cE_N}\sup_{0\le t\le T}\Cr_{A,j}(t)]$ is the same for all $N-\ell$ labels $j\notin S$, and the second bound of Eq.~\eqref{app:syk-eq-hight-sum} gives
\[
\E\Bigl[\one_{\cE_N}\sup_{0\le t\le T}\Cr_{A,j}(t)\Bigr]\le\frac{4\ell\bigl(e^{2B_\beta T}-1\bigr)}{N-\ell}.
\]
By continuity, $t_{j,c}\le T$ if and only if $\sup_{[0,T]}\Cr_{A,j}\ge c$. Markov's inequality on $\cE_N$ and $\Pr(\cE_N^c)\le2e^{-N}$ prove Eq.~\eqref{app:syk-eq-hight-cross}.

\emph{Unregularized OTOC.} For $A=\gamma_u$, $K_j(t)=\{\gamma_u(t),\gamma_j\}$, and for $A=\ii\gamma_u\gamma_v$, $K_j(t)=\ii[A(t),\gamma_j]$. By Lemma~\ref{app:syk-lem-even}(i), $\norm{[H,K_j(t)]}\le G$ on $\cE_N$, with $G=4256J$ and $6384J$ respectively. If $\Cu_{A,j}(t)\ge c$, Lemma~\ref{app:lem-transfer} gives $\Cr_{A,j}(t)\ge c_{\beta,G}$ at the same time. On $\cE_N$, therefore, $t^{\mathrm u}_{j,c}\le T$ implies $\sup_{[0,T]}\Cr_{A,j}\ge c_{\beta,G}$, and the argument of (ii) with $c_{\beta,G}$ in place of $c$ proves Eq.~\eqref{app:syk-eq-hight-u}. At $\beta=0$, the Gibbs state is maximally mixed and $C^{\mathrm r}=C^{\mathrm u}$.
\end{proof}

\section{Double-averaged defect, cavity reduction and Gaussian clipping}\label{app:sec-dyn}

This appendix establishes the static estimate for random spins and reduces the fixed-source problem to a Gaussian source with bounded spectrum. We use the ensemble of Eq.~\eqref{eq:spins}, with fixed $k$ and $d$. All probes are deterministic Hermitian contractions unless stated otherwise. Constants may depend on $k,d,\beta J$ and on a fixed filter, but are independent of $N$, the source and target sites, and the distributions of the interaction labels within the stated assumptions.

\subsection{Boundary normal derivative and swap energy}\label{app:spin-sec-boundary}

Let $B_a$ be mutually commuting Hermitian contractions with weights $w_a\ge0$, $\sum_aw_a=1$. Use these probes as both sources and targets in Proposition~\ref{prop:double}, with $\tau=+1$. Write $R=\rho^{1/2}$, $h=\beta/4$ and $\Delta=\norm{\cJ(R)}_2^2-F_2(0)$.

\begin{lemma}[Boundary derivative]\label{app:spin-lem-boundary}
For every finite-dimensional Hamiltonian,
\begin{equation}
\begin{aligned}
0\le\Delta&\le\frac{\beta}{4\pi}\mathcal E_{\mathrm{swap}},\\
\mathcal E_{\mathrm{swap}}
&=2\sum_{a,b}w_aw_b\Tr\bigl(R[B_a,[H,B_b]]R B_aB_b\bigr)\ge0.
\end{aligned}\label{app:spin-eq-swap}
\end{equation}
\end{lemma}

\begin{proof}
At the upper boundary point, commutativity gives
\[
F_2(\ii h)=\sum_{a,b}w_aw_b\Tr(RB_aB_bR B_aB_b)=\norm{\cJ(R)}_2^2.
\]
The bounded harmonic function $u(z)=\norm{\cJ(R)}_2^2-\Re F_2(z)$ is nonnegative on the strip by Proposition~\ref{prop:double}. Its two boundary values agree; denote them by $b(t)=u(t\pm\ii h)$. Since $b\ge0$ and $b(0)=0$, smoothness gives $b'(0)=0$ and $b(t)=O(t^2)$ near zero.

Put $L=2h=\beta/2$ and translate the strip to $0<v<L$. The Poisson formula at its center and the inward normal derivative at the lower boundary give
\begin{align}
\Delta&=\frac1L\int_{\mathbb R}b(t)\operatorname{sech}\frac{\pi t}{L}\,\dd t,\label{app:spin-eq-center-kernel}\\
d_\partial&=\frac\pi{L^2}\int_{\mathbb R}b(t)
\frac{\cosh(\pi t/L)}{\sinh^2(\pi t/L)}\,\dd t.\label{app:spin-eq-normal-kernel}
\end{align}
Indeed, the sum of the two Poisson kernels at height $v$ is
\[
K_v(t)=\frac{\sin(\pi v/L)}L
\frac{\cosh(\pi t/L)}{\sinh^2(\pi t/L)+\sin^2(\pi v/L)}.
\]
For $0<v<L/2$, $K_v(t)/v$ is bounded by the kernel in Eq.~\eqref{app:spin-eq-normal-kernel}. Multiplication by $b(t)=O(t^2)$ makes this bound integrable at zero, and boundedness of $b$ gives integrability at infinity. Dominated convergence proves the normal-derivative formula. The ratio of the center kernel to the normal-derivative kernel is $(L/\pi)\tanh^2(\pi t/L)$, so
\begin{equation}
\Delta\le\frac\beta{2\pi}d_\partial.\label{app:spin-eq-normal-comparison}
\end{equation}

For $y=\rho^{1/4}$, differentiation at the upper boundary yields
\[
\left.\partial_\eta\bigl(yB_a(\ii\eta)y\bigr)\right|_{\eta=h}
=-R[H,B_a].
\]
The upper and lower inward derivatives agree. With $P_{ab}=B_aB_b$, differentiation of the two source insertions gives
\[
d_\partial=-2\sum_{a,b}w_aw_b\Re\Tr(R[H,B_a]B_bR P_{ab}).
\]
Expand $[B_a,[H,B_b]]=B_aHB_b+B_bHB_a-P_{ab}H-HP_{ab}$. Interchanging $a,b$ in the first two terms and using cyclicity and $[H,R]=0$ in the last two terms gives
\[
d_\partial=\sum_{a,b}w_aw_b\Tr(R[B_a,[H,B_b]]R P_{ab})
=\tfrac12\mathcal E_{\mathrm{swap}}.
\]
Equation~\eqref{app:spin-eq-normal-comparison} proves the result, including positivity of the full sum.
\end{proof}

\subsection{Thermal response and the double-averaged defect}\label{app:spin-sec-response}

For fixed operators $C,O$ define
\[
\Phi(H;C,O)=\Tr(\rho_H^{1/2}C\rho_H^{1/2}O),
\qquad \rho_H=e^{-\beta H}/\Tr e^{-\beta H}.
\]
Along $H_s=H_*+sD$, differentiation of the normalized square root gives
\begin{equation}
\frac{\dd\rho_s^{1/2}}{\dd s}
=-\beta\int_0^{1/2}\rho_s^uD\rho_s^{1/2-u}\,\dd u
+\frac\beta2\Tr(\rho_sD)\rho_s^{1/2}.\label{app:spin-eq-root-response}
\end{equation}
Writing $R_s=\rho_s^{1/2}$, the two square-root derivatives give
\[
\begin{aligned}
\Phi_s'={}&-\beta\int_0^{1/2}
\Tr(\rho_s^uD\rho_s^{1/2-u}C R_sO)\,\dd u\\
&-\beta\int_0^{1/2}
\Tr(R_sC\rho_s^uD\rho_s^{1/2-u}O)\,\dd u\\
&+\beta\Tr(\rho_sD)\Phi_s.
\end{aligned}
\]
In each integrand the nonnegative thermal exponents sum to one. Since $\norm{\rho_s^u}_{1/u}=1$, Schatten H\"older bounds each trace by $\norm D\norm C\norm O$. The two integrals have total length one, and the normalization term has the same bound. Hence
\begin{equation}
|\Phi_s'|\le2\beta\norm D\norm C\norm O.\label{app:spin-eq-response}
\end{equation}

\begin{proposition}[Centered interaction estimate]\label{app:spin-prop-defect}
Let $H=\sum_eD_e$ have independent centered Hermitian labels, and let the commuting probes $B_a$ have deterministic supports $S_a$. Put
\[
v_e=\E\norm{D_e}^2,\qquad
p_e=\sum_{a:S_a\cap e\ne\varnothing}w_a.
\]
Then
\begin{equation}
\E\Delta\le\frac{4\beta^2}{\pi}\sum_ev_ep_e^2.\label{app:spin-eq-contact-bound}
\end{equation}
For the ensemble of Eq.~\eqref{eq:spins}, with one probe on each site and weights $1/N$,
\begin{equation}
\E\Delta\le\frac{4\beta^2J^2}{\pi N}.\label{app:spin-eq-single-defect}
\end{equation}
\end{proposition}

\begin{proof}
Condition on all labels except $D_e$, and set $H_*=H-D_e$, $C_e=[B_a,[D_e,B_b]]$, $O=B_aB_b$. Independence and centering give $\E_e\Phi(H_*;C_e,O)=0$. Keeping $C_e$ fixed along $H_*+sD_e$, we have
\[
\E_e\Phi(H;C_e,O)
=\E_e\int_0^1\frac{\dd}{\dd s}
\Phi(H_*+sD_e;C_e,O)\,\dd s.
\]
Equation~\eqref{app:spin-eq-response} and $\norm{C_e}\le4\norm{D_e}$ give
\begin{equation}
\bigl|\E\Phi(H;C_e,O)\bigr|\le8\beta v_e.\label{app:spin-eq-centered-response}
\end{equation}
The Jacobi identity and $[B_a,B_b]=0$ imply $C_e=0$ unless $e$ intersects both $S_a$ and $S_b$. Summing Eq.~\eqref{app:spin-eq-centered-response} with the pair weights and applying Lemma~\ref{app:spin-lem-boundary} proves Eq.~\eqref{app:spin-eq-contact-bound}. For single-site probes $p_e=k/N$, and
\[
\sum_ev_ep_e^2\le\binom Nk b_N^2\frac{k^2}{N^2}
=\frac{J^2}{N}\prod_{\ell=1}^{k-1}\left(1-\frac\ell N\right)
\le\frac{J^2}{N}.
\]
\end{proof}

\subsection{Multi-site probe families}\label{app:spin-sec-frame}

Choose a Hermitian basis on one site with $\Tr(Q^aQ^b)=\delta_{ab}$, $Q^0=I/\sqrt d$, and $0\le a<d^2$. These matrices have norm at most one. On $n$ sites, the complete $r$-site family consists of
\[
Q_S^\nu=\bigotimes_{j\in S}Q^{\nu_j},\qquad |S|=r,
\]
with equal weights $[\binom nr d^{2r}]^{-1}$. Denote its double-averaged defect by $\Delta_r$.

\begin{lemma}[Complete local frame]\label{app:spin-lem-frame}
Remove one site from the ensemble of Eq.~\eqref{eq:spins}, let $n=N-1$, and retain the coefficient $b_N$ on the remaining interactions. For fixed $r$ and $n\ge2r$,
\begin{equation}
\E\Delta_r\le\frac{4r^2}{n}+\frac{4\beta^2J^2r^2}{\pi n}.\label{app:spin-eq-frame-defect}
\end{equation}
\end{lemma}

\begin{proof}
For Hermitian contractions $A,B$, write the pair contribution to the defect as
\[
d_\rho(A,B)=\Tr(RABRBA)-\Tr(yAyByAyB),
\qquad y=\rho^{1/4}.
\]
Each trace has modulus at most one by Schatten H\"older. The defect of a weighted family is $\sum_{a,b}w_aw_b d_\rho(B_a,B_b)$.

Independently of the couplings, choose one index $\xi_j$ uniformly from $\{0,\ldots,d^2-1\}$ at each site. For fixed $\xi$, the probes $B_S^\xi=\bigotimes_{j\in S}Q^{\xi_j}$ commute, with weights $\binom nr^{-1}$. When $S,T$ are disjoint, their basis indices are independent and uniform, so
\[
\E_\xi d_\rho(B_S^\xi,B_T^\xi)
=d^{-4r}\sum_{\nu,\mu}d_\rho(Q_S^\nu,Q_T^\mu).
\]
For fixed $S$, a union bound over its $r$ sites gives the fraction of intersecting support pairs:
\[
\chi_{n,r}=1-\frac{\binom{n-r}{r}}{\binom nr}\le\frac{r^2}{n}.
\]
The two pair contributions can differ by at most four on these intersecting pairs. Thus, sample by sample,
\[
|\Delta_r-\E_\xi\Delta_\xi|\le4\chi_{n,r}.
\]
For a bath interaction, its contact weight is at most $rk/n$. The normalization of Eq.~\eqref{eq:spins} gives
\[
\sum_{e\subset\mathrm{bath}}v_ep_e^2
\le\frac{J^2r^2}{n}\left(\frac nN\right)^{k-1}
\le\frac{J^2r^2}{n}.
\]
Apply Proposition~\ref{app:spin-prop-defect} to the commuting family and then average over $\xi$.
\end{proof}

We also use $T^a=\sqrt d Q^a$, so $T^0=I$, $d^{-1}\Tr(T^aT^b)=\delta_{ab}$ and $\sum_a(T^a)^2=d^2I$. Replacing $Q_S^\nu$ by $T_S^\nu$ multiplies the defect and the double-averaged squared commutator by $d^{2r}$.

\subsection{Thermal moments and local work}\label{app:spin-sec-moments}

\begin{lemma}[Positive thermal moments]\label{app:spin-lem-moments}
Condition on a background $H_0$. Under this conditional law, let $H=H_0+\sum_eV_e$, where the independent labels are either centered Hermitian matrices with $\norm{V_e}\le b_e$, or scalar Gaussian directions $V_e=g_eL_e$ with $\norm{L_e}\le b_e$. The bounds $b_e$ and Gaussian directions $L_e$ are fixed under this conditioning. Let $W=\sum_eM_e$ be Hermitian, where $M_e$ is a centered linear function of the same label and has norm, or Gaussian direction norm, at most $2b_e$. If $v=\sum_eb_e^2\le\bar v$, then
\begin{align}
\E\Tr(\rho_H e^{zW})
&\le e^{3\beta^2v+8\beta v|z|+8vz^2},\qquad z\in\mathbb R,\label{app:spin-eq-linear-moment}\\
\E\Tr\left(\rho_H e^{W^2/(256\bar v)}\right)
&\le3e^{4\beta^2\bar v},\qquad \bar v>0.\label{app:spin-eq-square-moment}
\end{align}
\end{lemma}

\begin{proof}
Put $Z_0=\Tr e^{-\beta H_0}$, $R_Z=Z/Z_0$ and
\[
Y(z)=Z_0^{-1}\Tr(e^{-\beta H}e^{zW})\ge0.
\]
Expand the first exponential by Duhamel's formula:
\[
\begin{aligned}
e^{-\beta H}
={}&\sum_{q\ge0}(-1)^q\sum_{e_1,\ldots,e_q}
\int_{\substack{t_j\ge0\\t_0+\cdots+t_q=\beta}}
e^{-t_0H_0}V_{e_1}e^{-t_1H_0}\cdots\\[-2pt]
&\hspace{42mm}\cdots V_{e_q}e^{-t_qH_0}\,\dd\mathbf t .
\end{aligned}
\]
Here $t_0=\beta-\sum_{j=1}^qt_j$ and $\dd\mathbf t=\dd t_1\cdots\dd t_q$, so the simplex has volume $\beta^q/q!$. Expand $e^{zW}$ as well. Schatten H\"older bounds each trace by $Z_0$ times the product of the insertion norms, because the background heat times sum to $\beta$. An insertion from $W$ costs at most twice one from $H-H_0$. Summing over their positions combines the two lengths into
\[
x=\beta+2|z|,\qquad
\sum_{q+l=m}\frac{\beta^q(2|z|)^l}{q!\,l!}
=\frac{x^m}{m!}.
\]
For a bounded label, a total multiplicity one vanishes by centering, while multiplicity $m\ge2$ costs at most $b_e^m$. For a Gaussian direction, odd multiplicities vanish and multiplicity $2j$ costs at most $(2j-1)!!\,b_e^{2j}$. The respective generating functions are
\[
\begin{aligned}
h_e(x)&=1+\sum_{m\ge2}\frac{(b_ex)^m}{m!}
=e^{b_ex}-b_ex,\\
h_e^{\mathrm g}(x)&=\sum_{j\ge0}
\frac{(2j-1)!!(b_ex)^{2j}}{(2j)!}
=e^{b_e^2x^2/2}.
\end{aligned}
\]
Independence permits the product over labels. Since $e^u-u\le e^{u^2}$ for $u\ge0$, the absolutely summed expansion gives
\[
\E Y(z)\le e^{v(\beta+2|z|)^2}.
\]
For $Y(z)^2$, expand the two traces together, keeping the same random labels in both. Multiplicities are counted across the two traces, and the combined length is $2\beta+4|z|$. The same generating functions consequently give
\begin{equation}
\E Y(z)^2\le e^{4v(\beta+2|z|)^2}.\label{app:spin-eq-replica-moment}
\end{equation}
Peierls' inequality gives $R_Z\ge\exp[-\beta\Tr(\rho_{H_0}\sum_eV_e)]$. Conditional on $H_0$, the scalar variables $\Tr(\rho_{H_0}V_e)$ are centered and independent, with bounds $b_e$ or Gaussian variances at most $b_e^2$. Hoeffding's bound and the Gaussian generating function therefore give $\E R_Z^{-2}\le e^{2\beta^2v}$. Cauchy--Schwarz now yields
\[
\begin{aligned}
\E\Tr(\rho_H e^{zW})
&=\E[R_Z^{-1}Y(z)]\\
&\le\bigl(\E R_Z^{-2}\,\E Y(z)^2\bigr)^{1/2}\\
&\le e^{\beta^2v+2v(\beta+2|z|)^2},
\end{aligned}
\]
which is Eq.~\eqref{app:spin-eq-linear-moment}.

For a standard real Gaussian $g$, functional calculus gives $e^{\lambda W^2}=\E_g e^{\sqrt{2\lambda}gW}$. With $\lambda=(256\bar v)^{-1}$, Eq.~\eqref{app:spin-eq-linear-moment} bounds its thermal expectation by
\[
e^{3\beta^2\bar v}\E_g
\exp\left(\beta\sqrt{\bar v/2}|g|+g^2/16\right)
\le3e^{4\beta^2\bar v}.
\]
All spectral integrands are positive, so Tonelli's theorem justifies this last average; when $v=0$, $W=0$ and the first bound holds trivially.
\end{proof}

For a deterministic local unitary $U$, let $W_U=U^\dagger HU-H$. Taking the interactions meeting its support as the labels in Lemma~\ref{app:spin-lem-moments} gives
\begin{equation}
\norm{[H,U]}_{\rho_H}^2=\Tr(\rho_HW_U^2),\qquad
\E\Tr(\rho_He^{\lambda W_U^2})\le C,\label{app:spin-eq-local-work}
\end{equation}
for fixed support size and some $\lambda>0$ independent of $N$. In the bounded model a unitary on at most $\ell$ sites has variance budget at most $\ell J^2/k$. Every deterministic local Hermitian contraction is a convex combination of at most $2^{d^\ell}$ Hermitian involutions: diagonalize it and represent each eigenvalue in $[-1,1]$ as the mean of a sign. This decomposition is independent of the couplings. Convexity of the squared norm extends the energy bound to these contractions. In particular,
\begin{equation}
\E\bar\nu^2\le\sum_aw_a\E \nu_{B_a}^2\le C.\label{app:spin-eq-average-energy}
\end{equation}
Together with Lemma~\ref{app:spin-lem-frame} and Proposition~\ref{prop:double}, this gives an expected bath double-averaged OTOC at most $CN^{-1/2}e^{\pi|t|/\beta}$ for every fixed support size.

\begin{lemma}[Gaussian operator moments]\label{app:spin-lem-gaussian}
Let $X=\sum_\ell g_\ell L_\ell$, with independent standard real Gaussians and Hermitian directions. If $\sum_\ell L_\ell^2\le s^2I$, then, for integers $p\ge1$ and $\lambda\ge0$,
\begin{equation}
\E X^{2p}\le4^pp!s^{2p}I,\qquad
\E e^{\lambda X^2}\le\frac{I}{1-4\lambda s^2}\quad(4\lambda s^2<1).
\label{app:spin-eq-gaussian-operator}
\end{equation}
\end{lemma}

\begin{proof}
Let $\mathcal H_q$ be the degree-$q$ Gaussian Hermite space tensored with the physical Hilbert space. Multiplication by $X$ is $\mathcal C+\mathcal C^\dagger$, where $\mathcal C$ raises the Hermite degree. The matrix column
\[
\mathcal L\psi=(L_\ell\psi)_\ell,\qquad
\norm{\mathcal L\psi}^2
=\langle\psi,\textstyle\sum_\ell L_\ell^2\psi\rangle
\le s^2\norm\psi^2
\]
has norm at most $s$. On $\mathcal H_q$, creation applies this column and symmetrizes the extra tensor leg, with normalization $\sqrt{q+1}$. Hence
\[
\norm{\mathcal C|_{\mathcal H_q}}\le s\sqrt{q+1},
\qquad
\norm{\mathcal C^\dagger|_{\mathcal H_q}}\le s\sqrt q.
\]
The outputs of distinct homogeneous degrees are orthogonal for each of these two maps. On degrees at most $q$, multiplication by $X$ thus has norm at most $s(\sqrt{q+1}+\sqrt q)$ and raises the maximal degree by one. Starting with a constant vector $\psi$ and multiplying $p$ times gives
\[
\E\norm{X^p\psi}^2
\le s^{2p}\prod_{q=0}^{p-1}(\sqrt{q+1}+\sqrt q)^2\norm\psi^2
\le4^pp!s^{2p}\norm\psi^2.
\]
Since $X$ is Hermitian, $\E\norm{X^p\psi}^2=\langle\psi,\E X^{2p}\psi\rangle$, proving the operator inequality. For $\lambda\ge0$, summing these positive terms gives
\[
\E e^{\lambda X^2}
\le\sum_{p\ge0}(4\lambda s^2)^p I
=\frac{I}{1-4\lambda s^2}.
\]
\end{proof}

\subsection{Source isolation and comparison of states}\label{app:spin-sec-states}

From this point through the parameter choice in Appendix~\ref{app:sec-cond}, take $J=1$. Relabel the fixed source as site $0$, and write
\begin{equation}
H_o=K+V_o,\qquad
K=b_N\sum_{e:0\notin e}H_e,\qquad
V_o=b_N\sum_{e:0\in e}H_e.\label{app:spin-eq-cavity}
\end{equation}
The bath reference state on the full space is
\[
\rho_K=\frac{e^{-\beta K}}{\Tr e^{-\beta K}}
=\frac{I_0}{d}\otimes\rho_{\mathrm{bath}}.
\]
For deterministic target weights $w_j\ge0$, $\sum_jw_j=1$, define
\begin{equation}
p_{\eta,H'}^x(w)
=\E\sup_{0\le t\le T_N}\sum_jw_j C_{\eta,H';j}^x(t),
\qquad x\in\{\mathrm r,\mathrm u\}.\label{app:spin-eq-window-definition}
\end{equation}
Here the commutator is $Q_j(t)=\ii[e^{\ii tH'}A_0e^{-\ii tH'},B_j]$ and $C_{\eta,H';j}^{\mathrm u}=\Tr(\eta Q_j^2)$, $C_{\eta,H';j}^{\mathrm r}=\Tr(\eta^{1/2}Q_j\eta^{1/2}Q_j)$. We omit $w$ for uniform weights. Below, $S(\tau\Vert\sigma)=\Tr\tau(\log\tau-\log\sigma)$ denotes the relative entropy.

\begin{lemma}[Quench entropy]\label{app:spin-lem-entropy}
Let $V$ be Hermitian, $f$ real with $|f(x)|\le|x|$, $D=f(V)$, $H_f=K+D$, and $\rho_f$ its Gibbs state. Fix $\lambda>0$ and put $\ell_V=\log\Tr(\rho_Ke^{\lambda V^2})$ and $a=\beta/\sqrt\lambda$. For every state $\tau$,
\begin{equation}
\Tr(\tau V^2)\le\lambda^{-1}[S(\tau\Vert\rho_K)+\ell_V].\label{app:spin-eq-variational}
\end{equation}
If $\tau_t=e^{\ii tH_f}\tau e^{-\ii tH_f}$ and $h_t=S(\tau_t\Vert\rho_K)$, then
\begin{align}
\sqrt{h_t+\ell_V}&\le\sqrt{h_0+\ell_V}+a,\label{app:spin-eq-quench-entropy}\\
S(\rho_f\Vert\rho_K)+S(\rho_K\Vert\rho_f)
&\le2a\sqrt{\ell_V}+a^2.\label{app:spin-eq-two-entropies}
\end{align}
\end{lemma}

\begin{proof}
The Gibbs variational formula followed by Golden--Thompson gives
\[
\lambda\Tr(\tau V^2)-S(\tau\Vert\rho_K)
\le\log\Tr e^{\log\rho_K+\lambda V^2}
\le\ell_V.
\]
Conservation of $H_f$ and unitary invariance of entropy give
\[
h_t-h_0=\beta\Tr[(\tau-\tau_t)D]
\le a(\sqrt{h_0+\ell_V}+\sqrt{h_t+\ell_V}).
\]
Factor the difference of squares to obtain Eq.~\eqref{app:spin-eq-quench-entropy}. For the two Gibbs states, the exact identity
\[
S(\rho_f\Vert\rho_K)+S(\rho_K\Vert\rho_f)
=\beta\Tr[(\rho_K-\rho_f)D]
\]
gives a scalar quadratic inequality. If its left side is $s$, then $S(\rho_f\Vert\rho_K)\le s$ and Eq.~\eqref{app:spin-eq-variational} imply
\[
s\le a\bigl(\sqrt{\ell_V}+\sqrt{s+\ell_V}\bigr).
\]
Dividing the difference of squares by $\sqrt{s+\ell_V}+\sqrt{\ell_V}$ yields $\sqrt{s+\ell_V}\le\sqrt{\ell_V}+a$, proving Eq.~\eqref{app:spin-eq-two-entropies}.
\end{proof}

For $V_o$, conditioning on the bath and expanding only the centered source labels gives $\E\Tr(\rho_Ke^{zV_o})\le e^{vz^2}$ with $v\le1/k$. Gaussian integration as above supplies a fixed positive square-exponential parameter. For Gaussian sources the stronger operator bound of Lemma~\ref{app:spin-lem-gaussian} applies. Thus the expectation of the right side of Eq.~\eqref{app:spin-eq-two-entropies} is bounded uniformly for these sources and for every clipping function with $f^2\le V^2$.

For a target involution $B$, set $W_B=BKB-K$ and $\ell_B=\log\Tr(\rho_Ke^{\lambda W_B^2})$, decreasing $\lambda$ if necessary. Write $\ell=\max(\ell_V,\ell_B)$ and $h(\tau)=S(\tau\Vert\rho_K)$. A target operation obeys
\[
\begin{aligned}
h(B\tau B)-h(\tau)&=\beta\Tr(\tau W_B)
\le a\sqrt{h(\tau)+\ell},\\
\sqrt{h(B\tau B)+\ell}
&\le\sqrt{h(\tau)+\ell}+a/2.
\end{aligned}
\]
The second line follows by completing the square. A segment of $H_f$ evolution increases this square root by at most $a$, by Eq.~\eqref{app:spin-eq-quench-entropy}. Bath evolution and source-site unitaries preserve it. Thus a unitary word $R'$ containing at most $m$ quench segments and target operations satisfies
\[
\sqrt{h(R'\eta R'^\dagger)+\ell}
\le\sqrt{h(\eta)+\ell}+ma.
\]
For $X=D$ or $W_B$, a left unitary $L'$ drops out of the thermal norm, and the variational bound gives
\[
\begin{aligned}
\norm{L'XR'}_\eta^2
&=\Tr(R'\eta R'^\dagger X^2)\\
&\le\lambda^{-1}
\bigl(\sqrt{h(\eta)+\ell}+ma\bigr)^2.
\end{aligned}
\]
For $\eta=\rho_K$ the initial entropy is zero; for $\eta=\rho_f$ it is at most $2a\sqrt{\ell_V}+a^2$. For fixed $m$, both bounds are at most $C_m(1+\ell)$, uniformly in all propagation times.

Apply this estimate to $Q_j(t)=\ii[UA_0U^\dagger,B]$, where $U=e^{\ii tH_f}$ and the probes are first taken to be involutions. In the reference state, use
\[
[K,e^{\ii tH_f}]=-[D,e^{\ii tH_f}],\quad
[K,A_0]=0,\quad [K,B]=BW_B.
\]
For example,
\[
\begin{aligned}
[K,UA_0U^\dagger B]
={}&[K,U]A_0U^\dagger B
 +UA_0[K,U^\dagger]B\\
&+UA_0U^\dagger BW_B.
\end{aligned}
\]
Expanding the two propagator commutators gives five terms, each with one field insertion and a bounded number of unitary letters. The other word in $Q_j$ has the same count. In the state $\rho_f$, differentiate with respect to $H_f$ instead: the propagator derivatives vanish, while
\[
[H_f,A_0]=[D,A_0],\qquad
[H_f,B]=BW_B+[D,B].
\]
Each differentiated word again has five single-insertion terms. The preceding estimate and Cauchy--Schwarz bound both squared commutator norms by $C(1+\ell)$, sample by sample and uniformly in $t$. The source moments and Lemma~\ref{app:spin-lem-moments} give $\E\ell\le C$. Consequently,
\begin{equation}
\begin{aligned}
\E\sum_jw_j\sup_t\norm{[K,Q_j(t)]}_{\rho_K}^2&\le C,\\
\E\sum_jw_j\sup_t\norm{[H_f,Q_j(t)]}_{\rho_f}^2&\le C.
\end{aligned}
\label{app:spin-eq-word-energy}
\end{equation}
Lemma~\ref{app:spin-lem-moments} bounds the local-work moments for each target; their weighted sum has the same constant. The fixed convex decompositions of the probes extend Eq.~\eqref{app:spin-eq-word-energy} from involutions to contractions.

\begin{lemma}[State and regularization comparisons]\label{app:spin-lem-state-comparison}
For the original or Gaussian source and $H_f=K+f(V)$ as above, the four statements
\begin{equation}
\begin{aligned}
p_{\rho_K,H_f}^{\mathrm r}(w)\to0
&\iff p_{\rho_K,H_f}^{\mathrm u}(w)\to0\\
&\iff p_{\rho_f,H_f}^{\mathrm u}(w)\to0\\
&\iff p_{\rho_f,H_f}^{\mathrm r}(w)\to0
\end{aligned}\label{app:spin-eq-four-curves}
\end{equation}
are equivalent. The comparisons are uniform in the clipping function and in deterministic target weights.
\end{lemma}

\begin{proof}
For $\eta=e^{-\beta L}/\Tr e^{-\beta L}$ and Hermitian $Q$, splitting energy gaps at $\Omega>0$ gives
\begin{equation}
\Tr(\eta Q^2)\le\cosh(\beta\Omega/2)\Tr(\eta^{1/2}Q\eta^{1/2}Q)
+\frac{\norm{[L,Q]}_\eta^2}{\Omega^2}.\label{app:spin-eq-spectral-transfer}
\end{equation}
Indeed, in an eigenbasis of $L$, let $p_m=e^{-\beta E_m}/Z$. Hermiticity of $Q$ gives
\[
\begin{aligned}
\Tr(\eta Q^2)&=\sum_{m,n}\frac{p_m+p_n}{2}|Q_{mn}|^2,\\
C_\eta^{\mathrm r}(Q)&=\sum_{m,n}\sqrt{p_mp_n}|Q_{mn}|^2,\\
\norm{[L,Q]}_\eta^2
&=\sum_{m,n}\frac{p_m+p_n}{2}(E_m-E_n)^2|Q_{mn}|^2.
\end{aligned}
\]
On $|E_m-E_n|\le\Omega$, the ratio of the first two weights is at most $\cosh(\beta\Omega/2)$; on the complement, the first sum is bounded by the third divided by $\Omega^2$. This proves Eq.~\eqref{app:spin-eq-spectral-transfer}. Take the window supremum and use Eq.~\eqref{app:spin-eq-word-energy}, with $L=K$ for $\eta=\rho_K$ and $L=H_f$ for $\eta=\rho_f$. If the regularized expectation is $p<1$, choosing $\Omega=\beta^{-1}\log(1/p)$ gives an unregularized expectation at most
\begin{equation}
\sqrt p+\frac{C\beta^2}{\log^2(1/p)}.\label{app:spin-eq-regularization-map}
\end{equation}
For $p=0$ use faithfulness of the Gibbs state. The reverse implication follows from $C^{\mathrm r}\le C^{\mathrm u}$.

For the change of state, choose a maximizing time for the new state's cost on a finite grid, breaking ties by the order of the grid, and set
\[
E_\omega=\frac14\sum_jw_jQ_j(t_*(\omega))^2,\qquad 0\le E_\omega\le I.
\]
The same effect is measured in the two states of sample $\omega$. Write $p_\omega=\Tr(\eta_{\mathrm{new}}E_\omega)$ and $q_\omega=\Tr(\eta_{\mathrm{old}}E_\omega)$. The binary relative entropy
\[
d(p\Vert q)=p\log\frac pq+(1-p)\log\frac{1-p}{1-q}
\]
obeys, by data processing and the log-sum inequality,
\[
d(\E p_\omega\Vert\E q_\omega)
\le\E d(p_\omega\Vert q_\omega)
\le\E S(\eta_{\mathrm{new}}\Vert\eta_{\mathrm{old}}).
\]
Finally, $d(p\Vert q)\ge p\log(1/q)-\log2$. Thus
\[
\bar p\log(1/\bar q)\le\E S(\eta_{\mathrm{new}}\Vert\eta_{\mathrm{old}})+\log2,
\]
where $\bar p=\E\Tr(\eta_{\mathrm{new}}E_\omega)$ and $\bar q=\E\Tr(\eta_{\mathrm{old}}E_\omega)$. If the old window cost is $p$, then $\bar q\le p/4$. Equation~\eqref{app:spin-eq-two-entropies} therefore bounds the new unregularized window cost by
\begin{equation}
\min\left\{4,\frac{4(C+\log2)}{\log(4/p)}\right\}.\label{app:spin-eq-state-map}
\end{equation}
Both entropy directions are available. Increasing finite grids to a dense set, continuity and bounded convergence give the full-window comparison.
\end{proof}

\subsection{Gaussian replacement and spectral clipping}\label{app:spin-sec-replacements}

Condition on the bath. Expand each source label in the fixed tensor basis,
\[
H_e=\sum_\alpha x_{e,\alpha}T_e^\alpha,\qquad
\Sigma_e=\E(x_ex_e^{\mathsf T}),\qquad \Tr\Sigma_e\le1.
\]
Replace $x_e$ by an independent centered real Gaussian vector with covariance $\Sigma_e$. Diagonalizing these covariance matrices gives
\begin{equation}
V_g=\sum_\ell g_\ell L_\ell,\qquad H_g=K+V_g.\label{app:spin-eq-gaussian-source}
\end{equation}
The directions are deterministic and $g$ is independent of the bath. Matching the full covariance gives the two budgets
\begin{align}
\sum_\ell L_\ell^2&=\sum_{e\ni0}b_N^2\E H_e^2\le k^{-1}I,\label{app:spin-eq-operator-budget}\\
\sum_\ell\norm{L_\ell}^2&\le d^k/k=:\bar v_g.\label{app:spin-eq-direction-budget}
\end{align}
For the second bound, the operator norm of a local matrix is at most its local Hilbert--Schmidt norm, and the sum of the latter squared over covariance directions is $d^k\Tr\Sigma_e$.

\begin{lemma}[Replacement in the bath state]\label{app:spin-lem-lindeberg}
For $T_N=O(\log N)$ and deterministic target weights,
\begin{equation}
|p_{\rho_K,H_o}^{\mathrm r}(w)-p_{\rho_K,H_g}^{\mathrm r}(w)|=o(1),\label{app:spin-eq-lindeberg}
\end{equation}
uniformly over the probes and the weights.
\end{lemma}

\begin{proof}
For $|t|\le T$, the Dyson expansion along $H'+uD$ gives
\[
\norm{\partial_u^m\alpha_t^{H'+uD}(A_0)}\le(2T\norm D)^m.
\]
With $\rho_K$ fixed, differentiating the four thermal words of a regularized cost and using Schatten H\"older gives
\[
|\partial_u^m C_{\rho_K}(H'+uD,t)|\le4(4T)^m\norm D^m,
\qquad m=1,2,3.
\]
These bounds also hold for every convex target average. On an $M$-point time grid, replace its maximum by
\[
F_\eta(H')=\eta\log\sum_{a=1}^M
\exp\bigl(C_{\rho_K}(H',t_a)/\eta\bigr).
\]
Its error is at most $\eta\log M$. Along a direction $D$, put $c_a(u)=C_{\rho_K}(H'+uD,t_a)$ and $\pi_a=e^{c_a/\eta}/\sum_be^{c_b/\eta}$. Brackets below denote the average with weights $\pi_a$. Differentiating $\pi_a'=\eta^{-1}\pi_a(c_a'-\langle c'\rangle)$ gives
\[
\begin{aligned}
F_\eta'&=\langle c'\rangle,\\
F_\eta''&=\langle c''\rangle+\eta^{-1}\operatorname{Var}_\pi(c'),\\
F_\eta'''&=\langle c'''\rangle
+3\eta^{-1}\operatorname{Cov}_\pi(c',c'')\\
&\quad+\eta^{-2}\langle(c'-\langle c'\rangle)^3\rangle.
\end{aligned}
\]
All weights sum to one, so the derivative bounds on $c_a$ imply, independently of $M$,
\begin{equation}
|\partial_D^3F_\eta|
\le CT^3(1+\eta^{-1}+\eta^{-2})\norm D^3.\label{app:spin-eq-softmax}
\end{equation}
Replace the source labels one at a time, conditioning on the bath and all other source labels. If $D=b_NH_e$ or $b_NH_e^{\mathrm g}$ is the label being inserted, Taylor's formula gives
\[
\begin{aligned}
F_\eta(H'+D)={}&F_\eta(H')+\partial_DF_\eta(H')
 +\tfrac12\partial_D^2F_\eta(H')+\mathcal R_D,\\
|\mathcal R_D|&\le CT^3(1+\eta^{-1}+\eta^{-2})\norm D^3.
\end{aligned}
\]
The linear derivative has the same expectation for the two labels. The quadratic derivative is a bilinear form in their real coordinates, so matching $\Sigma_e$ cancels its expectation as well. Each Gaussian coordinate vector has $\Tr\Sigma_e\le1$, whence its Euclidean third moment is bounded by a numerical constant. The local matrix norm is at most $d^{k/2}$ times that Euclidean norm. Therefore
\[
\sum_{e\ni0}b_N^3\bigl(\E\norm{H_e}^3+\E\norm{H_e^g}^3\bigr)
\le C_{k,d}N^{-(k-1)/2}.
\]
Let $r=k-1$, and choose $\eta=N^{-r/8}$ and $M\asymp N^{k+1}$. Summing the Taylor remainders and the two soft-maximum errors gives
\[
C T^3N^{-r/4}+C N^{-r/8}\log N.
\]
To pass from the grid to the interval, differentiation in time gives $|\partial_tC_{\rho_K}|\le16\norm{H'}$. The triangle inequality over the interaction labels yields
\[
\E\norm{H'}\le Cb_N\binom Nk\le C N^{(k+1)/2},
\]
uniformly along the replacement. A uniform grid of spacing $O(T/M)$ therefore contributes at most $CTN^{-(k+1)/2}$. All three errors tend to zero for $T=O(\log N)$. The Taylor and soft-maximum bounds hold with constants independent of the conditioned bath $K$, while the grid-discretization error is controlled in expectation by the unconditional bound $\E\norm{H'}\le CN^{(k+1)/2}$, so averaging over the bath proves the claim.
\end{proof}

Fix an even $C^4$ function $\chi:\mathbb R\to[0,1]$, equal to one on $[-1,1]$ and supported on $[-2,2]$, with the Fourier moments specified in Appendix~\ref{app:spin-sec-filter}. For $\Lambda\ge1$, put
\begin{equation}
\begin{gathered}
f_\Lambda(x)=x\chi(x/\Lambda),\qquad D_\Lambda=f_\Lambda(V_g),\\
E_\Lambda=V_g-D_\Lambda,\qquad H_\Lambda=K+D_\Lambda.
\end{gathered}\label{app:spin-eq-clipping}
\end{equation}
Then $\norm{D_\Lambda}\le2\Lambda$ and $D_\Lambda^2\le V_g^2$. Denote the Gibbs states of $H_g,H_\Lambda$ by $\rho_g,\rho_\Lambda$.

The operator budget in Eq.~\eqref{app:spin-eq-operator-budget} and Lemma~\ref{app:spin-lem-gaussian} give
\begin{equation}
\lambda_{\mathrm{ref}}=k/8,\qquad
\E_g e^{\lambda_{\mathrm{ref}}V_g^2}\le2I.\label{app:spin-eq-reference-moment}
\end{equation}
The direction budget in Eq.~\eqref{app:spin-eq-direction-budget} and Lemma~\ref{app:spin-lem-moments} give a moment in the source-dependent Gibbs state,
\begin{equation}
\lambda_{\mathrm{act}}=(256\bar v_g)^{-1},\qquad
\E\Tr(\rho_g e^{\lambda_{\mathrm{act}}V_g^2})\le C.\label{app:spin-eq-actual-moment}
\end{equation}
For $u\ge0$, the spectral Markov bound gives
\[
\E\Tr(\rho_g\one_{V_g^2>u})\le Ce^{-\lambda_{\mathrm{act}}u}.
\]
Since $E_\Lambda^2\le V_g^2\one_{V_g^2>\Lambda^2}$, integrate this tail using
\[
x\one_{x>a}=a\one_{x>a}+\int_a^\infty\one_{x>u}\,\dd u
\qquad (x\ge0).
\]
This yields
\begin{equation}
\E r_\Lambda\le C(1+\Lambda^2)e^{-\lambda_{\mathrm{act}}\Lambda^2},
\qquad r_\Lambda=\Tr(\rho_gE_\Lambda^2).\label{app:spin-eq-actual-tail}
\end{equation}
Peierls and Golden--Thompson bound the partition-function ratio by
\[
\beta\Tr(\rho_gE_\Lambda)
\le\log(Z_\Lambda/Z_g)
\le\log\Tr(\rho_g e^{\beta E_\Lambda}).
\]
The relative entropy has the exact form
\[
S(\rho_g\Vert\rho_\Lambda)
=\log(Z_\Lambda/Z_g)-\beta\Tr(\rho_gE_\Lambda).
\]
Using $\log x\le x-1$ and $\beta|E_\Lambda|\le e^{\beta|E_\Lambda|}-I$ in functional calculus, we obtain
\[
S(\rho_g\Vert\rho_\Lambda)
\le2\Tr\bigl(\rho_g(e^{\beta|E_\Lambda|}-I)\bigr).
\]
The clipping error vanishes on $|V_g|\le\Lambda$ and obeys $|E_\Lambda|\le|V_g|$. With $\lambda=\lambda_{\mathrm{act}}$, the scalar inequality $\beta|x|\le\lambda x^2/2+\beta^2/(2\lambda)$ gives
\[
e^{\beta|E_\Lambda|}-I
\le e^{\beta^2/(2\lambda)-\lambda\Lambda^2/2}e^{\lambda V_g^2}.
\]
Equation~\eqref{app:spin-eq-actual-moment} thus bounds the expected relative entropy by $Ce^{-\lambda\Lambda^2/2}$. Pinsker's inequality and Cauchy--Schwarz then give
\begin{equation}
\E\delta_\Lambda\le Ce^{-\lambda_{\mathrm{act}}\Lambda^2/4},
\qquad \delta_\Lambda=\norm{\rho_g-\rho_\Lambda}_1.\label{app:spin-eq-clipped-state}
\end{equation}

\begin{lemma}[Stationary curve comparison]\label{app:spin-lem-stationary}
Let $H,\widetilde H$ have Gibbs states $\rho,\widetilde\rho$ at the same inverse temperature. Set $r=\Tr[\rho(\widetilde H-H)^2]$, $\delta=\norm{\rho-\widetilde\rho}_1$ and $\nu_A=\norm{[H,A]}_\rho$. For deterministic Hermitian contractions $A,B_j$, any target weights, $T>0$ and $\eta>0$,
\begin{equation}
\begin{aligned}
\sup_{|t|\le T}|\Cbar_{\rho,H}^{\mathrm r}(t)-\Cbar_{\widetilde\rho,\widetilde H}^{\mathrm r}(t)|
\le{}&16\eta \nu_A+8(\sqrt8+1)\sqrt\delta\\
&+8(1+e^{\beta^2/(8\eta^2)})T\sqrt r.
\end{aligned}\label{app:spin-eq-stationary-curve}
\end{equation}
\end{lemma}

\begin{proof}
Write $U=e^{-\ii tH}$ and $\widetilde U=e^{-\ii t\widetilde H}$. Duhamel's formula and $[U,\rho]=0$ imply
\[
\norm{(\widetilde U-U)\rho^{1/2}}_2\le |t|\sqrt r.
\]
The adjoint difference has the same squared norm:
\[
\begin{aligned}
\norm{\widetilde U-U}_\rho^2
&=2-2\Re\Tr(\rho U^\dagger\widetilde U),\\
\norm{\widetilde U^\dagger-U^\dagger}_\rho^2
&=2-2\Re\Tr(\rho U\widetilde U^\dagger).
\end{aligned}
\]
The two traces have equal real parts by cyclicity and $[U,\rho]=0$.

To control the thermal conjugate $\rho^{-1/2}A_\eta\rho^{1/2}$, smooth the source with the centered Gaussian density of variance $\eta^2$:
\[
A_\eta=\int_{\mathbb R}\frac{e^{-u^2/(2\eta^2)}}{\sqrt{2\pi}\eta}
\alpha_u^H(A)\,\dd u.
\]
Stationarity gives $\norm{\alpha_u^H(A)-A}_\rho\le|u|\nu_A$, and the mean absolute value of the Gaussian is at most $\eta$. Hence $\norm{A-A_\eta}_\rho\le\eta \nu_A$. If $k_\eta$ is its density, shifting the Gaussian contour gives
\[
\rho^{-1/2}A_\eta\rho^{1/2}
=\int_{\mathbb R}k_\eta(u+\ii\beta/2)\alpha_u^H(A)\,\dd u.
\]
The integrand is entire in finite dimension and the Gaussian decays at both horizontal ends. Since
$|k_\eta(u+\ii\beta/2)|=e^{\beta^2/(8\eta^2)}k_\eta(u)$, we obtain
\[
\norm{\rho^{-1/2}A_\eta\rho^{1/2}}
\le e^{\beta^2/(8\eta^2)}=:K_\eta.
\]
In the difference of the two evolutions of $A_\eta$, expand
\[
\widetilde U^\dagger A_\eta(\widetilde U-U)
+(\widetilde U^\dagger-U^\dagger)A_\eta U.
\]
The first term costs $|t|\sqrt r$ in the $\rho$ norm. For the second, use $[U,\rho]=0$ to write
\[
\begin{aligned}
&(\widetilde U^\dagger-U^\dagger)A_\eta U\rho^{1/2}\\
&\quad=(\widetilde U^\dagger-U^\dagger)\rho^{1/2}
(\rho^{-1/2}A_\eta\rho^{1/2})U,
\end{aligned}
\]
which costs $K_\eta|t|\sqrt r$. For $E=A-A_\eta$, $\norm E\le2$ and stationarity of $\widetilde\rho$ gives
\[
\Tr(\rho\widetilde U^\dagger E^2\widetilde U)
\le\Tr(\widetilde\rho E^2)+4\delta
\le\Tr(\rho E^2)+8\delta.
\]
Thus the full source-evolution difference has $\rho$ norm at most $2\eta \nu_A+\sqrt{8\delta}+(1+K_\eta)T\sqrt r$.

For Hermitian contractions $A,B$ and $R=\rho^{1/2}$, the regularized squared commutator is
\[
\begin{aligned}
C_\rho^{\mathrm r}(A,B)
={}&\Tr(RABRBA)+\Tr(RBARAB)\\
&-\Tr(RABRAB)-\Tr(RBARBA).
\end{aligned}
\]
Subtract this expression at $A_1,A_2$, and telescope the two occurrences of $A$ in each word. There are eight terms, each containing $E=A_1-A_2$ next to a thermal factor $R$. For example, for $\ell=1,2$,
\[
|\Tr(REBR A_\ell B)|
\le\norm{RE}_2\norm{BR A_\ell B}_2
\le\norm E_\rho.
\]
The other terms obey the same bound by cyclicity and adjunction. Thus
\[
|C_\rho^{\mathrm r}(A_1,B)-C_\rho^{\mathrm r}(A_2,B)|
\le8\norm{A_1-A_2}_\rho.
\]
For a fixed commutator of norm at most two, changing the state costs at most $8\sqrt\delta$, since $\norm{\rho^{1/2}-\widetilde\rho^{1/2}}_2^2\le\delta$. Combining the two bounds proves Eq.~\eqref{app:spin-eq-stationary-curve}, also after a convex target average.
\end{proof}

Apply Lemma~\ref{app:spin-lem-stationary} to $H_g,H_\Lambda$. The source energy in $\rho_g$ is bounded in expectation by Lemma~\ref{app:spin-lem-moments} and Eq.~\eqref{app:spin-eq-direction-budget}. For fixed $\eta>0$, Eqs.~\eqref{app:spin-eq-actual-tail} and~\eqref{app:spin-eq-clipped-state} bound the expectation of the right side of Eq.~\eqref{app:spin-eq-stationary-curve} by
\[
C\eta+Ce^{-\lambda_{\mathrm{act}}\Lambda^2/8}
+C_\eta T(1+\Lambda)e^{-\lambda_{\mathrm{act}}\Lambda^2/2}.
\]
The last two terms vanish if $T=O(\log N)$ and $\Lambda^2/\log\log N\to\infty$. Taking first $N\to\infty$ and then $\eta\downarrow0$, and using $|\sup a-\sup b|\le\sup|a-b|$, proves
\begin{equation}
\begin{gathered}
|p_{\rho_g,H_g}^{\mathrm r}(w)-p_{\rho_\Lambda,H_\Lambda}^{\mathrm r}(w)|\to0
\quad\text{if}\\
T_N=O(\log N),\quad\frac{\Lambda^2}{\log\log N}\to\infty.
\end{gathered}
\label{app:spin-eq-stationary-clipping}
\end{equation}
Lemma~\ref{app:spin-lem-state-comparison} applies at both ends, uniformly in $\Lambda$.

\section{Bath frequency localization and double-insertion bootstrap}\label{app:sec-meas}

We continue in units $J=1$ and use the bath and Gaussian source of Appendix~\ref{app:sec-dyn}. The frequency field is $\Xi=\ii[K,V_g]$. This appendix bounds its thermal moments on a bath event of exponentially high probability, then controls the second bath derivative of the quench propagators in the states generated by finite operator words. The constants depend on fixed $k,d,\beta$ and the chosen event exponent. The case $d=1$ is immediate, so take $d\ge2$.

Filtering a quench cocycle $Q$ on bath-frequency scale $G$ in Appendix~\ref{app:sec-cond} leaves a $G^{-2}$ remainder governed by $\ad_K^2Q$, which brings down two source insertions clipped at $\Lambda$. Via a logarithmic multiplier bootstrap (Lemmas~\ref{app:spin-lem-multiplier} and~\ref{app:spin-lem-two-insertions}), Corollary~\ref{app:spin-cor-filter-tail} reduces the $\Lambda$-growth of $\norm{\ad_K^2Q}_{W\rho_KW^\dagger}$ on the high-probability bath event from linear to $\sqrt{1+\Lambda}$, up to integrable logarithmic-moment and polynomial word-order factors. This gain allows a sublogarithmic cutoff $G=o(\log N)$ with $\Lambda^2\propto G$, keeping the modular amplification $N^{o(1)}$ while the filter error in the square-root window bound vanishes over $T\asymp\log N$ (Proposition~\ref{app:spin-prop-window}).

\subsection{The rectangular gradient and its variance bounds}\label{app:spin-sec-gradient}

Put $r=k-1$, $n=N-1$ and $b=b_N$. Using the complete tensor basis of Appendix~\ref{app:spin-sec-frame}, define the rectangular operator
\begin{equation}
\begin{aligned}
\mathsf R(x)\psi&=\bigl(b\,\ii[K,T_S^\nu]\psi\bigr)_{S,\nu},\\
\mathsf A(x)&=\mathsf R(x)^\dagger\mathsf R(x)
=b^2\sum_{S,\nu}(\ii[K,T_S^\nu])^2.
\end{aligned}\label{app:spin-eq-gradient}
\end{equation}
Here $|S|=r$, and $\Tr_e$ denotes the trace over the $k$ sites of $e$. Expand each bath label as $H_e=\sum_\alpha x_{e,\alpha}T_e^\alpha$, so $\norm{x_e}_2^2=d^{-k}\Tr_e(H_e^2)\le1$. Then $\mathsf R=\sum_{e,\alpha}x_{e,\alpha}\mathsf R_{e,\alpha}$, with rows
\begin{equation}
\mathsf R_{e,\alpha}[S,\nu]=b^2\ii[T_e^\alpha,T_S^\nu].\label{app:spin-eq-gradient-coefficients}
\end{equation}

\begin{lemma}[Gradient budgets]\label{app:spin-lem-gradient}
For constants $w,v$ depending only on $k,d$,
\begin{equation}
\sum_{e,\alpha}\mathsf R_{e,\alpha}\mathsf R_{e,\alpha}^\dagger
\le\frac wN I,\qquad
\E\Tr(\eta\mathsf A)\le v\label{app:spin-eq-gradient-budgets}
\end{equation}
for every deterministic state $\eta$. One may take
\[
w=\frac{4r^2}{k}d^{3(k+r)},\qquad v=\frac{4r}{k^2}d^{3r}.
\]
\end{lemma}

\begin{proof}
A row $(S,\nu)$ is zero unless $S\cap e\ne\varnothing$. Choosing one intersection site in $e$, then the remaining $r-1$ sites of $S$, bounds the number of nonzero rows for fixed $e,\alpha$ by
\[
L=k\binom{n-1}{r-1}d^{2r}.
\]
Conversely, for fixed $S,\nu$, choose one intersection site in $S$ and the remaining $k-1=r$ sites of $e$. The number of pairs $e,\alpha$ meeting this row is at most
\[
M=r\binom{n-1}{r}d^{2k}.
\]
Since $\norm{T_e^\alpha}\le d^{k/2}$ and $\norm{T_S^\nu}\le d^{r/2}$, every coefficient row has squared norm at most $a^2=4b^4d^{k+r}$.

For a row-space vector $\zeta=(\zeta_{S,\nu})$, Cauchy--Schwarz first within each label and then the row count give
\[
\begin{aligned}
\sum_{e,\alpha}\norm{\mathsf R_{e,\alpha}^\dagger\zeta}^2
&\le L a^2\sum_{e,\alpha}
\sum_{\substack{S,\nu\\S\cap e\ne\varnothing}}\norm{\zeta_{S,\nu}}^2\\
&\le LMa^2\norm\zeta^2.
\end{aligned}
\]
Using $b^2=r!/(kN^r)$ and $\binom{n-1}{j}\le N^j/j!$,
\[
LMa^2
\le\frac{4r^2}{kN}d^{3(k+r)}=\frac wN.
\]
This proves the row-operator bound.

In $\E\mathsf A$, the cross terms of distinct bath labels vanish by independence and centering. For a fixed $S,\nu$, there are at most $r\binom{n-1}{r}$ interacting edges, and $\norm{[H_e,T_S^\nu]}^2\le4d^r$. Thus
\[
\E\Tr(\eta\mathsf A)
\le4b^4d^{3r}\binom nr\,r\binom{n-1}{r}
\le\frac{4r}{k^2}d^{3r}.
\]
\end{proof}

\subsection{Block concentration and finite product-state tests}\label{app:spin-sec-frequency}

For a real coefficient vector $v=(v_{e,\alpha})$, the row-operator bound gives
\[
\begin{aligned}
\norm{\mathsf R(v)^\dagger\zeta}
&\le\norm v_2
\left(\sum_{e,\alpha}\norm{\mathsf R_{e,\alpha}^\dagger\zeta}^2\right)^{1/2}\\
&\le\sqrt{w/N}\,\norm v_2\norm\zeta.
\end{aligned}
\]
Thus $\norm{\mathsf R(v)}\le\sqrt{w/N}\norm v_2$. For a deterministic state $\eta$, the function $f_\eta(x)=\norm{\mathsf R(x)\eta^{1/2}}_2$ is convex, and
\[
|f_\eta(x)-f_\eta(y)|
\le\norm{\mathsf R(x-y)}\norm{\eta^{1/2}}_2
\le\sqrt{w/N}\norm{x-y}_2.
\]
Its subgradient blocks $g_e(x)$ therefore satisfy $\sum_e\norm{g_e(x)}_2^2\le w/N$. The convex subgradient inequality and $\norm{x_e-y_e}_2\le2$ on each label ball imply
\begin{equation}
\begin{aligned}
f_\eta(x)-f_\eta(y)&\le2\sum_e\norm{g_e(x)}_2\one_{x_e\ne y_e},\\
\sum_e4\norm{g_e(x)}_2^2&\le4w/N.
\end{aligned}\label{app:spin-eq-block-difference}
\end{equation}
Talagrand's inequality for independent coordinates~\cite[Theorem~4.20 and Remark~4.25]{vanHandel2016}, applied with each matrix label as one coordinate, and $\E f_\eta\le\sqrt v$, gives the convenient bound
\begin{equation}
\Pr(f_\eta\ge\sqrt{2v}+t)\le2e^{-Nt^2/(16w)}.
\label{app:spin-eq-fixed-state-concentration}
\end{equation}

Every term of $\mathsf A$ is supported on $S$ and two $k$-site edges meeting $S$, a union of at most $r+2(k-1)=3r$ sites. This fixed support degree allows the operator norm to be controlled by a finite family of product-state tests. To construct these tests, let $r_0=d^2-1$, $q=2r_0$ and $\vartheta=(dr_0)^{-1}$. For $1\le a\le r_0$, $s=\pm1$, put
\[
\eta_{a,s}=\frac{I+sT^a/\sqrt d}{d},\qquad
M_{a,s}=\frac{I+sT^a/\sqrt d}{2r_0}.
\]
They are positive, $\Tr\eta_{a,s}=1$, and $\sum_{a,s}M_{a,s}=I$. Expanding in the orthonormal basis gives
\begin{equation}
\sum_{a,s}\Tr(M_{a,s}\omega)\eta_{a,s}
=\vartheta\omega+(1-\vartheta)I/d=:\Phi_\vartheta(\omega).
\label{app:spin-eq-measure-prepare}
\end{equation}
With $\Phi_0(X)=\Tr(X)I/d$, the channel $\Phi_t=t\,\mathrm{id}+(1-t)\Phi_0$ for $0\le t\le\vartheta$ has the same prepared states and effects
\[
M_{a,s}^{(t)}
=\frac t\vartheta M_{a,s}
+\left(1-\frac t\vartheta\right)\frac Iq.
\]
Indeed, $q^{-1}\sum_{a,s}\eta_{a,s}=I/d$. For an arbitrary $n$-site state $\omega$, including an entangled state, the tensor channel consequently has the explicit form
\[
\begin{aligned}
\Phi_t^{\otimes n}(\omega)&=\sum_z p_z(\omega,t)\eta_z,\\
\eta_z&=\bigotimes_{j=1}^n\eta_{z_j},\qquad
p_z(\omega,t)=\Tr\left(\omega\bigotimes_{j=1}^nM_{z_j}^{(t)}\right).
\end{aligned}
\]
All $p_z$ are nonnegative and sum to one. There are $q^n$ prepared product states, independent of $\omega,t$ and the couplings.

For an integer $p\ge1$, let $A$ be Hermitian with tensor support degree at most $p$. Decompose $A=\sum_{j=0}^pA_j$ by the number of traceless tensor legs. Since $\Phi_t(I)=I$ and $\Phi_t(T^a)=tT^a$ for $a\ne0$, we have $\Phi_t^{\otimes n}(A)=\sum_{j=0}^pt^jA_j$. At nodes $t_\ell=\ell\vartheta/p$, Lagrange interpolation at $t=1$ gives
\begin{equation}
\begin{aligned}
A&=\sum_{\ell=0}^pc_\ell\Phi_{t_\ell}^{\otimes n}(A),\qquad
c_\ell=\prod_{j\ne\ell}\frac{1-t_j}{t_\ell-t_j},\\
\norm A&\le C_{p,d}\max_z|\Tr(\eta_zA)|,
\qquad C_{p,d}=\sum_{\ell=0}^p|c_\ell|.
\end{aligned}\label{app:spin-eq-interpolation}
\end{equation}
To obtain the norm bound, test the first identity on an arbitrary state $\omega$. Self-adjointness of $\Phi_t$ for the trace pairing gives
\[
|\Tr(\omega A)|
\le\sum_{\ell=0}^p|c_\ell|
\left|\Tr\bigl(\Phi_{t_\ell}^{\otimes n}(\omega)A\bigr)\right|
\le C_{p,d}\max_z|\Tr(\eta_zA)|.
\]
Taking the supremum over $\omega$ proves the second statement.

Since $\mathsf A$ has support degree at most $3r$, Eq.~\eqref{app:spin-eq-interpolation} applies with $p=3r$.

\begin{proposition}[Bath frequency event]\label{app:spin-prop-frequency}
For every fixed $\gamma>0$, there are constants $v_\Xi,\lambda_\Xi>0$ and an event $\mathcal G_N$ determined by the bath alone such that
\begin{equation}
\begin{gathered}
\Pr(\mathcal G_N^c)\le2e^{-\gamma N},\\
\sum_\ell(\ii[K,L_\ell])^2\le v_\Xi I\quad\text{on }\mathcal G_N,\\
\E_g e^{\lambda_\Xi\Xi^2}\le2I\quad\text{on }\mathcal G_N,
\qquad \lambda_\Xi=(8v_\Xi)^{-1}.
\end{gathered}\label{app:spin-eq-frequency-event}
\end{equation}
\end{proposition}

\begin{proof}
Apply Eq.~\eqref{app:spin-eq-fixed-state-concentration} to the $q^n$ product states with $t=4\sqrt{w(\log q+\gamma)}$. A union bound and Eq.~\eqref{app:spin-eq-interpolation} give a bath event of the stated probability on which
\[
\mathsf A\le A_{\mathrm{cap}}I,\qquad
A_{\mathrm{cap}}=C_{3r,d}\bigl(\sqrt{2v}+4\sqrt{w(\log q+\gamma)}\bigr)^2.
\]
For a source edge $e=\{0\}\cup S$, write its tensor index as $(a,\nu)$ and set
$C_{e,a,\nu}=bT_0^a\,\ii[K,T_S^\nu]$.
Diagonalizing the source covariance expresses its frequency directions as linear combinations of these Hermitian matrices. Since $0\le\Sigma_e\le I$, for every vector $\psi$,
\[
\begin{aligned}
&\sum_{\alpha,\alpha'}(\Sigma_e)_{\alpha\alpha'}
\langle C_{e,\alpha}\psi,C_{e,\alpha'}\psi\rangle\\
&\hspace{20mm}\le\sum_\alpha\norm{C_{e,\alpha}\psi}^2.
\end{aligned}
\]
Thus the covariance contraction is an operator inequality. Sum over $e$, and use $\sum_a(T_0^a)^2=d^2I$ and the commutation of source and bath factors, to obtain
\[
\sum_\ell(\ii[K,L_\ell])^2
\le d^2I_0\otimes\mathsf A\le d^2A_{\mathrm{cap}}I.
\]
Set $v_\Xi=d^2A_{\mathrm{cap}}$. Conditional on this bath, $\Xi=\sum_\ell g_\ell\ii[K,L_\ell]$ is a Gaussian matrix series. Lemma~\ref{app:spin-lem-gaussian} proves the last assertion.
\end{proof}

\subsection{The filter and a logarithmic multiplier}\label{app:spin-sec-filter}

One fixed choice of the cutoff in Eq.~\eqref{app:spin-eq-clipping} is $\chi=\one_{[-3/2,3/2]}*\nu$, where $\nu$ is the convolution of six uniform probability densities on $[-1/12,1/12]$. With the convention $\chi(x)=\int\check\chi(u)e^{\ii ux}\,\dd u$,
\begin{equation}
\check\chi(u)=\frac{\sin(3u/2)}{\pi u}
\left(\frac{\sin(u/12)}{u/12}\right)^6.
\label{app:spin-eq-filter-kernel}
\end{equation}
This function has finite absolute moments through order two. Direct integration, splitting at $|u|=1$ and $|u|=12$, gives
\[
\int|\check\chi(u)|\,\dd u\le27/\pi,
\qquad \int u^2|\check\chi(u)|\,\dd u\le1800/\pi.
\]
For $Y_G=\chi(\ad_K/G)Y$, spectral decomposition in the stationary state $\rho_K$ gives
\begin{equation}
\norm{Y-Y_G}_{\rho_K}^2
\le G^{-2}\norm{[K,Y]}_{\rho_K}^2.\label{app:spin-eq-stationary-filter}
\end{equation}
The Fourier multiplier of $\rho_K^sY_G\rho_K^{-s}$ is $h_z(x)=e^{-zx}\chi(x)$, with $z=s\beta G$. In our Fourier convention,
\[
|\check h_z(u)|\le\frac{\norm{h_z}_1}{2\pi},
\qquad
u^2|\check h_z(u)|\le\frac{\norm{h_z''}_1}{2\pi}.
\]
Split the absolute integral at $|u|=1$. Using $\norm\chi_1=3$, $\norm{\chi'}_1\le2$ and $\norm{\chi''}_1\le24$ gives
\[
\int|\check h_z(u)|\,\dd u
\le\frac{\norm{h_z}_1+\norm{h_z''}_1}{\pi}
\le\frac{e^{2|z|}}{\pi}(27+4|z|+3z^2).
\]
The Fourier representation as an integral of bath conjugations therefore yields
\begin{equation}
\begin{aligned}
\norm{\rho_K^sY_G\rho_K^{-s}}&\le\mathcal M_\chi(s\beta G)\norm Y,\\
\mathcal M_\chi(z)&=\frac{e^{2|z|}}\pi(27+4|z|+3z^2).
\end{aligned}\label{app:spin-eq-modular-filter}
\end{equation}
In particular, the half-power multiplier is at most $11e^{2\beta G}$.

\begin{lemma}[Logarithmic multiplier]\label{app:spin-lem-multiplier}
Let $X$ be Hermitian, $\lambda>0$, $Y\ne0$, and $\theta>0$. Write $\ell_X=\log\Tr(\rho_Ke^{\lambda X^2})$ and $z=\norm Y_{\rho_K}^2$. Then
\begin{equation}
\norm{XY}_{\rho_K}^2
\le\theta\norm{[K,Y]}_{\rho_K}^2
+C_{\beta,\lambda}z\left[\ell_X+1+\log\frac{\norm Y^2}{z}+\theta^{-1}\right].
\label{app:spin-eq-log-multiplier}
\end{equation}
\end{lemma}

\begin{proof}
Put $P_x=\one_{X^2>x}$, $L_Y=\norm Y$ and $e_Y=\norm{[K,Y]}_{\rho_K}^2$. Positive spectral Markov gives $\Tr(\rho_KP_x)\le e^{\ell_X-\lambda x}$. Split $Y$ into $Y_G$ and its remainder. For the first term, write $Y_G\rho_K^{1/2}=\rho_K^{1/2}(\rho_K^{-1/2}Y_G\rho_K^{1/2})$ and use Eq.~\eqref{app:spin-eq-modular-filter}; for the second use Eq.~\eqref{app:spin-eq-stationary-filter}. This gives
\[
\Tr(\rho_KY^\dagger P_xY)
\le242L_Y^2e^{4\beta G+\ell_X-\lambda x}+2e_Y/G^2.
\]
At $G=\lambda x/(8\beta)$, the right side is
\[
242L_Y^2e^{\ell_X-\lambda x/2}
+\frac{128\beta^2e_Y}{\lambda^2x^2}.
\]
The same trace is bounded by $z$. Integrate its minimum with this last bound over $x\ge0$, splitting at
\[
a=\frac{2[\ell_X+1+\log(484L_Y^2/z)]}{\lambda}
+\frac{128\beta^2}{\theta\lambda^2}.
\]
The low interval costs $az$, the exponential tail costs at most $z/(\mathrm e\lambda)$, and the $x^{-2}$ tail costs at most $\theta e_Y$. This proves Eq.~\eqref{app:spin-eq-log-multiplier}.
\end{proof}

\subsection{Words with two insertions}\label{app:spin-sec-words}

Fix a bath in $\mathcal G_N$ and a source realization. Write $V=V_g$, $D=f_\Lambda(V)$ and $\Xi=\ii[K,V]$. Let $\mathcal G_0$ be the group generated by bath evolutions and source-site unitaries. Its elements commute with $K$ and $\rho_K$. For $G_a\in\mathcal G_0$, set
\[
V_a=G_aVG_a^\dagger,\qquad D_a=f_\Lambda(V_a),\qquad
\Xi_a=G_a\Xi G_a^\dagger.
\]
For one target, use the fixed finite set of involutions in its convex decomposition. The allowed unitary words have letters from $\mathcal G_0$, $e^{\ii t(K+D_a)}$, these target operations, and $e^{\ii uV_a}$, with arbitrary real $t,u$. Their order is
\begin{equation}
\begin{aligned}
\operatorname{ord}(W)
={}&\#(\text{quench segments})+\#(\text{target operations})\\
&+\sum_{\text{source kicks}}|u|.
\end{aligned}\label{app:spin-eq-word-order}
\end{equation}
Letters in $\mathcal G_0$ have order zero.

Choose a positive $\lambda$ no larger than the parameters for the source, target work and frequency moments. Define
\begin{equation}
\begin{aligned}
\ell_V&=\log\Tr(\rho_Ke^{\lambda V^2}),\qquad
\ell_\Xi=\log\Tr(\rho_Ke^{\lambda\Xi^2}),\\
\ell_B&=\max_B\log\Tr(\rho_Ke^{\lambda W_B^2}),\qquad
\ell_* =\max(\ell_V,\ell_B,\ell_\Xi).
\end{aligned}\label{app:spin-eq-log-moments}
\end{equation}
The maximum over $B$ is within the decomposition of this target. Conjugation by $\mathcal G_0$ preserves these reference moments. For a kick state $\tau_u=e^{\ii uV_a}\tau e^{-\ii uV_a}$, unitary invariance of its entropy and $\Xi_a=\ii[K,V_a]$ give
\[
\partial_uS(\tau_u\Vert\rho_K)=\beta\Tr(\tau_u\Xi_a).
\]
Applying the variational bound to $\Xi_a$ gives
\[
|\partial_uS(\tau_u\Vert\rho_K)|
\le\beta\sqrt{[S(\tau_u\Vert\rho_K)+\ell_*]/\lambda}.
\]
Thus a kick of duration $u$ increases $\sqrt{S(\tau\Vert\rho_K)+\ell_*}$ by at most $\beta|u|/(2\sqrt\lambda)$. A target operation costs at most $\beta/(2\sqrt\lambda)$, and Lemma~\ref{app:spin-lem-entropy} bounds a quench segment by $\beta/\sqrt\lambda$, uniformly in its duration. Starting at $\rho_K$ and summing along a word proves, for $\operatorname{ord}(W)\le m$,
\begin{equation}
\begin{aligned}
S(W\rho_KW^\dagger\Vert\rho_K)&\le h_m,\\
h_m&=(\sqrt{\ell_*}+m\beta/\sqrt\lambda)^2-\ell_*,\\
Z_m&=(h_m+\ell_*)/\lambda.
\end{aligned}\label{app:spin-eq-word-entropy}
\end{equation}
The expectation of each of $D_a^2,W_B^2,\Xi_a^2$ in this word state is at most $Z_m$ by Eq.~\eqref{app:spin-eq-variational}.

The exact source-kick derivative is
\begin{equation}
[K,e^{\ii uV_a}]
=\int_0^u e^{\ii(u-v)V_a}\Xi_a e^{\ii vV_a}\,\dd v.
\label{app:spin-eq-kick-derivative}
\end{equation}
Its total variation is $|u|$, and $|u-v|+|v|=|u|$ on the oriented interval. For $f(x)=x\chi(x)$, write $f(x)=\int\widehat f(u)e^{\ii ux}\,\dd u$. Equation~\eqref{app:spin-eq-filter-kernel}, differentiated once, implies
\[
M_j:=\int |u|^j|\widehat f(u)|\,\dd u<\infty,
\qquad j=0,1,2.
\]
Since $\widehat f_\Lambda(u)=\Lambda^2\widehat f(\Lambda u)$, Eq.~\eqref{app:spin-eq-kick-derivative} and the kick entropy bound give
\begin{equation}
\begin{aligned}
\norm{[K,D_a]}_{W\rho_KW^\dagger}^2&\le L_m,\\
L_m&=\left[M_1\sqrt{(h_m+\ell_*)/\lambda}
+\frac{\beta M_2}{4\lambda\Lambda}\right]^2.
\end{aligned}\label{app:spin-eq-inner-insertion}
\end{equation}
Indeed, for $\tau=W\rho_KW^\dagger$, the right sub-kick in Eq.~\eqref{app:spin-eq-kick-derivative} gives
\[
\norm{\Xi_a}_{e^{\ii vV_a}\tau e^{-\ii vV_a}}
\le\sqrt{(h_m+\ell_*)/\lambda}+\frac{\beta|v|}{2\lambda}.
\]
The left sub-kick drops out of the norm. Minkowski's inequality, first in $v$ and then in the source Fourier parameter, therefore gives
\[
\begin{aligned}
\norm{[K,D_a]}_\tau
&\le\int|\widehat f_\Lambda(u)|
\left[|u|\sqrt{(h_m+\ell_*)/\lambda}
+\frac{\beta u^2}{4\lambda}\right]\dd u\\
&=M_1\sqrt{(h_m+\ell_*)/\lambda}
+\frac{\beta M_2}{4\lambda\Lambda}.
\end{aligned}
\]
Squaring proves Eq.~\eqref{app:spin-eq-inner-insertion}.

\begin{lemma}[Two insertions]\label{app:spin-lem-two-insertions}
For $m,\Lambda\ge1$, let
\begin{equation}
\begin{gathered}
\mathfrak M_m=\sup\norm{XUD_bW}_{\rho_K}^2,
\quad X\in\{D_a,W_B,\Xi_a\},\\
\operatorname{ord}(U)+\operatorname{ord}(W)\le m.
\end{gathered}
\label{app:spin-eq-two-insertion-definition}
\end{equation}
Then
\begin{equation}
\mathfrak M_m\le C(1+\ell_*)^2(1+m^4)(1+\Lambda).
\label{app:spin-eq-two-insertion-bound}
\end{equation}
The constant is independent of the propagation times, kick parameters, $m,\Lambda$ and $N$.
\end{lemma}

\begin{proof}
Apply Lemma~\ref{app:spin-lem-multiplier} to $XY$ with $Y=UD_bW$. Its parameters satisfy $z=\norm Y_{\rho_K}^2\le Z_m$ and $\norm Y\le2\Lambda$. Differentiate the full inner word,
\begin{equation}
[K,Y]=[K,U]D_bW+U[K,D_b]W+UD_b[K,W].
\label{app:spin-eq-full-word-derivative}
\end{equation}
For a quench segment, $[K,e^{\ii t(K+D_a)}]=-[D_a,e^{\ii t(K+D_a)}]$. A target operation contributes $BW_B$, and a kick contributes Eq.~\eqref{app:spin-eq-kick-derivative}.

Consider the three derivative classes separately. If a letter in the left word is differentiated, write $U=P U_aR$ for a quench letter $U_a=e^{\ii t(K+D_a)}$. Its two contributions are
\[
-P D_aU_aR D_bW,\qquad P U_aD_aR D_bW.
\]
Removing the left unitaries leaves $D_a$ as the outer insertion, with $D_b$ still inside. A left target derivative similarly leaves $W_B$, and a left kick derivative leaves $\Xi_a$. Each term is in the supremum defining $\mathfrak M_m$.

For a quench letter in the right word, write $W=P U_aR$. After removing the left unitary $U$, the two terms are
\[
-D_bP D_aU_aR,\qquad D_bP U_aD_aR.
\]
Here $D_b$ is the outer insertion and $D_a$ the inner one, so these terms also belong to $\mathfrak M_m$. The coefficient variation of these two classes together is at most $2m$.

For a target derivative in the right word, the corresponding term is $UD_bPBW_BR$. Its squared norm is at most
\[
4\Lambda^2\norm{W_BR}_{\rho_K}^2\le4\Lambda^2Z_m.
\]
A right kick derivative has the form
$UD_bP e^{\ii(u-v)V_a}\Xi_a e^{\ii vV_a}R$
and satisfies the same estimate, with $\Xi_a$ tested in the right-word state. The total variation of right target and kick derivatives is at most $m$. Finally, the middle term $U[K,D_b]W$ is bounded by $L_m$.

For a sum or integral with coefficient measure $\mu$, use
$\norm{\int Z\,\dd\mu}^2\le|\mu|(\mathbb R)\int\norm Z^2\,\dd|\mu|$.
The total coefficient variation is at most $3m+1$, giving
\begin{equation}
\norm{[K,Y]}_{\rho_K}^2
\le(3m+1)(2m\mathfrak M_m+4m\Lambda^2Z_m+L_m).
\label{app:spin-eq-insertion-recursion}
\end{equation}
All remaining words have order at most $m$ by Eq.~\eqref{app:spin-eq-kick-derivative}.

We choose $\theta$ to absorb $\mathfrak M_m$ while keeping the remaining dependence on $\Lambda$ linear. Put $a_m=2m(3m+1)$ and choose $\theta=[2a_m(1+\Lambda)]^{-1}$. Inserting Eq.~\eqref{app:spin-eq-insertion-recursion} into Eq.~\eqref{app:spin-eq-log-multiplier}, the coefficient of $\mathfrak M_m$ is $[2(1+\Lambda)]^{-1}\le1/4$. The supremum is finite for each sample: all word factors are unitary, $\norm{D_b}\le2\Lambda$, and the norms of $D_a,\Xi_a$ are unchanged by their conjugations. Thus this term can be moved to the left.

The other derivative coefficients satisfy
\[
\theta(3m+1)4m\Lambda^2=\frac{\Lambda^2}{1+\Lambda},
\qquad
\theta(3m+1)=\frac1{4m(1+\Lambda)}.
\]
The $\theta^{-1}$ term is at most $C m^2(1+\Lambda)Z_m$. For the logarithm use
\[
z\log(4\Lambda^2/z)\le Z_m\log(4\Lambda^2)+1/\mathrm e.
\]
Since $\ell_X\le\ell_*$, the resulting bound is
\[
\mathfrak M_m\le C\left[
\bigl(\ell_*+1+\log(4\Lambda^2)+m^2(1+\Lambda)\bigr)Z_m
+L_m+1\right].
\]
Finally,
\[
Z_m+L_m\le C(1+\ell_*)(1+m^2),\qquad
a_m\le8m^2,
\]
and $\log(4\Lambda^2)\le C(1+\Lambda)$. These bounds give Eq.~\eqref{app:spin-eq-two-insertion-bound}.
\end{proof}

\begin{corollary}[Quench derivatives and filter tails]\label{app:spin-cor-filter-tail}
Let $Q$ be a quench cocycle $e^{\ii t(K+D_a)}e^{-\ii tK}$, its reverse-order form, or a bath conjugate of either, and let $\operatorname{ord}(W)\le m$. Then
\begin{align}
\norm{\ad_K^2Q}_{W\rho_KW^\dagger}
&\le C(1+\ell_*)(1+m^2)\sqrt{1+\Lambda},\label{app:spin-eq-second-derivative}\\
\norm{Q-Q_G}_{W\rho_KW^\dagger}
&\le\frac{C(1+\ell_*)(1+m^2)\sqrt{1+\Lambda}}{G^2},\label{app:spin-eq-word-filter-tail}
\end{align}
where $Q_G=\chi(\ad_K/G)Q$. The same bounds hold with $Q,Q_G$ replaced by $Q^\dagger,Q_G^\dagger$.
\end{corollary}

\begin{proof}
For $U=e^{\ii t(K+D_a)}$,
\begin{equation}
\ad_K^2U=-[[K,D_a],U]+[D_a,[D_a,U]].
\label{app:spin-eq-quench-second-identity}
\end{equation}
For $\tau=W\rho_KW^\dagger$, the terms $[K,D_a]U$ and $U[K,D_a]$ are bounded by $\sqrt{L_{m+1}}$: in the first term the right word is $UW$, and in the second the left $U$ drops out. The second commutator expands as
\[
[D_a,[D_a,U]]=D_a^2U-2D_aUD_a+UD_a^2.
\]
These are double-insertion words of order at most $m+1$, so
\[
\norm{\ad_K^2U}_{W\rho_KW^\dagger}
\le2\sqrt{L_{m+1}}+4\sqrt{\mathfrak M_{m+1}}.
\]
The bounds on $L_{m+1}$ and $\mathfrak M_{m+1}$ prove Eq.~\eqref{app:spin-eq-second-derivative}. Bath factors commute with $K$ and preserve the word class. Negative times and reversing the order of the bath and quench factors give the assertion for $Q^\dagger$.

Use the integral representation
\[
Q_G=\int\check\chi(u)e^{\ii uK/G}Qe^{-\ii uK/G}\,\dd u.
\]
The zeroth moment is one and the first moment is zero. Taylor's formula gives the exact remainder
\[
Q-Q_G=\frac1{G^2}\int u^2\check\chi(u)
\int_0^1(1-v)\alpha_{vu/G}^K(\ad_K^2Q)\,\dd v\,\dd u.
\]
For any word state $\tau$, $\norm{\alpha_s^K(X)}_\tau
=\norm X_{\alpha_{-s}^K(\tau)}$. Rotating the state by bath evolution leaves its word order unchanged. Minkowski's inequality and Eq.~\eqref{app:spin-eq-second-derivative} therefore bound the remainder by
\[
\frac{C(1+\ell_*)(1+m^2)\sqrt{1+\Lambda}}{2G^2}
\int u^2|\check\chi(u)|\,\dd u.
\]
This proves Eq.~\eqref{app:spin-eq-word-filter-tail}; the same calculation applies to $Q^\dagger$.
\end{proof}

The random constants in these estimates are integrable uniformly in $N$. For the finite involution family $\mathcal I_j$ of target $j$,
\[
e^{\ell_*}\le e^{\ell_V}+e^{\ell_\Xi}
+\sum_{B\in\mathcal I_j}\Tr(\rho_Ke^{\lambda W_B^2}).
\]
Equations~\eqref{app:spin-eq-reference-moment} and~\eqref{app:spin-eq-frequency-event} bound the conditional source and frequency terms on $\mathcal G_N$, and Eq.~\eqref{app:spin-eq-local-work} bounds each target-work term. Since $|\mathcal I_j|\le2^d$, this gives
\begin{equation}
\E[\one_{\mathcal G_N}e^{\ell_*}]\le C.\label{app:spin-eq-log-moment-average}
\end{equation}
This estimate is applied to each target before taking its weighted average.

Finally, whenever a clipped factor $D_a$ inside a finite word is expanded via its Fourier representation $D_a = \int \widehat{f}_\Lambda(u)e^{\ii u V_a}\,\dd u$, the word order increases by $|u|$ and the bound in Eq.~\eqref{app:spin-eq-word-filter-tail} grows quadratically in $|u|$. Integrating over $u$ is then controlled at cost $O(\Lambda)$ by
\begin{equation}
\int|\widehat f_\Lambda(u)|(1+u^2)\,\dd u
=\Lambda M_0+M_2/\Lambda\le C\Lambda.
\label{app:spin-eq-fourier-cost}
\end{equation}

\section{Filtered Duhamel comparison and the scrambling window}\label{app:sec-cond}

This appendix combines the bath double average with the quench estimates to prove Theorem~\ref{thm:spins}. We first estimate the linear Gaussian reference, then compare it with the fixed-source dynamics of $H_\Lambda$. The state and Hamiltonian comparisons of Appendix~\ref{app:sec-dyn} complete the passage to the original model.

\subsection{The linear Gaussian reference}\label{app:spin-sec-reference}

Let $r=k-1$, $n=N-1$ and $M_r=\binom nr$. In the bath Gibbs state define
\begin{equation}
\mathcal B_r(s)=\frac{1}{M_r^2d^{4r}}
\sum_{S,T,\nu,\mu}C_{\rho_{\mathrm{bath}}}^{\mathrm r}
\bigl(T_S^\nu(s),T_T^\mu\bigr),
\label{app:spin-eq-bath-average}
\end{equation}
where both supports have size $r$ and the evolution is generated by $K$. Lemma~\ref{app:spin-lem-frame}, Eq.~\eqref{app:spin-eq-average-energy} and Proposition~\ref{prop:double}, after the fixed rescaling from $Q$ to $T$, imply
\begin{equation}
\E\sup_{|s|\le T}\mathcal B_r(s)
\le CN^{-1/2}e^{\pi T/\beta}.\label{app:spin-eq-bath-window}
\end{equation}
Indeed, the proposition bounds the cost by $(4+\beta\bar\nu)\sqrt{\Delta_r}\,e^{\pi T/\beta}$ for the contraction frame. Cauchy--Schwarz in disorder, $\E\Delta_r=O(1/N)$ and $\E\bar\nu^{\,2}\le C$ give Eq.~\eqref{app:spin-eq-bath-window}.

\begin{lemma}[Linear source reference]\label{app:spin-lem-reference}
Let $V_g(s)=e^{\ii sK}V_ge^{-\ii sK}$. For arbitrary deterministic single-site target contractions,
\begin{equation}
\E\frac1N\sum_{j\ne0}C_{\rho_K}^{\mathrm r}(V_g(s),B_j)
\le CN^{-1/2}e^{\pi|s|/\beta}.\label{app:spin-eq-reference}
\end{equation}
If the coupling law is invariant under site permutations fixing $0$, the same bound, up to $N/(N-1)$, holds for each fixed bath target.
\end{lemma}

\begin{proof}
Conditional on the bath, put
$Z_{e,\alpha}=\rho_K^{1/4}\ii[T_e^\alpha(s),B_j]\rho_K^{1/4}$.
The regularized cost is $b_N^2\norm{\sum_{e,\alpha}x_{e,\alpha}^{\mathrm g}Z_{e,\alpha}}_2^2$. Independence of the source labels and $\Sigma_e\le I$ give
\[
\begin{aligned}
&\E_g\left\|\sum_{e,\alpha}x_{e,\alpha}^{\mathrm g}Z_{e,\alpha}\right\|_2^2\\
&\quad=\sum_{e,\alpha,\alpha'}(\Sigma_e)_{\alpha\alpha'}
\langle Z_{e,\alpha},Z_{e,\alpha'}\rangle
\le\sum_{e,\alpha}\norm{Z_{e,\alpha}}_2^2.
\end{aligned}
\]
For $e=\{0\}\cup S$, write $T_e^\alpha=T_0^aT_S^\nu$. The source factor contributes
$d^{-1}\Tr[(T^a)^2]=1$ to each squared norm. Summing over its $d^2$ indices yields
\begin{equation}
\begin{aligned}
\E_g\frac1N\sum_{j\ne0}C_{\rho_K}^{\mathrm r}(V_g(s),B_j)
\le\frac{b_N^2d^2}{N}\sum_{S,\nu,j\ne0}
C_{\rho_{\mathrm{bath}}}^{\mathrm r}(T_S^\nu(s),B_j).
\end{aligned}\label{app:spin-eq-reference-contraction}
\end{equation}
For each target, expand $B_j=\sum_ac_{j,a}T_j^a$. Since $\sum_a|c_{j,a}|^2=d^{-1}\Tr B_j^2\le1$, Hilbert-space Cauchy--Schwarz gives
\[
C_{\rho_{\mathrm{bath}}}^{\mathrm r}(T_S^\nu(s),B_j)
\le\sum_{a\ne0}
C_{\rho_{\mathrm{bath}}}^{\mathrm r}(T_S^\nu(s),T_j^a).
\]
To apply the bath double-average bound, we embed the single-site targets in the same $r$-site family as the sources. Every nonidentity one-site basis target occurs $m_r=\binom{n-1}{r-1}$ times in the complete $r$-site target family, with identity on the other legs. These are distinct indexed terms, so positivity implies
\[
\sum_{j,a\ne0}C_{\rho_{\mathrm{bath}}}^{\mathrm r}(T_S^\nu(s),T_j^a)
\le\frac1{m_r}\sum_{T,\mu}
C_{\rho_{\mathrm{bath}}}^{\mathrm r}(T_S^\nu(s),T_T^\mu).
\]
Summing the source frame and using $M_r/m_r=n/r$ yields
\begin{equation}
\E_g\frac1N\sum_{j\ne0}C_{\rho_K}^{\mathrm r}(V_g(s),B_j)
\le b_N^2M_r\frac nN\frac{d^{4r+2}}r\,\mathcal B_r(s).
\label{app:spin-eq-reference-frame}
\end{equation}
The normalization $b_N^2M_r\le1/k$ and Eq.~\eqref{app:spin-eq-bath-window} prove Eq.~\eqref{app:spin-eq-reference}.

For a fixed bath site $j_0$, copy its local matrix $B$ to every bath site and use this auxiliary target family in Eq.~\eqref{app:spin-eq-reference}. The joint law of $(K,V_g)$ is invariant under permutations fixing $0$: the original labels transform covariantly, and centered Gaussian laws are determined by the full covariance tensors used in their replacement. The reference costs for these copied targets have the same expectation. Hence
\begin{equation}
\begin{aligned}
\E C_{\rho_K}^{\mathrm r}(V_g(s),B_{j_0})
&=\frac1{N-1}\sum_{j\ne0}\E C_{\rho_K}^{\mathrm r}(V_g(s),B_j)\\
&\le\frac N{N-1}CN^{-1/2}e^{\pi|s|/\beta}.
\end{aligned}\label{app:spin-eq-fixed-reference}
\end{equation}
\end{proof}

The operator moment of Eq.~\eqref{app:spin-eq-reference-moment} controls the clipping error at each reference time. In fact, positive spectral integration gives
\begin{equation}
\begin{gathered}
\E_g(V_g(s)-f_\Lambda(V_g(s)))^2
\le\varepsilon_\Lambda^{\mathrm{ref}}I,\\
\varepsilon_\Lambda^{\mathrm{ref}}
=2(\Lambda^2+\lambda_{\mathrm{ref}}^{-1})e^{-\lambda_{\mathrm{ref}}\Lambda^2}.
\end{gathered}\label{app:spin-eq-reference-tail}
\end{equation}
Expanding the commutator of this tail with $B_j$, its unregularized squared norm is at most $4\varepsilon_\Lambda^{\mathrm{ref}}$ after averaging $g$. Here the operator bound is tested against $\rho_K$ and $B_j\rho_KB_j$, both fixed in this average. The regularized norm is smaller. The squared triangle inequality consequently gives
\begin{equation}
\E_g C_{\rho_K}^{\mathrm r}(D_\Lambda(s),B_j)
\le2\E_g C_{\rho_K}^{\mathrm r}(V_g(s),B_j)
+8\varepsilon_\Lambda^{\mathrm{ref}}.
\label{app:spin-eq-clipped-reference}
\end{equation}

\subsection{Duhamel expansion and filter errors}\label{app:spin-sec-duhamel}

Let $S(t)=e^{-\ii tK}e^{\ii tH_\Lambda}$ and $W(t)=e^{\ii tH_\Lambda}e^{-\ii tK}$. Since $[K,A_0]=0$,
\begin{equation}
\alpha_t^{H_\Lambda}(A_0)=W(t)A_0W(t)^\dagger.\label{app:spin-eq-quench-source}
\end{equation}
The equation $\dot W(t)=\ii W(t)D_\Lambda(t)$ implies
\[
\frac{\dd}{\dd t}\bigl([W(t),B_j]W(t)^\dagger\bigr)
=\ii W(t)[D_\Lambda(t),B_j]W(t)^\dagger.
\]
Integrate from zero and multiply on the right by $W(t)$. For a bath target this gives
\begin{equation}
[W(t),B_j]=\ii\int_0^tW(s)[D_\Lambda(s),B_j]
\alpha_t^K(S(t-s))\,\dd s.\label{app:spin-eq-duhamel}
\end{equation}
Here $W(s)^\dagger W(t)=\alpha_t^K(S(t-s))$. Substituting into
\begin{equation}
\begin{aligned}
[WA_0W^\dagger,B_j]
={}&[W,B_j]A_0W^\dagger\\
&-WA_0W^\dagger[W,B_j]W^\dagger
\end{aligned}\label{app:spin-eq-five-cocycles}
\end{equation}
gives two integral words. More explicitly, put
$R_j(s)=[D_\Lambda(s),B_j]$ and $P_{s,t}=\alpha_t^K(S(t-s))$. Their integrands, apart from the factor $\ii$, are
\[
\begin{aligned}
&W(s)R_j(s)P_{s,t}A_0W(t)^\dagger,\\
&-W(t)A_0W(t)^\dagger W(s)R_j(s)P_{s,t}W(t)^\dagger.
\end{aligned}
\]
They contain three and five cocycles, respectively, one source operation and one reference commutator. The reference time remains $s\in[0,t]$.

\begin{proposition}[Weighted window comparison]\label{app:spin-prop-window}
For deterministic target weights, put
\[
\mathcal C_w(s)=\E\sum_{j\ne0}w_jC_{\rho_K}^{\mathrm r}(V_g(s),B_j),
\qquad q_\Lambda(w)=p_{\rho_K,H_\Lambda}^{\mathrm r}(w).
\]
For $G>0$, $\Lambda\ge1$ and $T=T_N$,
\begin{equation}
\begin{aligned}
\sqrt{q_\Lambda(w)}\le{}&C\mathcal M_\chi(\beta G/4)^5
\int_0^T\bigl(\sqrt{\mathcal C_w(s)}
+\sqrt{\varepsilon_\Lambda^{\mathrm{ref}}}\bigr)\,\dd s\\
&+\frac{CT\Lambda\sqrt{1+\Lambda}}{G^2}
+4e^{-\gamma N/2}+2\sqrt{w_0}.
\end{aligned}\label{app:spin-eq-weighted-window}
\end{equation}
The constant is uniform over the probes and weights.
\end{proposition}

\begin{proof}
For each cocycle in Eq.~\eqref{app:spin-eq-five-cocycles}, substitute $Q_G=\chi(\ad_K/G)Q$ and telescope the product. To express the factors surrounding each $Q-Q_G$ as integrals of admissible unitary words, expand the reference commutator into two products, represent $D_\Lambda(s)$ by its Fourier integral, and represent the other filtered factors by the time integral in Corollary~\ref{app:spin-cor-filter-tail}. In these error terms, also expand $A_0$ and $B_j$ in their fixed convex combinations of involutions. The filtered-factor integrals have bounded total variation. Each resulting word is $LE_QR'$, with $E_Q=Q-Q_G$ and unitary words $L,R'$ from Appendix~\ref{app:spin-sec-words}. The squared thermal norms of $L E_Q R'$ and its adjoint are
\begin{equation}
\begin{aligned}
\norm{LE_QR'}_{\rho_K}^2
&=\Tr(R'\rho_KR'^\dagger E_Q^\dagger E_Q),\\
\norm{(LE_QR')^\dagger}_{\rho_K}^2
&=\Tr(L^\dagger\rho_KL E_QE_Q^\dagger).
\end{aligned}\label{app:spin-eq-error-directions}
\end{equation}
Use Eq.~\eqref{app:spin-eq-word-filter-tail} for each of these two norms. For any operator $X$, the regularized norm satisfies
\[
\norm{\rho_K^{1/4}X\rho_K^{1/4}}_2^2
\le\tfrac12\bigl(\norm X_{\rho_K}^2+\norm{X^\dagger}_{\rho_K}^2\bigr),
\]
by the arithmetic--geometric mean inequality in the eigenbasis of $\rho_K$. The word order is at most $m_0+|u|$, with $m_0$ fixed: the cocycle count is at most five, bath rotations have order zero, and the sole source Fourier kick has order $|u|$. Equation~\eqref{app:spin-eq-fourier-cost} therefore performs the source Fourier integral at cost $C\Lambda$.

Let $\ell_{*,j}$ denote the logarithmic moment for target $j$, and let $\mathcal E_j(t)$ be its total filter-error commutator. The estimates just obtained are uniform in the propagation times and give, on $\mathcal G_N$,
\[
\sup_{0\le t\le T}
\norm{\rho_K^{1/4}\mathcal E_j(t)\rho_K^{1/4}}_2
\le\frac{CT\Lambda\sqrt{1+\Lambda}}{G^2}(1+\ell_{*,j}).
\]
By Eq.~\eqref{app:spin-eq-log-moment-average},
$\E[\one_{\mathcal G_N}\sum_jw_j(1+\ell_{*,j})^2]\le C$.
Thus the root-mean-square error for the weighted window is bounded by
\begin{equation}
\mathrm{Err}_{\mathrm{filter}}
\le CT\Lambda\sqrt{1+\Lambda}/G^2.\label{app:spin-eq-full-filter-error}
\end{equation}
For a fully filtered integrand, write its factors as $L_G R_j(s)R_G$ and set $y_K=\rho_K^{1/4}$. Then
\[
y_KL_G R_j(s)R_Gy_K
=(y_KL_Gy_K^{-1})(y_KR_j(s)y_K)(y_K^{-1}R_Gy_K).
\]
Each cocycle in the outside factors costs at most $\mathcal M_\chi(\beta G/4)$ by Eq.~\eqref{app:spin-eq-modular-filter}. The source operation commutes with $\rho_K$ and has cost one. Hence
\[
\norm{y_KL_G R_j(s)R_Gy_K}_2
\le\mathcal M_\chi(\beta G/4)^5
\sqrt{C_{\rho_K}^{\mathrm r}(D_\Lambda(s),B_j)}.
\]

For each sample, take the direct-sum norm over targets with weights $w_j$. The two filtered integrands and the triangle inequality bound the window supremum of this norm by
\[
C\mathcal M_\chi(\beta G/4)^5
\int_0^T
\left(\sum_{j\ne0}w_j C_{\rho_K}^{\mathrm r}(D_\Lambda(s),B_j)\right)^{1/2}\dd s.
\]
The resulting integrand is independent of the terminal time $t$. On the good-bath event $\mathcal{G}_N$, Minkowski's inequality in the disorder probability space brings the $L^2$ norm inside the time integral. Equation~\eqref{app:spin-eq-clipped-reference} and positivity of the integrand, which allows dropping the indicator $\one_{\mathcal G_N}$ after taking the source expectation, then bound that norm by $C(\sqrt{\mathcal C_w(s)}+\sqrt{\varepsilon_\Lambda^{\mathrm{ref}}})$, giving the first line of Eq.~\eqref{app:spin-eq-weighted-window}. On the complement $\mathcal{G}_N^c$, the raw fixed-source OTOC is at most $4$, contributing at most $2\Pr(\mathcal{G}_N^c)^{1/2} \le 2\sqrt{2}e^{-\gamma N/2} \le 4e^{-\gamma N/2}$. The source-site target contributes at most $4w_0$ to the cost. The error estimates are uniform over the finite convex decompositions, while the filtered main term retains the original $A_0$ and $B_j$. Combining the estimates proves Eq.~\eqref{app:spin-eq-weighted-window} for the original contractions and their reference cost $\mathcal C_w(s)$.
\end{proof}

For uniform weights, Lemma~\ref{app:spin-lem-reference} gives
\begin{equation}
\begin{aligned}
\sqrt{q_\Lambda}\le{}&CT\mathcal M_\chi(\beta G/4)^5
\left[N^{-1/4}e^{\pi T/(2\beta)}
+\sqrt{\varepsilon_\Lambda^{\mathrm{ref}}}\right]\\
&+\frac{CT\Lambda\sqrt{1+\Lambda}}{G^2}
+4e^{-\gamma N/2}+\frac2{\sqrt N}.
\end{aligned}\label{app:spin-eq-uniform-window}
\end{equation}
Under the permutation hypothesis, take $w_j=\delta_{j,j_0}$ with $j_0\ne0$ and use Eq.~\eqref{app:spin-eq-fixed-reference}. The same bound holds without the last term, after a fixed change of $C$.

\subsection{Choice of scales}\label{app:spin-sec-scales}

Let $L_N=\log(N+e)$ and choose
\begin{equation}
G_N=L_N^{9/10},\qquad
\Lambda_N^2=\kappa G_N,\qquad
\kappa>40\beta/k.\label{app:spin-eq-scales}
\end{equation}
This choice keeps $G_N=o(\log N)$, so the reference decay dominates the modular factor, while $\log N/G_N^{5/4}\to0$ suppresses the filter error (any exponent in $(4/5,1)$ suffices for fixed $\epsilon>0$).

At $T=(1-\epsilon)[\beta/(2\pi)]\log N$, the reference factor in Eq.~\eqref{app:spin-eq-uniform-window} is $N^{-\epsilon/4}$. Since $\mathcal M_\chi(\beta G/4)^5$ is a fixed polynomial in $G$ times $e^{5\beta G/2}$,
\begin{equation}
\begin{aligned}
T\mathcal M_\chi(\beta G_N/4)^5N^{-\epsilon/4}
&=\exp\bigl[-(\epsilon/4)\log N\\
&\qquad+O(L_N^{9/10})+O(\log L_N)\bigr]\to0.
\end{aligned}
\label{app:spin-eq-reference-decay}
\end{equation}
For the clipping tail, $\lambda_{\mathrm{ref}}=k/8$ and
\begin{equation}
T\mathcal M_\chi(\beta G_N/4)^5\sqrt{\varepsilon_{\Lambda_N}^{\mathrm{ref}}}
\le\operatorname{poly}(L_N)
e^{-(\lambda_{\mathrm{ref}}\kappa-5\beta)G_N/2}\to0.
\label{app:spin-eq-clipping-decay}
\end{equation}
The filter error is
\begin{equation}
\frac{T\Lambda_N\sqrt{1+\Lambda_N}}{G_N^2}
=O(L_N/G_N^{5/4})=O(L_N^{-1/8}).\label{app:spin-eq-filter-decay}
\end{equation}
The bad-bath and source-site terms also vanish. Moreover, $\Lambda_N^2/\log\log N\to\infty$, as required by Eq.~\eqref{app:spin-eq-stationary-clipping}. Thus $p_{\rho_K,H_\Lambda}^{\mathrm r}\to0$ on this window, both for uniform target weights and for a fixed target under the permutation hypothesis.

The same choice allows $\epsilon_N=L_N^{-1/20}$. The exponent in Eq.~\eqref{app:spin-eq-reference-decay} then becomes $-\tfrac14L_N^{19/20}+O(L_N^{9/10})+O(\log L_N)$, which tends to minus infinity. All preceding comparisons are uniform for $T_N\le[\beta/(2\pi)]\log N$.

To restore units, use
\[
\widehat K=K/J,\quad\widehat V=V/J,\quad
\widehat\Xi=\Xi/J^2,\quad
\widehat\beta=\beta J,\quad\widehat t=Jt.
\]
Equation~\eqref{app:spin-eq-scales} becomes
\begin{equation}
G_N=JL_N^{9/10},\qquad
\Lambda_N^2=\kappa J^2L_N^{9/10},\qquad
\kappa>40\beta J/k,\label{app:spin-eq-physical-scales}
\end{equation}
and the physical time window remains $(1-\epsilon)[\beta/(2\pi)]\log N$.

\subsection{Return to the same-sample Gibbs state}\label{app:spin-sec-completion}

We now prove Theorem~\ref{thm:spins}. Start with the convergence $p_{\rho_K,H_\Lambda}^{\mathrm r}(w)\to0$ established above. In the following implications every displayed quantity tends to zero on the same time window.

First, Lemma~\ref{app:spin-lem-state-comparison} uses $K$ for the spectral conversion and $S(\rho_\Lambda\Vert\rho_K)$ for the state comparison, giving
\[
p_{\rho_K,H_\Lambda}^{\mathrm r}(w)
\Longrightarrow p_{\rho_K,H_\Lambda}^{\mathrm u}(w)
\Longrightarrow p_{\rho_\Lambda,H_\Lambda}^{\mathrm u}(w)
\Longrightarrow p_{\rho_\Lambda,H_\Lambda}^{\mathrm r}(w).
\]
The stationary clipping comparison, Eq.~\eqref{app:spin-eq-stationary-clipping}, gives $p_{\rho_g,H_g}^{\mathrm r}(w)\to0$. Applying the same state-comparison lemma with $H_g$ as the spectral generator in $\rho_g$, and the reverse entropy $S(\rho_K\Vert\rho_g)$, yields
\[
p_{\rho_g,H_g}^{\mathrm r}(w)
\Longrightarrow p_{\rho_g,H_g}^{\mathrm u}(w)
\Longrightarrow p_{\rho_K,H_g}^{\mathrm u}(w)
\Longrightarrow p_{\rho_K,H_g}^{\mathrm r}(w).
\]
Lemma~\ref{app:spin-lem-lindeberg} now changes the source law at fixed $\rho_K$, giving $p_{\rho_K,H_o}^{\mathrm r}(w)\to0$. Finally, the spectral conversion with $K$ and the entropy $S(\rho_o\Vert\rho_K)$ give
\[
p_{\rho_K,H_o}^{\mathrm r}(w)
\Longrightarrow p_{\rho_K,H_o}^{\mathrm u}(w)
\Longrightarrow p_{\rho_o,H_o}^{\mathrm u}(w)
\Longrightarrow p_{\rho_o,H_o}^{\mathrm r}(w).
\]
Here $H_o$ has its original bounded-label distribution and $\rho_o=e^{-\beta H_o}/\Tr e^{-\beta H_o}$ is the Gibbs state of that same realization. The bad-bath contribution was included in Proposition~\ref{app:spin-prop-window} before these comparisons.

For uniform weights this proves $p_{\rho_o,H_o}^x\to0$, $x\in\{\mathrm r,\mathrm u\}$. Markov's inequality gives, for every fixed $c>0$,
\begin{equation}
\Pr\left(\sup_{0\le t\le T_N}\frac1N\sum_jC_{ij}^x(t)\ge c\right)
\le c^{-1}p_{\rho_o,H_o}^x\longrightarrow0.
\label{app:spin-eq-final-crossing}
\end{equation}
This is Eq.~\eqref{eq:main-spins}. Under the permutation hypothesis, Eq.~\eqref{app:spin-eq-fixed-reference} and Proposition~\ref{app:spin-prop-window} give the same initial convergence for $w_j=\delta_{j,j_0}$. All comparisons above hold for these weights, so Eq.~\eqref{app:spin-eq-final-crossing} also holds with the target average replaced by $C_{ij_0}^x(t)$. This proves the fixed-pair assertion.

\section{Numerical evaluation of the static defects}\label{app:sec-numerics}

\emph{SYK model.} Figure~\ref{fig:double}(c) evaluates the SYK Hamiltonian $H_N$ of Eq.~\eqref{eq:syk} on $\mathbb C^D$ ($D=2^{N/2}$) at $\beta J=2$, with equal source and target weights $1/N$ on all $N$ Majorana operators $\gamma_i$ ($N^2$ ordered pairs, including $i=j$) and $\tau=-1$. Because every quartic product $\Gamma_E$ is even, $H_N$ decomposes into two fermion-parity blocks of dimension $2^{N/2-1}$. Both sectors share the single sample partition function $Z=\Tr e^{-\beta H_N}$, and the disorder average is taken after normalizing each sample's Gibbs state $\rho=e^{-\beta H_N}/Z$. Since $N^{-1}\sum_{i=1}^N g_i=N^{-1}\sum_{i=1}^N\Tr(R\gamma_iR\gamma_i)=m$, the mean single-source defect obeys $\bar\delta=N^{-1}\sum_{i=1}^N\delta_i=m^2+\sigma-F_2(0)=\Delta+\sigma-\sigma^2$, with $\norm{\cJ(R)}_2^2=m^2+\sigma^2$.

For $N=8,10,\dots,24$, we retain the full spectra of both parity blocks ($64$ independent disorder samples per size through $N=20$, and $32$ at $N=22,24$) and evaluate $\cJ(R)$ and $F_2(0)=-N^{-2}\sum_{i,j=1}^N\Tr(y\gamma_iy\gamma_jy\gamma_iy\gamma_j)$ by exact sums over all $N$ Majorana modes. For $N=26,28,30,32,34,36$ ($16$ samples per size), we preassemble both parity blocks of $H_N$ in compressed sparse row format (checking Hermiticity and $\Tr(H_N^2)/D=(6J^2/N^3)\sum_{|E|=4}g_E^2$) and estimate their extremal eigenvalues using Krylov iterations, expanding the spectral interval $[E_{\min},E_{\max}]$ by a $5\%$ safety margin on each side. A common energy shift $H_N\mapsto H_N-E_{\min}I$ is used in the propagation and cancels from normalized correlators. Imaginary-time actions $e^{-u(H_N-E_{\min}I)}$ ($u=\beta/4,\beta/2$) are applied to batches of probe vectors via Chebyshev polynomial recurrence~\cite{Weisse2006Kernel} on $[E_{\min},E_{\max}]$ with modified-Bessel tail tolerance $2\times10^{-13}$ (a shared recurrence evaluates $\beta J\in\{1,2,4\}$, with $\beta J=2$ displayed).

The larger-$N$ calculations use $64$ independent $\mathbb Z_4$ random-phase trace vectors~\cite{Iitaka2004Random} $r\in\{1,\ii,-1,-\ii\}^D/\sqrt D$ per sample, processed in batches of $8$. Setting $q=e^{-\beta H_N/2}r$ gives $\E_r\norm{q}^2=Z/D$, while the exact $N$-mode channel vector $c=N^{-1}\sum_{i=1}^N\gamma_ie^{-\beta H_N/2}\gamma_ir$ and centered residual $d=c-q$ determine $m=1+\E_r[\Re(q^\dagger d)]/\E_r\norm{q}^2$ and $\sigma^2=\E_r\norm{d}^2/\E_r\norm{q}^2-(m-1)^2$. To evaluate $F_2(0)$ without $O(N^2)$ propagations, for each $r$ we draw an isotropic orthonormal pair of real unit vectors $a,b\in\mathbb R^N$ ($a\cdot b=0$) and set $\gamma_u=\sum_{i=1}^N u_i\gamma_i$ ($u\in\{a,b\}$); decomposing an independent unit vector in $\mathbb R^N$ along $a$ and its orthogonal complement yields the exact finite-$N$ stratification
\begin{equation}
\begin{aligned}
F_2(0)={}&-\left(1-\frac1N\right)\E_{a,b}\Tr\bigl(y\gamma_ay\gamma_by\gamma_ay\gamma_b\bigr)\\
&-\frac1N\E_a\Tr\bigl((y\gamma_a)^4\bigr),
\end{aligned}
\label{app:eq-syk-crossed-estimator}
\end{equation}
whose unnormalized traces are evaluated on $r$ by batched $e^{-\beta H_N/4}$ propagations and divided by the same-sample estimate of $Z/D$. Delete-group jackknife resampling across the trace probes quantifies within-sample ratio bias and probe standard errors, and paired comparisons against full-spectrum exact diagonalization at $N=16,20,24$ (two disorder samples per size and $256$ trace probes per temperature) validate the stochastic estimator within statistical uncertainty. At $N=24$ and $\beta J=2$, four additional independent runs of $256$ probes per disorder sample give pooled two-sample estimates of $\Delta$ and $\bar\delta$ that differ from exact diagonalization by $0.68$ and $0.94$ probe standard errors, respectively.

For each observable, the plotted standard error takes the larger of the empirical standard error across independent disorder samples and the propagated within-sample probe standard error. Dashed $N^{-1/2}$ and $N^{-1}$ guides have fixed exponents with amplitudes anchored at the $N=24$ means; they are slope references rather than fitted curves.

\emph{Random spins.} Figure~\ref{fig:spins}(b) uses the ensemble of Eq.~\eqref{eq:spins} with $k=d=2$ and
\begin{equation}
H_{ij}=\frac13\bigl(\epsilon_{ij}Z_iZ_j+\eta_{ij}X_i+\xi_{ij}X_j\bigr),\qquad i<j,
\label{app:eq-numerical-spin-model}
\end{equation}
where all signs are independent and equiprobable. Each label is centered and has norm at most one. The source and target probes are $Z_i$. For each sample, conjugation by a product of $Z_i$ maps the Hamiltonian to $H_Z-\sum_i h_iX_i$, where $H_Z$ is diagonal and $h_i\ge0$. This conjugation leaves the probes fixed. In this basis $y=\rho^{1/4}$ has nonnegative matrix entries, and the four imaginary-time cuts have the normalized weight
\begin{equation}
p_{\mathrm{wl}}(s_0,s_1,s_2,s_3)
=y_{s_0s_1}y_{s_1s_2}y_{s_2s_3}y_{s_3s_0}.
\label{app:eq-four-cut-weight}
\end{equation}
Its sum is $\Tr y^4=1$. Write $q(s,s') = N^{-1}\sum_i z_i(s)z_i(s')$, where $z_i(s) \in \{\pm 1\}$ is the value of $Z_i$ in configuration $s$, and set $q_0 = q(s_0, s_2)$ and $q_1 = q(s_1, s_3)$. Expanding the traces in this basis gives
\begin{equation}
\begin{aligned}
m&=\langle q_0\rangle_{\mathrm{wl}},\qquad
\norm{\cJ(R)}_2^2=\langle q_0^2\rangle_{\mathrm{wl}},\\
F_2(0)&=\langle q_0q_1\rangle_{\mathrm{wl}},
\end{aligned}
\label{app:eq-worldline-moments}
\end{equation}
where the brackets denote the worldline average of the same sample. A quarter-period shift interchanges $q_0$ and $q_1$, so
\begin{equation}
\Delta=\frac12\langle(q_0-q_1)^2\rangle_{\mathrm{wl}}.
\label{app:eq-worldline-defect}
\end{equation}
Thus $\Delta$ is measured as a positive squared difference. The single-source average follows from $N^{-1}\sum_i g_i=m$:
\begin{equation}
\sigma^2=\langle q_0^2\rangle_{\mathrm{wl}}-m^2,
\qquad \bar\delta=\Delta+\sigma-\sigma^2.
\label{app:eq-worldline-single}
\end{equation}

We sample the periodic worldlines using full-worldline conditional heat-bath updates~\cite{Krzakala2008Path}. For $L$ slices (with $L$ a multiple of $4$, so that $\rho_L^{1/4} = S_L^{L/4}/(\Tr\,S_L^L)^{1/4}$), the symmetric transfer step is $S_L=e^{-\beta H_Z/(2L)}e^{(\beta/L)\sum_i h_iX_i}e^{-\beta H_Z/(2L)}$. The finite-step state is $\rho_L=S_L^L/\Tr S_L^L$. Equations~\eqref{app:eq-worldline-moments}--\eqref{app:eq-worldline-single} hold with $\rho_L$ in place of $\rho$, and we average over all time origins. Conditional on the other spins, a spin's entire periodic trajectory is sampled by two-state transfer matrices. A sweep updates every spin once in a random order.

The data use $\beta J=4$, $N=16,32,64,128,256,512,1024$ and 32 independent disorder samples at each size. Each sample has two independent chains, initialized with uniformly aligned spins and with random time-independent spins. After 1000 equilibration sweeps, each chain records 4000 sweeps in blocks of 100. Block jackknife estimates give $\sigma$ and $\bar\delta$, and the plotted standard errors are computed across the independent disorder samples. The main data use $L=64$. Calculations at $L=128,256$ and longer runs at $N=64,256,1024$ check the discretization and sampling errors. In matched-sample comparisons, increasing $L$ from $64$ to $128$ or $256$ changes $\Delta$ and $\bar\delta$ by at most $1.09$ and $2.07$ combined Monte Carlo standard errors, respectively. Direct matrix traces, finite-step transfer matrices and the independent-spin limit provide further checks. The disorder average is taken after the normalized thermal average of each sample.

\emph{Coexistence.} Figure~\ref{fig:spins}(c) uses the uniform Majorana averages of Appendix~\ref{app:sec-coexist}. We sum over the $n+1$ magnetization sectors, whose Gibbs weights at $\beta_c$ are proportional to $\binom n\ell\exp[\beta_cJn(2\ell/n-1)^4]$. Within a sector, the single-spin weights are $\ell/n$ and $1-\ell/n$. The two-spin weights are $\ell(\ell-1)/[n(n-1)]$, $(n-\ell)(n-\ell-1)/[n(n-1)]$ and $\ell(n-\ell)/[n(n-1)]$ for $(+,+)$, $(-,-)$ and each mixed configuration, respectively. These weights evaluate the overlaps $a_n,b_n,f_n$. We compute $\Delta$ from Eq.~\eqref{app:eq-coexist-exact}, accumulating $b_n-f_n$ with the positive expression in Eq.~\eqref{app:eq-coexist-positive}, and obtain the single-source defect from
\begin{equation}
\begin{aligned}
\sigma_n^2&=\frac1n+\left(1-\frac1n\right)b_n-a_n^2,\\
\delta&=\Delta+\sigma_n-\sigma_n^2.
\end{aligned}
\label{app:eq-coexist-finite-defect}
\end{equation}

\clearpage
\interlinepenalty=10000
\bibliography{references}
\end{document}